\documentclass[12pt]{article}
\usepackage{graphicx}
\usepackage{xcolor}
\usepackage{amsmath,amssymb,amsthm}
\numberwithin{equation}{section} 

\usepackage[top=3cm,bottom=3cm, left=3cm,right=3cm]{geometry}
\usepackage[breaklinks=true,linktocpage=true]{hyperref}
\usepackage{ascmac}
\usepackage{cite}

\newcommand{\link}{\,{\rm Link}\,}

\newcommand{\im}{\,{\rm Im}\,}
\renewcommand{\ker}{\,{\rm Ker}\,}

\newcommand{\ig}{\includegraphics}
\newcommand{\trr}{\triangleright}
\newcommand{\os}{\overset}

\newcommand{\corr}{\leftrightarrow}

\newcommand{\vevs}[1]{\langle #1 \rangle}

\newcommand{\er}[1]{Eq.~\eqref{#1}}
\newcommand{\ers}[1]{Eqs.~\eqref{#1}}
\newcommand{\qtq}[1]{\quad\text{#1}\quad}

\newcommand{\bb}{\mathbb}

\newcommand{\hph}{\hphantom}

\newcommand{\bs}{\boldsymbol}

\newcommand{\fr}{\frac}

\newcommand{\ul}{\underline}
\newcommand{\der}{\partial}

\renewcommand{\(}{\left(}
\renewcommand{\)}{\right)}

\renewcommand{\frak}{\mathfrak}

\newcommand{\wed}{\wedge}
\newcommand{\bmx}{\left(\begin{matrix}}
\newcommand{\emx}{\end{matrix}\right)}
\newcommand{\mtx}[1]{\bmx #1 \emx}

\begin{document}

\allowdisplaybreaks[4]
\begin{titlepage}
\hfill KEK-TH-2254, J-PARC-TH-0225
\vspace{-1em}
\def\thefootnote{\fnsymbol{footnote}}%
   \def\@makefnmark{\hbox
       to\z@{$\m@th^{\@thefnmark}$\hss}}%
 \vspace{3em}
 \begin{center}%
  {\Large
Global 3-group symmetry and 
't Hooft anomalies
\par
in axion electrodynamics
 \par
   }%
 \vspace{1.5em} 
  {Yoshimasa Hidaka,\footnote{hidaka@post.kek.jp}${}^{a,b,c}$
    Muneto Nitta,\footnote{nitta@phys-h.keio.ac.jp}${}^{d}$
and
    Ryo Yokokura\footnote{ryokokur@post.kek.jp}${}^{a,d}$
   \par
   }%
  \vspace{1em} 
${}^a${\small\it KEK Theory Center, Tsukuba 305-0801, Japan}
\par
${}^b$
{\small\it Graduate University for Advanced Studies (Sokendai), Tsukuba 305-0801, Japan}
\par
${}^c$
{\small\it RIKEN iTHEMS, RIKEN, Wako 351-0198, Japan}
\par
${}^d${\small\it Department of Physics \& Research and Education Center for Natural Sciences,
\par 
Keio University, Hiyoshi 4-1-1, Yokohama, Kanagawa 223-8521, Japan}
 \end{center}%
 \par
\vspace{1.5em}
\begin{abstract}
We investigate a higher-group structure of massless axion electrodynamics in $(3+1)$ dimensions.
By using the background gauging method, 
we show that the higher-form symmetries necessarily have
a global semistrict 3-group (2-crossed module) structure, and
exhibit 't Hooft anomalies of the 3-group.
In particular, we find a cubic mixed 't Hooft anomaly between 0-form and 1-form symmetries, 
which is specific to the higher-group structure.
\end{abstract}
\end{titlepage}
 \setcounter{footnote}{0}%
\def\thefootnote{$*$\arabic{footnote}}%
   \def\@makefnmark{\hbox
       to\z@{$\m@th^{\@thefnmark}$\hss}}%
\tableofcontents

\section{Introduction}

Axion electrodynamics has been widely investigated 
from particle physics and cosmology to condensed matter physics.
In particle physics, the axion has been introduced to resolve 
the strong CP problem, and it is a candidate of cold dark matter~\cite{Peccei:1977hh,Weinberg:1977ma,Wilczek:1977pj,Dine:1981rt,Zhitnitsky:1980tq,Kim:1979if,Shifman:1979if}
(see also Refs.~\cite{Kim:1986ax,Dine:2000cj,Peccei:2006as,Kawasaki:2013ae}
as a review).
In condensed matter physics, the axion electrodynamics can 
describe magneto-electric responses in topological matter \cite{Wilczek:1987mv,Qi:2008ew,Essin:2008rq} (see also Ref.~\cite{Hasan:2010xy} as a review).

One of the characteristic features of the axion electrodynamics 
is a topological coupling between the axion and photon.
This coupling originates from the chiral anomaly of massive 
Dirac fermions coupled with them, which modifies the electric Gauss law and the Maxwell-Amp\`ere 
law~\cite{Fischler:1983sc,Sikivie:1984yz,Wilczek:1987mv,Qi:2008ew,Essin:2008rq}.
Furthermore, this topological coupling leads to 
non-trivial effects on extended objects in the axion electrodynamics.
There are spatially or temporally extended objects such as 
magnetic monopoles, axionic domain walls, and axionic strings~\cite{Sikivie:1982qv,Vilenkin:1982ks,Davis:1986xc,Teo:2010zb}.
One of the characteristic effects is the Witten effect \cite{Witten:1979ey}
for the axion due to the modification of the electric Gauss law~\cite{Fischler:1983sc}.
If an axionic domain wall encloses a magnetic monopole,
electric charges are induced on the axionic domain wall~\cite{Sikivie:1984yz,Wilczek:1987mv}.
This domain wall enclosing the magnetic monopole is called 
a monopole bag~\cite{Kogan:1992bq,Kogan:1993yw}.

Another characteristic effect is the so-called anomalous Hall 
effect for the axion
due to the modification of the Maxwell-Amp\`ere law~\cite{Sikivie:1984yz,Wilczek:1987mv,Essin:2008rq,Qi:2008ew,Ferrer:2015iop,Yamamoto:2015maz,Ferrer:2016toh,Qiu:2016hzd,Bednik:2016phb}.
If a domain wall is placed in the electric flux background, 
electric currents are induced on the domain wall
whose direction is perpendicular to the electric flux.
This effect also arises in the presence of the axionic strings 
in the electric flux background~\cite{Qi:2008ew,Teo:2010zb,Wang:2012bgb}.
There are induced electric currents whose directions are perpendicular
to both of the electric flux and the gradient of the axion.
Since the electric current flows to the axionic strings,
this effect is related to the so-called anomaly inflow mechanism
of axionic strings~\cite{Callan:1984sa,Naculich:1987ci} (see also Refs.~\cite{AlvarezGaume:1983cs,Kaplan:1987kh,Manohar:1988gv,Harvey:1988in}). 
By these non-trivial phenomena of extended objects due to 
the topological coupling, 
the axion electrodynamics has also been investigated 
as a simple model of string theory~\cite{Townsend:1993wy,Izquierdo:1993st,Harvey:2000yg}.

What are the underlying structures for the above peculiar effects for
the extended objects?  One of the candidates is the notion of
symmetries, giving us non-perturbative, model-independent, and universal
understandings of physical phenomena.  In fact, the chiral symmetry and
its anomaly in the axion electrodynamics are essential notions to
understand these effects. However, its symmetry transformation acts on
only local fields rather than extended objects.  If we try to understand
the effects on extended objects by symmetries, it is plausible to
consider symmetries whose transformations act on extended objects.

Recently, the notion of symmetries has been generalized to 
ones for extended objects, 
called higher-form 
symmetries~\cite{Banks:2010zn,Kapustin:2014gua,Gaiotto:2014kfa}
(see also related topics \cite{Batista:2004sc,Pantev:2005zs,Pantev:2005wj,Pantev:2005rh, Nussinov:2006iva, Nussinov:2008aa, Nussinov:2009zz, Nussinov:2011mz,Distler:2010zg}).
For $p$-form symmetries ($p= 0,1,...,D$), 
charged objects are $p$-dimensional, where 
$D$ is the spacetime dimensions.
Symmetry generators acting on the 
charged objects are $(D-p-1)$-dimensional topological objects,
while 
the conventional symmetries can be understood as 0-form symmetries, 
since they act on local 0-dimensional objects, i.e., local fields.
Such higher-form symmetries give us new aspects of modern physics.
For example, we can understand photons in the pure Maxwell theory 
as Nambu-Goldstone bosons~\cite{Kovner:1992pu,Gaiotto:2014kfa,Lake:2018dqm}.
Such an interpretation has been generalized to 
non-relativistic cases as well~\cite{Yamamoto:2015maz, Ozaki:2016vwu, Sogabe:2019gif, Hidaka:2020ucc}.
Here, a charged object is a 1-dimensional Wilson loop
whose vacuum expectation value is finite in the Coulomb phase.
Another application of higher-form symmetries is that 
Abelian topologically ordered 
phases~\cite{Wen:1989iv,Wen:1990zza,Wen:1991rp,Hansson:2004wca}
can be regarded as broken phases of higher-form symmetries~\cite{Nussinov:2006iva,Nussinov:2009zz,Gaiotto:2014kfa},
in which the charged object is a worldline of an anyon.
One can further classify phases of gauge theories 
based on those symmetries~\cite{Gaiotto:2017yup,Gaiotto:2017tne,Tanizaki:2017qhf,Tanizaki:2017mtm,Komargodski:2017dmc,Hirono:2018fjr,Hirono:2019oup,Hidaka:2019jtv,Anber:2019nze,Misumi:2019dwq,Anber:2020xfk,Anber:2020gig,Furusawa:2020kro}.
Thus, it becomes possible to understand 
phenomena of extended objects in terms of higher-form symmetries.

A more elegant description of higher-form symmetries can be 
given by so-called {\it higher-groups}~\cite{Sharpe:2015mja},
which are extensions of conventional groups describing 
ordinary (0-form) symmetries.
Here, higher $n$-groups are given by a set of $n$ groups
with maps between them.
For example, a 2-group is given by a set of two groups $(G, H)$, 
a map $H \to G$, and actions of $G$ on $G$ and $H$.
The higher-groups have been recently applied to 
various aspects of theoretical physics, 
such as higher gauge theories where charged objects are 
extended~\cite{Baez:2002jn, Baez:2003fs, Baez:2004in, Baez:2005qu, Martins:2009evc, Baez:2010ya, Ho:2012nt, Saemann:2013pca, Wang:2013dwa, Soncini:2014zra, Jurco:2014mva, Jurco:2016qwv, Samann:2019nuj,Radenkovic:2019qme},
effective theories of gapped or gapless phases of topological 
matter~\cite{Kapustin:2013uxa, Kapustin:2013qsa, Bhardwaj:2016clt, deAlmeida:2017dhy, Delcamp:2018wlb, Delcamp:2019fdp, Thorngren:2020aph, Wen:2018zux,Hsin:2019fhf,Hsin:2020nts},
deformation of current algebra 
for tensorial currents~\cite{Cordova:2018cvg}
and quantum chromodynamics~\cite{Tanizaki:2019rbk}.
Therefore,
higher-form symmetries and higher-group structures 
may provide us with new understandings of the effects 
in the axion electrodynamics.

In the previous paper by the present authors~\cite{Hidaka:2020iaz},
it was shown that the axion electrodynamics possesses 
a 0-form symmetry, an electric 1-form symmetry, 
a magnetic 1-form symmetry, and a 2-form symmetry.
Here, the 0-form symmetry is a shift symmetry of the axion.
The electric and magnetic 1-form symmetries 
are related to conservation laws of electric and magnetic fluxes, 
respectively.
The 2-form symmetry is the conservation law of the 
winding number of axionic strings.
Furthermore, we found that the 
higher-form symmetries can have a semistrict 3-group (or 2-crossed module) 
structure by analyzing the 
correlation functions of the symmetry generators.
We hereafter abbreviate the semistrict 3-group 
to the 3-group for simplicity. 
The 3-group is a set of three groups $(G,H,L)$ with maps between 
them~\cite{CONDUCHE1984155}.
One of the particular properties of the 3-group is
the presence of actions of $G$ on $G$, $H$, $L$.
Another property is that there must be a map from 
two elements in $H$ 
to $L$, which is called the Peiffer lifting.

In this paper, we investigate the higher-group structure 
in the $(3+1)$-dimensional axion electrodynamics in more detail 
by introducing  background gauge fields corresponding to the higher-form symmetries.
The background gauging enables us to describe the 't Hooft anomalies,
which are obstructions to dynamical gauging of global 
symmetries~\cite{tHooft:1979rat,Frishman:1980dq,Coleman:1982yg}.
Since the 't Hooft anomalies constrain possible phase structure of a given theory~\cite{Gaiotto:2017yup,Gaiotto:2017tne,Tanizaki:2018wtg} 
and describe anomalous phase factors in 
correlation functions of symmetry 
generators~\cite{Gaiotto:2014kfa,Benini:2018reh},
it is important to determine the 't Hooft anomalies 
for the axion electrodynamics.

There are at least two methods to establish the background gauging.
One is to establish gauged actions 
with couplings between background gauge fields 
and symmetry 
generators~\cite{Gaiotto:2014kfa,Cordova:2018cvg,Tanizaki:2019rbk}.
We show that 
a naive gauging violates the invariance under the 
transformations of the axion and photon, which 
should be avoided~\cite{Cordova:2018cvg,Benini:2018reh}.
The absence of apparent inconsistencies requires
modifications of the gauge transformation laws 
corresponding to the 3-group structure.
We should note that this gauging procedure is based on 
the higher-form symmetries and the gauge invariance, 
but it does not a priori assume the 3-group structure.

Next, we show that the
above background gauging with the modified gauge 
transformations can be sufficiently described by 
a 3-group gauge theory formulated 
in Refs.~\cite{Martins:2009evc,Saemann:2013pca,Wang:2013dwa}.
To this end, we establish the 3-group gauge theory 
for the axion electrodynamics.
Here, we assume the global 3-group structure in the axion electrodynamics. 
The gauge transformation laws of the background gauge fields 
are determined by the basic quantities of the 3-group 
rather than the gauge invariance of the axion and photon.
By comparing the gauge transformation laws and field strengths,
we confirm that these two independent methods 
result in the same physics.

As a consequence of the background gauging, 
we determine 't Hooft anomalies of the higher-form symmetries.
We find that there are three kinds of the 't Hooft anomalies.
One is a mixed 't Hooft anomaly for the axion, 
which prevents us from a simultaneous gauging of 
the 0- and 2-form symmetries.
The second is for the photon, 
which forbids a simultaneous gauging of the two 1-form symmetries.
These two anomalies are extensions of 
previously known anomalies for the axion and photon in 
the absence of the topological coupling~\cite{Gaiotto:2014kfa}.
The third is a cubic 't Hooft anomaly, 
that is so-called 2-group anomaly~\cite{Benini:2018reh},
which implies the obstruction to the simultaneous gauging 
of the 0-form and the 1-form symmetries.

This paper is organized as follows.
In section \ref{hf}, we review the axion electrodynamics 
and higher-form symmetries in this system in detail.
In section \ref{bganom}, we consider the 
background gauging of the higher-form symmetries 
that are consistent with the gauge invariance for the dynamical fields.
We further determine the 't Hooft anomalies of 
the higher-form symmetries.
In section \ref{bg3g}, we discuss the other gauging procedure, 
which is based on the 3-group gauge theory.
We show that both of the gauging methods give rise to the same results in the axion electrodynamics.
Finally, we summarize this paper in section \ref{sum}.
We give four appendices.
In appendix \ref{dual},
we give explicit forms of the 't Hooft loop and worldsheet of the 
axionic strings which are charged objects of the magnetic 1-form
symmetry and the 2-form symmetry, respectively.
We show detailed derivations of correlation 
functions used in this paper in appendix \ref{corr}.
In appendix \ref{3g}, we review the notions
of the 3-group, the Lie algebra of the 3-group, and the 
3-group gauge theories.
We also give an intuitive and diagrammatic expression 
of the 3-group in appendix \ref{diag}.

\section{Higher-form symmetries in axion electrodynamics\label{hf}}
In this section, we review the higher-form symmetries 
in $(3+1)$-dimensional axion electrodynamics~\cite{Hidaka:2020iaz}
in detail.
In particular, we carefully discuss the symmetry groups 
for the higher-form symmetries.
After giving an action of the massless axion electrodynamics,
we show the existence of the higher-form symmetries
by the equations of motion (EOM) and Bianchi 
identities of the axion and photon.
We also present the charged objects, symmetry generators, 
and symmetry groups for the higher-form symmetries.

\subsection{Action}
Here, we give an action of the massless 
electrodynamics, in which 
we regard the photon as a gauge field of $U(1)$ gauge symmetry,
and the axion as a circle valued pseudo-scalar field.
The action has the form~\cite{Wilczek:1987mv}
\begin{equation}
 S = -\int_{M_4}\( \fr{v^2}{2}
 |d\phi|^2
 +\fr{1}{2e^2} |da|^2
-\fr{N}{8\pi^2} \phi \, da \wed da
\).
\label{200628.1231}
\end{equation}
Here, $\phi$ is the axion, $a$ the photon, 
$v$ a decay constant of the axion, 
$e$ a coupling constant of the photon,
and $N$ an integer.
$|d\phi|^2$ and $|da|^2$ denote 
$d\phi \wed \star d\phi$ and  $da\wed \star da$,
where $\star$ is the Hodge star operator.
We refer to $M_4$ as a $(3+1)$-dimensional spacetime manifold, 
e.g., the Minkowski spacetime.
Throughout this paper, we assume that $M_4$ is a spin manifold 
such that the axion photon coupling term is well-defined.
The axion has a $2\pi$ periodicity 
at each point ${\cal P}$ in the spacetime:
\begin{equation}
 \phi({\cal P}) +2\pi \sim \phi({\cal P}).
\label{200217.1131}
\end{equation}
We have assumed that the mass-dimension of the scalar field is
normalized as $0$.
We regard the periodicity as a redundancy of the axion.
In other words, the redundancy can be understood as 
a $(-1)$-form gauge symmetry~\cite{Kapustin:2014gua,Cordova:2019jnf,Cordova:2019uob}.
An invariant operator under the redundant transformation in \er{200217.1131}
is 
a point operator,
\begin{equation}
I (q_{\phi E}, {\cal P}) := e^{i q_{\phi E} \phi ({\cal P})}, 
\label{200627.1251}
\end{equation}
rather than $\phi({\cal P})$ itself. 
Here, the invariance requires that $q_{\phi E}$ is an integer.
Although $I (q_{\phi E}, {\cal P})$ is a single-valued function,
$\phi ({\cal P})$ can be a multi-valued function on a closed loop  ${\cal C}$ with the winding number,
\begin{equation}
 \int_{\cal C} d\phi \in 2\pi \bb{Z}.
\label{191027.0030}
\end{equation}
Physically, the nonvanishing winding number implies the existence of a string object with a topologically quantized charge.

The photon is described by a $U(1)$ 1-form gauge field $a$,
which is transformed as 
\begin{equation}
 a \to a + d\lambda.
\label{200628.1244}
\end{equation}
Here, $\lambda$ is a $U(1)$ gauge parameter,
which satisfies $\lambda ({\cal P})+2\pi \sim \lambda(\cal P)$.
Since the gauge parameter is circle valued rather than $\bb{R}$ valued, 
the gauge parameter can have a winding number 
$\int_{\cal C} d\lambda \in 2\pi \bb{Z}$.
Such a transformation with a nonvanishing winding number is called 
a large gauge transformation.
An operator that is invariant under the large gauge transformation is 
a Wilson loop,
\begin{equation}
 W(q_{aE},{\cal C}) := e^{i q_{a E} \int_{\cal C} a},
\label{200627.2309}
\end{equation}
where a charge $q_{aE}$ should be an integer.
When ${\cal C}$ is a boundary of a surface ${\cal S_{C}}$, we can rewrite the Wilson loop by using the Stokes theorem as
\begin{equation}
  W(q_{aE},{\cal C})  = e^{i q_{aE} \int_{\partial\cal S_C} a}=e^{i q_{aE} \int_{\cal S_C} da}=e^{i q_{aE} \int_{\cal S_C} f},
\end{equation}
where $f=da$ is the field strength.
In general, the field strength is a globally well-defined closed two-form that may not be the exact form,
and it is quantized on a closed surface ${\cal S} $ as
\begin{equation}
\int_{\cal S} f \in 2\pi \bb{Z}.
\end{equation}
It physically means that there can be a magnetic 
monopole in the interior of 
${\cal S}$.  
This is nothing but the Dirac quantization condition.
Throughout this paper, we simply denote the field 
strength as $da$ and use the Dirac quantization condition 
on a closed surface as
\begin{equation}
\int_{\cal S} da \in 2\pi \bb{Z},
\end{equation}
bearing in mind that $a$ is not globally well-defined.
\subsection{Higher-form symmetries}

Here, we review
higher-form symmetries in this system~\cite{Hidaka:2020iaz}.
In the following, we show that 
there are four kinds of the higher-form symmetries:
$\bb{Z}_N$ 0-form,
electric $\bb{Z}_N$ 1-form, 
magnetic $U(1)$ 1-form,
and $U(1)$ 2-form symmetries.
They are associated with the EOM
or Bianchi identities of the axion and photon.

\subsubsection{{$\bb{Z}_N$} 0-form symmetry}

First, we consider the $\bb{Z}_N$ 0-form symmetry,
which is a shift symmetry of the axion.
The EOM of the axion, $v^2 d \star d\phi - \fr{N}{8\pi^2} da \wed da =0$
lead to the following closed current 3-form and 
conserved charge,
\begin{equation}
 j_{\phi E} := -v^2 \star d \phi - \fr{N}{8\pi^2} a \wed da,
\quad
Q_{\phi E}({\cal V}) := \int_{\cal V} j_{\phi E}.
\label{eq:QphiE}
\end{equation}
Here, ${\cal V}$ is a 3-dimensional closed subspace.
The charge $Q_{\phi E} ({\cal V})$ is topological: it is 
invariant under a small deformation 
${\cal V} \to {\cal V} \cup \der \Omega_0$
with a four-dimensional subspace $\Omega_0$, because of 
the Stokes theorem.
A gauge invariant observable given by the
 current $j_{\phi E}$ 
is the following unitary object,
\begin{equation}
 U_{\phi E} (e^{2\pi i n_\phi/N}, {\cal V})
 := e^{\fr{2\pi i n_\phi}{N} Q_{\phi E}({\cal V})},
\end{equation}
where $e^{2\pi i n_\phi/N} \in \bb{Z}_N$ parameterizes 
the topological object.
This object is topological meaning that 
\begin{equation}
 U_{\phi E} (e^{2\pi i n_\phi/N}, {\cal V} \cup \der \Omega_0 )
= 
 U_{\phi E} (e^{2\pi i n_\phi/N}, {\cal V} ).
 \label{200804.2241}
\end{equation}
Therefore, $U_{\phi E}$ is a topological unitary object.

One might think that the group parameterizing the symmetry 
is a continuous group such as $U(1)$ since 
there is a conserved current.
However, the symmetry group is restricted to 
$\bb{Z}_N$ by the large gauge invariance or the Dirac 
quantization condition of the $U(1)$ gauge field.
This is due to the fact 
that the current is not gauge invariant, and 
the conserved charge is not large gauge invariant.
Let us consider this problem in detail.
We consider a topological unitary object 
$ U_{\phi E} (e^{i\alpha_{\phi E}}, {\cal V})
 = e^{ i\alpha_{\phi E}  Q({\cal V})}$ with 
a real parameter $\alpha_{\phi E}$.
We focus on the gauge variant
 term $e^{-\fr{ i N \alpha_{\phi E}}{8\pi^2} \int_{\cal V} a \wed da}$
in $U_{\phi E} (e^{i\alpha_{\phi E}}, {\cal V})$
and try to define it by using a gauge invariant 
integrand~\cite{Dijkgraaf:1989pz,Witten:2003ya}.
We define this term by using an auxiliary 
4-dimensional subspace $\Omega_{\cal V}$ with a boundary 
$\der \Omega_{\cal V}  = {\cal V}$ as
\begin{equation}
e^{-\fr{ i N \alpha_{\phi E}}{8\pi^2} \int_{\cal V} a \wed da} 
:= 
e^{-\fr{ i N \alpha_{\phi E}}{8\pi^2} \int_{\Omega_{\cal V}} da \wed da}.
\label{200909.0110}
\end{equation} 
However, the integral has an ambiguity of the 
choice of $\Omega_{\cal V}$. 
We can also define 
$e^{-\fr{ i N \alpha_{\phi E}}{8\pi^2}\int_{\cal V} a \wed da} $ 
by using another 4-dimensional subspace $\Omega'_{\cal V}$ 
satisfying $\der \Omega'_{\cal V} = {\cal V}$ as
\begin{equation}
e^{-\fr{ i N \alpha_{\phi E}}{8\pi^2} \int_{\cal V} a \wed da} 
:= 
e^{-\fr{ i N \alpha_{\phi E}}{8\pi^2} \int_{\Omega'_{\cal V}} da \wed da} . 
\end{equation}
The difference should be invisible, 
so that we require the following condition,
\begin{equation}
 e^{-\fr{ i N \alpha_{\phi E}}{8\pi^2}\int_{\Omega} da \wed da} = 1,
\end{equation}
where $\Omega = \Omega_{\cal V} \cup (-\Omega'_{\cal V})$ is the
4-dimensional closed subspace, 
and $- \Omega'_{\cal V}$ is 
the 4-dimensional subspace $\Omega'_{\cal V}$ with an opposite 
orientation.
By the Dirac quantization condition, 
the integral is 
$\int_{\Omega} da \wed da \in 2 \cdot (2\pi)^2 \bb{Z}$ on a spin manifold. 
Therefore, the parameter $\alpha_{\phi E}$
should satisfy $e^{i\alpha_{\phi E}} \in \bb{Z}_N$.
\footnote{
This requirement is the same as the quantization 
of the Chern-Simons term in $(2+1)$ dimensions \cite{Henneaux:1986tt}.}

The charged object for the symmetry is the 0-dimensional point object 
in \er{200627.1251}, and therefore
this symmetry is a $\bb{Z}_N$ 0-form symmetry.
The symmetry transformation is generated 
by the topological unitary object and 
is expressed by the correlation function,
\begin{equation}
 \vevs{
U_{\phi E} (e^{2\pi i n_\phi /N},{\cal V}) 
I(q_{\phi E}, {\cal P})
}
= e^{2\pi i n_\phi q_{\phi E} \link ({\cal V, P})/N} 
\vevs{I(q_{\phi E}, {\cal P})}.
\label{200713.1906}
\end{equation}
Here, the symbol `$\vevs{}$' denotes a vacuum expectation value (VEV),
and $\link ({\cal V,P}) \in \bb{Z}$ is a linking number of 
${\cal V}$ and ${\cal P}$.
In appendix \ref{corr0}, we show the derivation in detail.

\subsubsection{Electric {$\bb{Z}_N$} 1-form symmetry}

Second, we show a $\bb{Z}_N$ 1-form symmetry originated from 
the EOM of the photon, 
$-\fr{1}{e^2} d \star da + \fr{N}{4\pi^2} d\phi \wed da =0$,
which would imply the conservation of electric fluxes 
modified by the axion.
The closed current 2-form, conserved charge, and 
topological unitary object are given by
\begin{equation}
 j_{aE} = \fr{1}{e^2}  \star da - \fr{N}{4\pi^2} \phi da,
\quad
Q_{aE} ({\cal S}) = \int_{\cal S} j_{aE}, 
\quad
U_{aE} (e^{2\pi i n_a/N}, {\cal S})
 = e^{\fr{2\pi i n_a}{N} Q_{aE} ({\cal S})},
\end{equation}
respectively.
The topological unitary object $U_{aE}$ is parameterized by 
a $\bb{Z}_N$ group instead of a $U(1)$ group due to the gauge variant 
integrand $\phi da $.
The restriction on the group can be shown as follows.
We try to define the integral of the gauge variant term 
$e^{-\fr{ i\alpha_{a E} N}{4\pi^2} \int_{\cal S} \phi da}$ 
in a gauge invariant way,
where $\alpha_{aE}$ is a real parameter that will be determined 
by the large gauge invariance.
We define the integral by using a 3-dimensional subspace ${\cal V_S}$
as 
\begin{equation}
 e^{-\fr{ i\alpha_{a E} N}{4\pi^2}  \int_{\cal S} \phi da} 
=   e^{-\fr{ i\alpha_{a E} N}{4\pi^2} \int_{\cal V_S} d\phi  \wed da}. 
\end{equation}
The condition that the integral does not depend on the auxiliary 
subspace ${\cal V_S}$ leads to
\begin{equation}
   e^{-\fr{ i\alpha_{a E} N}{4\pi^2} \int_{\cal V} d\phi  \wed da} =1,
\end{equation}
where ${\cal V}$ is a 3-dimensional closed subspace.
Since $\int_{\cal V} d\phi  \wed da \in (2\pi)^2 \bb{Z}$, 
the parameter $\alpha_{a E}$ should be 
chosen as $e^{i\alpha_{a E}} \in \bb{Z}_N$.

The charged object for the symmetry is 
a Wilson loop in \er{200627.2309}.
The transformation law is given by 
\begin{equation}
 \vevs{U_{aE}(e^{2\pi i n_a/ N},{\cal S}) 
W(q_{aE}, {\cal C})}
= e^{2\pi i n_a q_{aE} \link ({\cal S,C})
/N}\vevs{W(q_{aE}, {\cal C})}. 
\end{equation}
The derivation is shown in appendix \ref{corr1}.
Since the charged object is a 1-dimensional object, 
the symmetry is a $\bb{Z}_N$ 1-form symmetry.
We refer to this 1-form symmetry as the electric $\bb{Z}_N$ 
1-form symmetry, since the symmetry is related to a conservation 
of the electric fluxes.
\subsubsection{Magnetic {$U(1)$} 1-form symmetry}
Third, we discuss a 1-form symmetry
due to the Bianchi identity of the photon, $dda =0$.
The corresponding closed current 2-form, 
conserved charge, and symmetry generator are given by
\begin{equation}
 j_{aM} = \fr{1}{2\pi} da,
 \quad
Q_{aM}({\cal S} ) = \int_{\cal S} j_{aM},
\quad
U_{aM} (e^{i\alpha_{a}}, {\cal S}) = e^{i\alpha_{a} Q({\cal S})},
\end{equation}
respectively.
The charged object is an 't Hooft loop $T(q_{aM},{\cal C})$, 
which is a closed worldline 
of a magnetic monopole.
Here, $q_{aM}$ is an integer by the Dirac quantization condition.
Note that the explicit form of the 't Hooft loop is shown in appendix \ref{dual}.

If the worldline of the monopole ${\cal C}$ is linked with 
a surface ${\cal S}$ of the charge $Q_{aM} ({\cal S})$, 
the charge detects the monopole 
charge $q_{aM}$ as 
$Q_{aM} ({\cal S}) = \fr{1}{2\pi}\int_{\cal S} da = q_{aM} \link ({\cal S,C})$. 
In terms of the correlation function of
the 't Hooft loop and the symmetry generator,
this property can be expressed as a $U(1)$ transformation 
of the 't Hooft loop by $U_{aM}$:
\begin{equation}
 \vevs{U_{aM} (e^{i\alpha_a}, {\cal S}) T(q_{aM}, {\cal C})} 
= e^{i\alpha_{a}q_{aM}\link ({\cal S,C})} \vevs{T(q_{aM}, {\cal C})}.
\end{equation}
Since the charged object is a 1-dimensional object, 
the symmetry is a $U(1)$ 1-form symmetry.
Hereafter, we refer to this $U(1)$ 1-form symmetry
as the magnetic $U(1)$ 1-form symmetry, 
since it is related to the conservation law of the magnetic fluxes.
\subsubsection{{$U(1)$} 2-form symmetry}
Finally, we consider a $U(1)$ 2-form symmetry
originated from the Bianchi identity of the axion, $dd\phi =0$.
The corresponding current 1-form, conserved charge,
and symmetry generator are given by  
\begin{equation}
 j_{\phi M} = \fr{1}{2\pi} d\phi, 
\quad
Q_{\phi M} ({\cal C}) = \int_{\cal C} j_{\phi M},
\quad
U_{\phi M} (e^{i\alpha_{\phi }},{\cal C})
 = e^{i\alpha_{\phi } Q({\cal C})},
\end{equation}
respectively.
Here, $e^{i\alpha_{\phi }} \in U(1)$ parameterizes 
the symmetry generator.
The charged object for the symmetry generator is 
a worldsheet of the axionic string denoted as 
$V(q_{\phi M},{\cal S})$, 
where ${\cal S}$ is a 2-dimensional closed subspace.
In the presence of the axionic string, the winding number
 of the axion becomes 
$\int_{\cal C} d\phi  = 2\pi q_{\phi M} \link({\cal C, S})$.
Note that the explicit form of the worldsheet of the axionic string 
is shown in appendix \ref{dual}.

We can regard this as a symmetry transformation of 
the worldsheet of the axionic string, since the axionic string is 
a source of a topological object $Q_{\phi M}({\cal S})$.
In terms of the correlation function,
the transformation law is given by
\begin{equation}
 \vevs{U_{\phi M} (e^{i\alpha_{\phi }},{\cal C})
V(q_{\phi M},{\cal S})}
= 
e^{i\alpha_{\phi } q_{\phi M} \link({\cal C,S})} \vevs{
V(q_{\phi M},{\cal S})}.
\end{equation}
We summarize the higher-form symmetries of 
the massless axion electrodynamics introduced in this section 
in Table \ref{tab:aph}.
\begin{table}[t]
\begin{center}
 \begin{tabular}[t]{c|c|c|c}
\hline\hline
Form & Symmetry generator & Charged object & Group
\\
\hline
0-form $U_{\phi E}$ & 
$e^{ \fr{2\pi i n_{\phi}}{N} \int_{\cal V} 
(-v^2 \star d\phi - \fr{N}{8\pi^2} a \wed da)
}
$ & $e^{i q_{\phi E}\phi ({\cal P})}$& $\bb{Z}_N$
\\
\hline
1-form $U_{a E}$ & 
$e^{ \fr{2\pi i n_{a}}{N} \int_{\cal S} 
(\fr{1}{e^2}\star da -\fr{N}{4\pi^2} \phi da)
}
$ & $e^{i q_{a E}\int_{\cal C} a }$& $\bb{Z}_N$
\\
\hline
1-form $U_{a M}$ & 
$e^{ \fr{i \alpha_{a }}{2\pi } \int_{\cal S} da
}
$ & $T(q_{aM}, {\cal C})$& $U(1)$
\\
\hline
2-form $U_{\phi M}$ & 
$e^{ \fr{i \alpha_{\phi}}{2\pi } \int_{\cal C} d\phi
}
$ & $V(q_{\phi M}, {\cal S})$& $U(1)$
\\

\hline\hline
 \end{tabular}
 \end{center}
\caption{Higher-form symmetries of the massless axion electrodynamics.}
\label{tab:aph}
\end{table}

\section{Background gauging and 't Hooft anomalies\label{bganom}}

In this section, we consider the background gauging of the 
higher-form symmetries.
We couple the action of the axion electrodynamics 
with the background gauge fields corresponding to the 
higher-form symmetries following Ref.~\cite{Gaiotto:2014kfa}.
We show that the invariance of the gauged action 
under the gauge transformations of the axion and photon 
(up to $2\pi \bb{Z}$) 
leads to modifications of the gauge transformation laws
of the background gauge fields.

\subsection{Gauging {$\bb{Z}_N$} 0-form symmetry}
First, we couple a background gauge field of the 
$\bb{Z}_N$ 0-form symmetry,
which is introduced as a $U(1)$ gauge field
with a constraint~\cite{Banks:2010zn}.
Although this constraint is already known,
we here show the derivation of the 
constraint explicitly in our case for self-containedness.
We also note a relation between the background 
gauge field and the symmetry generator.

\subsubsection{Constraint on background gauge field}
First, let us derive the constraint on the background gauge field.
The constraint is required by the invariance
under the $U(1)$ gauge transformation of the photon, 
or equivalently, by the fact that the global symmetry is not $U(1)$ but 
$\mathbb{Z}_N$.
At the linearized level, the background gauging could be 
done by adding a coupling of the conserved current with 
a background 1-form gauge field $A^{\phi E}_1$,
\begin{equation}
 S_{0, {\rm lin.}} = 
S+ \int_{M_4} j_{\phi E} \wed A_1^{\phi E}
= S-\int_{M_4} \(v^2  \star d\phi +\fr{N}{8\pi^2} a \wed da \) \wed  A_1^{\phi E}
\label{200628.1240}
\end{equation} 
to the action in \er{200628.1231}. 
Here, $A^{\phi E}_1$ is a $U(1)$ gauge field that is transformed 
as
\begin{equation}
 A^{\phi E}_1 \to A^{\phi E}_1 + d\Lambda^{\phi E}_0,
\end{equation}
where $\Lambda^{\phi E}_0$ is a $U(1)$ 
0-form gauge parameter
that satisfies $\int_{\cal C} d\Lambda^{\phi E}_0 \in 2\pi \bb{Z}$ on a closed one-dimensional manifold $\cal C$.
However, the coupling in \er{200628.1240} is not 
invariant under the gauge transformation of 
the photon $a$ in \er{200628.1244}.
Since the gauge transformation of $a$ leads to
the term proportional to 
$\int_{M_4} d\lambda \wed da \wed A^{\phi E}_1$, 
the gauge invariance may be preserved if we impose the flat condition $dA^{\phi E}_1=0$, in which $A^{\phi E}_1$ is locally expressed as $\alpha dA_0^{\phi E}$.
Here, $\alpha $ is a parameter that 
will be determined below.

The gauge transformation of the coupling 
$\int_{M_4} j_{\phi E} \wed A_1^{\phi E}$
becomes a total derivative under the condition,
but this total derivative may not vanish under 
a large gauge transformation.
This problem is caused by the presence of the 
gauge variant integrand $ a \wed da$ in $S_{0,{\rm lin.}}$.
In order to discuss the large gauge invariance, 
we would like to define the term
$\fr{N}{8\pi^2}\int_{M_4} a \wed da \wed A^{\phi E}_1$
by using gauge invariant integrand.
We define the term $\fr{N}{8\pi^2}\int_{M_4} a \wed da \wed A^{\phi E}_1$
on an auxiliary 5-dimensional manifold $X_5$ 
 satisfying $\der X_5 = M_4$ as
\begin{equation}
 \fr{N}{8\pi^2}\int_{M_4} a \wed da \wed A^{\phi E}_1
 :=
\fr{N}{8\pi^2}\int_{X_5} da \wed da \wed A^{\phi E}_1
= \fr{N \alpha }{8\pi^2}\int_{X_5} da \wed da \wed d A^{\phi E}_0
\quad\text{mod}\, 2\pi.
\end{equation}
Hereafter, we omit ``mod $2\pi$'' when 
we discuss the definitions of actions by using 
5-dimensional manifolds. 
Note that this definition is a natural extension of 
the definition of the $(2+1)$-dimensional 
Chern-Simons term by using $(3+1)$-dimensional integral~\cite{Dijkgraaf:1989pz},
which we have already discussed in \er{200909.0110}.
While the integrand is manifestly invariant
under the gauge transformation of the photon $a$ in \er{200628.1244},
we have chosen the auxiliary space $X_5$.
The ambiguity of the choice of the auxiliary space 
does not exist if the following condition is satisfied:
\begin{equation}
 \fr{N \alpha }{8\pi^2}\int_{Z_5} da \wed da \wed d A^{\phi E}_0
\in 2\pi \bb{Z},
\label{200628.1427}
\end{equation} 
where $Z_5$ is a 5-dimensional manifold without boundaries.
Under the normalization $\int_{\cal C} dA^{\phi E}_0 \in 2\pi \bb{Z}$,
we have the condition $\alpha = 1/{N}$.
Therefore, the gauge field $A^{\phi E}_1$ should satisfy
\begin{equation}
 N A^{\phi E}_1 = dA^{\phi E}_0.
\label{200628.1443}
\end{equation}
As a consequence, the field strength of $A_1^{\phi E}$ vanishes:
\begin{equation}
 F_2^{\phi E} : =  dA_1^{\phi E} =0.
\label{200820.0900}
\end{equation}
We refer to the 1-form gauge field with this 
condition as the $\bb{Z}_N$ 1-form gauge field.
This construction is consistent with the fact that the $0$-form global symmetry is a finite group $\mathbb{Z}_N$, whose gauge field need to be a flat connection.

We have explained the gauging the $\bb{Z}_N$ 0-form symmetry 
at a linearized level of $A^{\phi E}_1$, and derived the 
condition of $A^{\phi E}_1$ in \er{200628.1443}.
We can further gauge the $\bb{Z}_N$ 0-form symmetry 
at a non-linear level, 
which can be done as in ordinary gauge theories.
We can couple the background gauge field 
to the action by replacing $d\phi $ with $d\phi - A^{\phi E}_1$.
Here, the axion is shifted under a gauge transformation of 
$A^{\phi E}_1$ as
\begin{equation}
 A^{\phi E}_1 \to A^{\phi E}_1 +d\Lambda_0^{\phi E},
\quad
A^{\phi E}_0 \to A^{\phi E}_0 + N \Lambda_0^{\phi E},
\quad
\phi \to \phi + \Lambda_0^{\phi E}.
\label{200820.0904}
\end{equation}
We can confirm that
the action with the background gauge field is invariant
under the gauge transformations of the axion and photon.
In order to make the gauge invariance manifest, 
we define a gauged action by using the 5-dimensional action as
\begin{equation}
 S_{0} = -\int_{M_4} 
\(\fr{v^2}{2}|d\phi -A^{\phi E}_1|^2 + \fr{1}{2e^2} |da|^2\)
+\fr{N}{8\pi^2} \int_{X_5} (d\phi - A^{\phi E}_1) \wed da \wed da
\quad\text{mod}\, 2\pi.
\end{equation}
The action, in particular the last term, does not depend on 
the choice of $X_5$, as a consequence of 
\begin{equation}
 \fr{N}{8\pi^2} \int_{Z_5} d\phi \wed da \wed da
\in 2\pi N \bb{Z},
\qtq{and}
 \fr{N}{8\pi^2} \int_{Z_5} A^{\phi E}_1 \wed da \wed da
\in 2\pi \bb{Z}.
\end{equation}
Therefore, the gauged action is invariant 
under the gauge transformations of dynamical fields.

\subsubsection{Background gauging as insertion of symmetry generators\label{bgphys}}
We can interpret the background gauging as a 
network of the symmetry generator in the spacetime~\cite{Gaiotto:2014kfa}, 
and the configuration of the symmetry generators is 
expressed by the background gauge field $A^{\phi E}_1$.
In particular, we can obtain the symmetry generator 
$U_{\phi E} (e^{2\pi i n_\phi/N},{\cal V})$
by choosing
$A^{\phi E}_1 = \fr{2\pi n_\phi}{N} \delta_1({\cal V})$.
Here, we have introduced the delta functional $p$-form such that, 
in $D$-dimensional spacetime $M_D$,
\begin{equation}
\int_{M_D}J\wedge \delta_{p}({\cal V}_{D-p})=\int_{{\cal V}_{D-p}} J
\label{eq:deltaFunctionalForm}
\end{equation}
for a $(D-p)$-form $J$ and a $(D-p)$-dimensional manifold ${\cal V}_{D-p}$.
In the viewpoint of the symmetry generator, 
the gauge transformation
$A^{\phi E}_1 \to A^{\phi E}_1 + d\Lambda^{\phi E}_0$ 
corresponds to a topological deformation 
${\cal V} \to {\cal V} \cup \der \Omega_0$
in \er{200804.2241} by choosing 
$\Lambda_0^{\phi E} = \frac{2 \pi n_\phi}{N} \delta_0(\Omega_0)$,
since $d \delta_0(\Omega_0) =  \delta_1(\der \Omega_0)$.
Note that the condition in \er{200628.1443} implies 
$A^{\phi E}_0 = 2\pi n_\phi \delta_0 (\Omega_{\cal V})$,
since 
$N A^{\phi E}_1 = 2\pi n_\phi d \delta_0 (\Omega_{\cal V})$.
Here, $\Omega_{\cal V}$ is a 4-dimensional subspace whose 
boundary is ${\cal V}$.

\subsection{Gauging electric {$\bb{Z}_N$} 1-form symmetry
and {$U(1)$} 2-form symmetry}

Next, 
we gauge the $\bb{Z}_N$ electric 1-form symmetry. 
As we see below, we need to gauge the $U(1)$ 2-form symmetry
simultaneously in order to preserve the the 
gauge invariance for the axion.
\subsubsection{Gauging electric {$\bb{Z}_N$} 1-form symmetry}

Here, we consider 
the gauging of the electric $\bb{Z}_N$ 1-form symmetry,
which 
can be done by introducing a 2-form gauge field $B^{aE}_2$.
Since the global symmetry is parameterized by 
the $\bb{Z}_N$ group, there is a similar constraint on $B^{aE}_2$.
At the linearized level, the coupling would be written as
$\int_{M_4} j_{aE} \wed B^{aE}_2$.
However, this is generally not invariant under the $2\pi$ shift of $\phi$
due to the term $\int_{M_4} \fr{N}{4\pi^2} \phi da \wed  B^{aE}_2 $,
and the deviation is $\fr{N}{2\pi}\int_{M_4}  da \wed B^{aE}_2$.
In order to derive the condition for $B^{aE}_2$ such that 
the coupling $\int_{M_4}  j_{aE} \wed B^{aE}_2 $ is gauge invariant, 
 we define the term $\fr{N}{4\pi^2}  \int_{M_4}   \phi da \wed B^{aE}_2$   
by using a 5-dimensional space as
\begin{equation}
\fr{N}{4\pi^2}  \int_{M_4}   \phi da \wed B^{aE}_2
=  \fr{N}{4\pi^2} \int_{X_5}  d\phi \wed  da \wed B^{aE}_2.
\end{equation}
The ambiguity of the choice of $X_5$ is absent 
if $NB^{aE}_2 = dB^{aE}_1$ with the normalization 
$\int_{\cal S} dB^{aE}_1\in 2\pi \bb{Z}$.
Therefore, we require that the 2-form gauge field is
constrained by the 1-form gauge field as 
\begin{equation}
 NB^{aE}_2 = dB^{aE}_1,
\end{equation}
which means that the field strength vanishes,
\begin{equation}
 H_3^{aE} : =  dB_2^{aE} =0.
\label{200820.0901}
\end{equation}
At the nonlinear level, the gauging could be done by 
replacing $da$ with $da - B^{aE}_2$.
The gauged action would be
\begin{equation}
\begin{split}
 S_{1E} &
= -\int_{M_4} \big(
\fr{v^2}{2} |d\phi|^2 
+ \fr{1}{2e^2}|da-B_2|^2 
\big)
+ \fr{N}{8\pi^2}\int_{M_4} \phi (da - B^{aE}_2) \wed (da -B^{aE}_2).
\end{split}
\label{200330.1906}
\end{equation}
The gauge transformation laws of $B_2^{aE}$, $B_1^{aE}$,
and $a$ are
\begin{equation}
 B^{aE}_2 \to  B^{aE}_2 + d\Lambda^{aE}_1, 
\quad
 B^{aE}_1 \to  B^{aE}_1 + N\Lambda^{aE}_1,
\quad
 a \to  a + \Lambda^{aE}_1.
\label{200820.0905}
\end{equation}
Here, $\Lambda^{aE}_1$ is a 1-form gauge parameter 
with the normalization $\int_{\cal S} d\Lambda^{aE}_1 \in 2\pi \bb{Z}$.
This action can lead to the coupling at the linearized level.
However, the non-linear term 
$\fr{N}{8\pi^2} \int_{M_4} \phi B^{aE}_2 \wed B^{aE}_2$
is not invariant under the $2\pi $ shift of $\phi$ up to $2\pi$.
In fact, the deviation is 
$\fr{N}{4\pi} \int_{M_4} B^{aE}_2 \wed B^{aE}_2 = 
\fr{1}{4\pi N} \int_{M_4} dB^{aE}_1 \wed dB^{aE}_1 \in \fr{2\pi}{N} \bb{Z}$.

We can discuss the problem by using the 5-dimensional action
whose integrand is manifestly gauge 
invariant 
(see, e.g., recent Refs.~\cite{Witten:2015aba,Yonekura:2018ufj,Witten:2019bou}).
We can define the gauged topological term 
in a 5-dimensional spacetime $X_5$ as
\begin{equation}
 \fr{N}{8\pi^2}\int_{M_4} \phi (da - B^{aE}_2) \wed (da -B^{aE}_2)  
= 
 \fr{N}{8\pi^2}\int_{X_5} 
d \phi \wed (da - B^{aE}_2) \wed (da -B^{aE}_2).  
\end{equation}
This action is manifestly invariant under the $2\pi $ shift 
of $\phi$, but we have chosen an auxiliary 5-dimensional spacetime $X_5$.
The gauged action suffers from the ambiguity of the choice of 
the spacetime:
\begin{equation}
 \fr{N}{8\pi^2}\int_{Z_5} 
d \phi \wed (da - B^{aE}_2) \wed (da -B^{aE}_2) 
\in \fr{2\pi}{N} \bb{Z}.
\end{equation}
Therefore, we cannot gauge the 1-form symmetry by itself.
\subsubsection{Gauging {$U(1)$} 2-form symmetry}
This problem can be resolved by gauging the $U(1)$ 2-form symmetry 
simultaneously.
This is because the problematic term $ \fr{N}{8\pi^2}\int_{X_5} 
d \phi \wed (da - B^{aE}_2) \wed (da -B^{aE}_2)$ 
is associated to a closed current 1-form 
$j_{\phi M} = \fr{1}{2\pi} d\phi$ 
of the $U(1)$ 2-form symmetry.

Before discussing the resolution,
we consider the gauging of the $U(1)$ 2-form symmetry independently.
We introduce a 3-form gauge field $C^{\phi M}_3$, which 
couples to the closed current of 
the $U(1)$ 2-form symmetry $j_{\phi M}$ as
\begin{equation}
 S_2 = 
S + 
\int_{M_4} j_{\phi M} \wed C^{\phi M}_3 = 
S + \fr{1}{2\pi}\int_{M_4} d\phi \wed C^{\phi M}_3.
\end{equation}
Here, the 3-form gauge field is normalized by the Dirac quantization 
condition 
\begin{equation}
 \int_{\Omega} dC_3 \in 2\pi \bb{Z},
\end{equation}
where $\Omega$ is a 4-dimensional closed subspace.
The gauge transformation law of the 3-form gauge field 
is 
\begin{equation}
 C^{\phi M}_3 \to 
 C^{\phi M}_3 + d\Lambda^{\phi M}_2,
\end{equation}
where $\Lambda^{\phi M}_2$ is a 2-form gauge parameter 
that is normalized as $\int_{\cal V} d\Lambda_2^{\phi M} \in 2\pi \bb{Z}$.
Since the gauge transformation of the 
coupling $\int_{M_4} j_{\phi M} \wed C^{\phi M}_3$
is a total derivative, 
the large gauge invariance of the coupling is nontrivial.
In order to show the large gauge invariance, 
we define the coupling on a 5-dimensional manifold as
\begin{equation}
 S_2 = S -  \fr{1}{2\pi} \int_{X_5} d\phi \wed dC^{\phi M}_3.
\label{200628.1709}
\end{equation}
The gauged action does not depend on the choice of $X_5$:
\begin{equation}
  \fr{1}{2\pi} \int_{Z_5} d\phi \wed dC^{\phi M}_3
 \in 2\pi \bb{Z}.
\end{equation}

We now resolve the problem of the gauging of the $\bb{Z}_N$
1-form symmetry.
We can cancel the problematic term 
$\fr{N}{8\pi^2} \int_{X_5} d\phi \wed B^{aE}_2 \wed B^{aE}_2$ 
in \er{200330.1906} by
modifying the field strength $dC^{\phi M}_3$ in \er{200628.1709}
as
\begin{equation}
 dC^{\phi M}_3 
\to G_4^{\phi M a E}
 =  
dC^{\phi M}_3  + \fr{N}{4\pi} B^{aE}_2\wed B^{aE}_2.
\label{200910.1735}
\end{equation}
The modification requires 
 an additional gauge transformation law
of $C^{\phi M}_3$
under $B^{aE}_2 \to B^{aE}_2 + d\Lambda^{aE}_1$,
\begin{equation}
 C^{\phi M}_3 \to
C^{\phi M}_3 + 
d\Lambda_2^{\phi M}
-
\fr{N}{2\pi} \Lambda^{aE}_1 \wed B^{aE}_2 
-
\fr{N}{4\pi} \Lambda^{aE}_1 \wed d\Lambda^{aE}_1. 
\label{200820.0907}
\end{equation}
Note that the additional transformation does not violate
the normalization of $C_3^{\phi M}$, 
\begin{equation}
 \int_{\Omega} 
d\(
d\Lambda_2^{\phi M}
-
\fr{N}{2\pi} \Lambda^{aE}_1 \wed B^{aE}_2 
-
\fr{N}{4\pi} \Lambda^{aE}_1 \wed d\Lambda^{aE}_1\)
\in 2\pi \bb{Z}.
\end{equation}
Eventually, the gauged action can be defined on the 5-dimensions as
\begin{equation}
\begin{split}
S_{1E,2} 
&= 
-\int_{M_4} \(\fr{v^2}{2} |d\phi|^2 +\fr{1}{2e^2} |da - B^{aE}_2|^2\)
 - \fr{1}{2\pi} \int_{X_5} d\phi \wed G_4^{\phi M aE}
\\
&
\quad
+
\fr{N}{8\pi^2}\int_{X_5} 
d \phi \wed (da - B^{aE}_2) \wed (da -B^{aE}_2)
.
\end{split}
\end{equation}
This gauged action has no ambiguity since 
the problematic term 
$\fr{N}{8\pi^2} \int_{X_5} d\phi \wed B^{aE}_2 \wed B^{aE}_2 $
is canceled out as
\begin{equation}
\begin{split}
&
  \fr{N}{8\pi^2}\int_{Z_5} 
d \phi \wed (da - B^{aE}_2) \wed (da -B^{aE}_2)
 - \fr{1}{2\pi} \int_{Z_5} d\phi \wed G^{\phi M aE}_4
\\
&
=
  \int_{Z_5} 
\(\fr{N}{8\pi^2} d \phi \wed da  \wed da
-
\fr{1}{4\pi^2}  d \phi \wed da \wed dB^{aE}_1
 - \fr{1}{2\pi} d\phi \wed dC^{\phi M}_3 \)
\in 2\pi \bb{Z}.
\end{split}
\end{equation}

\subsection{Gauging all symmetries\label{allgauge}}

We now gauge the $\bb{Z}_N$ 0-form,
electric $\bb{Z}_N$ 1-form, and $U(1)$ 2-form 
symmetries.
We show that we should simultaneously gauge the magnetic $U(1)$ 1-form 
symmetry in order to preserve the invariance 
under the gauge transformation of the photon.
In other words, the simultaneous gauging of 
the $\bb{Z}_N$ 0-form and $\bb{Z}_N$ 1-form symmetries 
requires the gauging all of the higher-form symmetries.

Let us try to gauge the $\bb{Z}_N$ 0-form, 
electric $\bb{Z}_N$ 1-form, and $U(1)$ 2-form symmetries.
We deform the action $S_{1E,2}$
by gauging the $\bb{Z}_N$ 0-form symmetry. 
The gauged action would be
\begin{equation}
\begin{split}
S_{0,1E,2} 
&= 
-\int_{M_4} 
\(\fr{v^2}{2} |d\phi - A^{\phi E}_1|^2 +\fr{1}{2e^2} |da - B^{aE}_2|^2\)
 - \fr{1}{2\pi} \int_{X_5} (d\phi - A^{\phi E}_1) \wed G_4^{\phi M aE}
\\
&
\quad
+
\fr{N}{8\pi^2}\int_{X_5} 
(d \phi - A^{\phi E}_1) \wed (da - B^{aE}_2) \wed (da -B^{aE}_2)
.
\end{split}
\label{200805.0244}
\end{equation}
However, this gauging depends on the choice of $X_5$,
or equivalently, the violation of the gauge invariance 
in the 4-dimensional action.
In terms of the 4-dimensional action, the 
violation of the gauge invariance may be seen as follows.
The problematic term is 
$\fr{N}{4\pi^2}\int_{M_4} 
A^{\phi E}_1 \wed a \wed B^{aE}_2$
in \er{200805.0244}
after the partial integration.
The deviation 
under the large gauge transformation of the photon
$a \to a +d\lambda$ with $\int_{\cal C} d\lambda \in 2\pi \bb{Z}$
is 
\begin{equation}
 \fr{N}{4\pi^2}\int_{M_4} 
A^{\phi E}_1 \wed d\lambda \wed B^{aE}_2
=  
\fr{1}{4\pi^2 N}\int_{M_4} 
d A^{\phi E}_0 \wed d\lambda \wed d B^{aE}_1 \in \fr{2\pi}{N} \bb{Z}.
\end{equation}

On the other hand, in the 5-dimensional action, 
the ambiguity of the choice of the 5-dimensional space can be
expressed as
\begin{equation}
\begin{split}
&
  \fr{N}{8\pi^2}\int_{Z_5} 
(d \phi - A_1^{\phi E}) \wed (da - B^{aE}_2) \wed (da -B^{aE}_2)
 - \fr{1}{2\pi} \int_{Z_5} (d \phi - A_1^{\phi E}) \wed G^{\phi M aE}_4
 \qtq{mod $2\pi$}
\\
&
=
  \int_{Z_5} 
\Big(\fr{N}{8\pi^2} (d \phi - A_1^{\phi E}) \wed da  \wed da
-
\fr{1}{4\pi^2}  (d \phi - A_1^{\phi E}) \wed da \wed dB^{aE}_1
\\
&\quad\hph{  \int_{Z_5} 
\Big(}
 - \fr{1}{2\pi} (d \phi - A_1^{\phi E}) \wed dC^{\phi M}_3 \Big)
 \qtq{mod $2\pi$}
\\
&
=
  \int_{Z_5} 
\Big(
\fr{1}{2\pi} A_1^{\phi E} \wed dC^{\phi M}_3
+
\fr{1}{4\pi^2 N} da \wed dA_0^{\phi E} \wed dB^{aE}_1
 \Big)
\qtq{mod $2\pi$}  .
\end{split}
\label{200628.1906}
\end{equation}
The first term 
$ \int_{Z_5} \fr{1}{2\pi } A_1^{\phi E} \wed dC^{\phi M}_3 
\in \fr{2\pi}{N}\bb{Z}$ 
in the last line a mixed 't~Hooft anomaly of the axion, 
which just expresses the fact that we cannot regard $A^{\phi E}_1$
and $C^{\phi M}_3$ as dynamical variables 
simultaneously~\cite{Gaiotto:2014kfa}.
However, the second term,
\begin{equation}
\int_{Z_5} \fr{1}{4\pi^2 N} da  \wed dA_0^{\phi E} \wed dB^{aE}_1
\in \fr{2\pi}{N} \bb{Z}, 
\label{200629.1224}
\end{equation} 
is problematic since it depends on the dynamical field 
(q-number) $a$.
We expect that we can eliminate the term by gauging the magnetic $U(1)$
1-form symmetry,
since the problematic term is proportional to the 
closed 2-form current of the magnetic $U(1)$ 1-form symmetry
$j_{aM} =  \fr{1}{2\pi} da$.

Before gauging the magnetic $U(1)$ 1-form symmetry in 
$S_{0,1E,2}$, we gauge it in the original action $S$
for simplicity.
We introduce a $U(1)$ 2-form gauge field $B^{a M}_2$ 
that is coupled to $j_{aM}$ as
$ S_{1M} = S+ \fr{1}{2\pi} \int_{M_4} da \wed B^{aM}_2$.
Here, the gauge transformation law of $B^{aM}_2$ is
\begin{equation}
 B^{aM}_2 \to B^{aM}_2 + d\Lambda^{aM}_1,
\end{equation}
where $\Lambda^{aM}_1$ is a $U(1)$ 1-form gauge parameter
normalized as $\int_{\cal S} d\Lambda^{aM}_1 \in 2\pi \bb{Z}$.
The normalization of $B_2^{aM}$ is 
$\int_{\cal V} dB^{aM}_2 \in 2\pi \bb{Z}$
by the Dirac quantization condition.
In order to make $S_{1M}$ manifestly invariant
 under the large gauge transformations,
we again define the coupling in the 5-dimensional space as 
\begin{equation}
 S_{1M} = S + \fr{1}{2\pi} \int_{X_5} da \wed dB^{aM}_2.
\label{200629.1222}
\end{equation}

Now, we gauge the magnetic $U(1)$ 1-form symmetry 
in $S_{0,1E,2}$
to eliminate the problematic term in \er{200628.1906}.
Since we have already gauged the electric $U(1)$ 1-form symmetry 
in $S_{0,1E,2}$, the photon $a$ is shifted under the 
gauge transformation of $B^{aE}_2$.
Thus, the field strength $da$ in \er{200629.1222} 
should be replaced with $da - B^{aE}_2$.
Including the term canceling the problematic term in \er{200629.1224},
we gauge the magnetic 1-form symmetry by introducing
 the following term,
\begin{equation}
\begin{split}
S_{0,1E,1M,2}
&
= S_{0,1E,2}
+
\fr{1}{2\pi}\int_{X_5} (da -B_2^{aE}) \wed 
\(d B^{a M}_2 - \fr{N}{2\pi} A_1^{\phi E} \wed B_2^{aE}\)
\\
& 
= 
-\int_{M_4} 
\(\fr{v^2}{2} |d\phi - A^{\phi E}_1|^2 +\fr{1}{2e^2} |da - B^{aE}_2|^2\)
 - \fr{1}{2\pi} \int_{X_5} (d\phi - A^{\phi E}_1) \wed dC_3^{\phi M}
\\
&
\quad
+
\fr{N}{8\pi^2}\int_{X_5} 
(d \phi - A^{\phi E}_1) \wed da  \wed da
-
\fr{N}{4\pi^2}\int_{X_5} 
d \phi \wed da  \wed B^{aE}_2
\\
&
\quad
+
\fr{1}{2\pi}\int_{X_5} da \wed B^{a M}_2 
-
\fr{1}{2\pi}\int_{X_5} B_2^{aE} \wed 
\(d B^{a M}_2 - \fr{N}{2\pi} A_1^{\phi E} \wed B_2^{aE}\).
\end{split}
\label{200716.1848}
\end{equation}
In order to make the gauged action gauge invariant, 
the gauge transformation law $B^{aM}_2$ should be modified as
\begin{equation}
 B_2^{aM} \to B_2^{aM} + d\Lambda_1^{aM} 
+ \fr{N}{2\pi} \Lambda_0^{\phi E} B^{aE}_2 
- \fr{N}{2\pi} (A_1^{\phi E} + d\Lambda^{\phi E}_0) \wed \Lambda^{aE}_1
\label{200820.0906}
\end{equation} 
with $A^{\phi E}_1 \to A^{\phi E}_1 + d\Lambda^{\phi E}_0 $
 and $B_2^{aE} \to B_2^{aE} + d\Lambda^{aE}_1$.
Note that the modified gauge transformations of 
$B^{aM}_2$ preserve the Dirac quantization condition of $B_2^{aM}$:
\begin{equation}
\int_{\cal V}d \(
d\Lambda_1^{aM}+ \fr{N}{2\pi} \Lambda_0^{\phi E} B^{aE}_2 
- \fr{N}{2\pi} (A_1^{\phi E} + d\Lambda^{\phi E}_0) \wed \Lambda^{aE}_1\)
\in 2\pi \bb{Z}.
\end{equation}
Accordingly, the gauge invariant
field strength for $B^{aM}_{2}$ is identified as
\begin{equation}
H_3^{aM, \phi E}
:= 
d B^{a M}_2 - \fr{N}{2\pi} A_1^{\phi E} \wed B_2^{aE} .
\label{200820.0911}
\end{equation}
By adding the term, the problematic
 operator-valued ambiguity in \er{200628.1906} is now absent,
\begin{equation}
  \begin{split}
&
  \fr{N}{8\pi^2}\int_{Z_5} 
(d \phi - A_1^{\phi E}) \wed (da - B^{aE}_2) \wed (da -B^{aE}_2)
 - \fr{1}{2\pi} \int_{X_5} (d \phi - A_1^{\phi E}) \wed G^{\phi M aE}_4
\\
&
+\fr{1}{2\pi}\int_{Z_5} (da -B_2^{aE}) \wed 
H^{a M, \phi E}_3
\qtq{mod $2\pi$} 
\\
&
=
 \fr{1}{2\pi }
  \int_{Z_5} 
  A_1^{\phi E} \wed dC^{\phi M}_3
-\fr{1}{2\pi}
\int_{Z_5} B_2^{aE} \wed 
dB_2^{a M}
\\
&
\quad
+\fr{N}{(2\pi)^2}
\int_{Z_5} A_1^{\phi E}  \wed B_2^{aE} \wed B^{aE}_2
\quad
\text{mod $2\pi$}   .
 \end{split}
\label{200820.0854}
\end{equation}
The remaining ambiguity in the right-hand side 
represents the 't~Hooft anomalies.

In summary, we have introduced the background gauge fields 
$(A^{\phi E}_1, B^{aE}_2, B^{aM}_2, C^{\phi M}_3)$,
whose action is given in Eq.~\eqref{200716.1848}.
The gauge transformation laws are given by
\ers{200820.0904}, 
\eqref{200820.0905},
\eqref{200820.0906},
and 
\eqref{200820.0907}, respectively.
The field strengths are determined by 
\ers{200820.0900}, 
\eqref{200820.0901}, 
\eqref{200820.0911}
and 
\eqref{200910.1735}, 
respectively.
In the next section, we derive the above gauge transformation laws 
and field strength from the viewpoint of the 
3-group gauge theory.

Thus, we have successfully gauged the higher-form symmetries.
Furthermore, we have obtained the 't Hooft anomalies
for the higher-form symmetries.
The first term in \er{200820.0854} 
is 
the mixed 't Hooft anomalies 
of the axion, which has been discussed previously.
The second term is that of the photon, 
which prohibits the simultaneous dynamical 
gauging of 
the pair of the electric $1$-form and magnetic 
$1$-form symmetries.
The third one can be identified as the so-called 
2-group anomaly~\cite{Benini:2018reh}.
This anomaly means the obstruction to the
simultaneous gauging of 
the $\bb{Z}_N$ 0-form symmetry and the electric $\bb{Z}_N$ 
1-form symmetry.

The existence of the 't Hooft anomalies forbids a trivial gapped vacuum.
In our case, this requirement is satisfied by the existence of the
massless axion and photon.
The existence of the 
massless axion corresponds to the existence of the 
mixed anomaly between the 0-form and 2-form symmetries.
Likewise,
the existence of the 
massless photon corresponds to the existence of the 
mixed anomaly between the electric and magnetic 1-form symmetries.
If we deform the system with preserving these higher-form symmetries, 
any trivial gapped vacuum is still forbidden.
For example, if a gapped vacuum is realized while 
preserving the symmetries, we can have topologically ordered phases,
whose ground states can be degenerated on a compact spatial manifold.

Further, the existence of the 2-group anomaly implies the 
existence of a fractionally charged particle where 
$A^{\phi E}_1 \wed B_2^{a E}$ is non-zero.
Physically, it means that we have a fractionally charged particle on the domain wall if we add the magnetic field through the domain wall,
which was proposed in Ref.~\cite{Sikivie:1984yz}.

In order to see this effect, we consider the following partition function
given by the gauged action in \er{200716.1848},
\begin{equation}
 Z [A_1^{\phi E}, B_2^{aE}, B_2^{a M} , C_3^{\phi M},X_5]
= {\cal N} \int {\cal D}[\phi, a] 
e^{iS_{0,1E, 1M, 2}[\phi, a, A_1^{\phi E}, B_2^{aE}, B_2^{a M} , C_3^{\phi M}, X_5]}.
\end{equation}
Here, ${\cal N}$ is a normalization factor such that 
$\vevs{1} =1$.
By setting 
$A^{\phi E}_1 = \fr{2\pi n_\phi}{N} \delta_1({\cal V})$,
$B_2^{a E} = \fr{2\pi n_a}{N} \delta_2({\cal S})$,
$B_2^{aM} =0$, and $C_3^{\phi M} =0$, 
we obtain
\begin{equation}
\begin{split}
  Z \left[\fr{2\pi n_\phi}{N} \delta_1({\cal V}),\fr{2\pi n_a}{N} \delta_2({\cal S}), 0,0,X_5\right]
&=
 {\cal N} \int {\cal D}[\phi, a] 
e^{-\fr{i n_a n_\phi}{N}
 \int_{M_4} a\wed  \delta_1({\cal V}) \wed \delta_2 ({\cal S})}
e^{iS[\phi, a]}
\\
&=
\vevs{e^{-\fr{i n_a n_\phi}{N}
 \int_{M_4} a\wed  \delta_1({\cal V}) \wed \delta_2 ({\cal S})}} ,
\end{split}
\end{equation}
where we have redefined 
$\phi - \fr{2\pi n_\phi}{N} \delta_0 (\Omega_{\cal V}) \to \phi$ and 
$a + \fr{2\pi n_a}{N} \delta_1 ({\cal V_S})  \to a$ in the path integral.
On the right-hand side, we have the term 
$e^{-\fr{i n_a n_\phi}{N}
 \int_{M_4} a\wed  \delta_1({\cal V}) \wed \delta_2 ({\cal S})}$
which is given by the term related to the 2-group anomaly 
$\fr{1}{2\pi} \int_{X_5} (da - B^{a E}_2) \wed (dB_2^{aM} - \fr{N}{2\pi} A_1^{\phi E} \wed B_2^{a E} )$.
Since $\delta_1({\cal V}) \wed \delta_2 ({\cal S})$ is a delta function 
3-form on the closed line ${\cal V} \cap {\cal S}$, 
the right-hand side implies the existence of the Wilson loop
on ${\cal V} \cap {\cal S}$ with the fractional charge 
$- \fr{n_a n_\phi }{N}$.
Note that this fractionally charged particle does not arise
as long as we do not consider the intersection of the 
background fields i.e., the symmetry generators.

\section{Global 3-group symmetry and its gauging in axion electrodynamics\label{bg3g}}

In this section, we derive the background gauging 
by a different approach based on the 3-group gauge theory.
First, we review the global 3-group symmetry 
by the structure of the correlation functions of the 
symmetry generators~\cite{Hidaka:2020iaz}.
The correlation functions give us ingredients of the 3-group.
Next, we establish the 3-group gauge theory,
which can be formulated for a given 3-group.
We should remark that 
this 3-group gauge theory is based on a mathematical procedure,
independent of the gauging based on the gauge invariance
of dynamical fields in the previous section.
We confirm that the gauging of this section coincides with 
the one in the previous section.

\subsection{Correlation functions of symmetry generators\label{3gcorr}}

We review the correlation functions of the symmetry 
generators~\cite{Hidaka:2020iaz}, 
which give us the group structure in the 
higher-form symmetries.
This is a natural generalization of current 
algebra in ordinary quantum field theories.
The details of the derivations are summarized in appendix \ref{corrsym}.
Note that we only consider the correlation functions of the symmetry
generators which are not intersected to each other.
Therefore, we do not have to consider fractionally charged objects 
due to the intersection of the symmetry generators 
in section \ref{allgauge}.

It has been shown that the correlation functions 
of the symmetry generators are 
not independent, but related to each other.
First, the correlation functions of 
the 0-form and electric 1-form symmetry generators 
induce a magnetic 1-form symmetry generator:
\begin{equation}
 \vevs{
U_{\phi E} (e^{2\pi i n_{\phi} /N}, {\cal V}) 
U_{aE} (e^{2\pi i n_a/N},{\cal S}) 
}
= \vevs{
U_{aE} (e^{2\pi i n_a/N},{\cal S}) 
U_{a M} (e^{-2 \pi i n_\phi n_a/N}, \Omega_{\cal V} \cap {\cal S})
},
\label{200805.1726}
\end{equation}
\begin{equation}
 \vevs{
U_{\phi E} (e^{2\pi i n_{\phi} /N}, {\cal V}) 
U_{aE} (e^{2\pi i n_a/N},{\cal S}) 
}
= \vevs{
U_{\phi E} (e^{2\pi i n_{\phi} /N}, {\cal V}) 
U_{a M} (e^{-2\pi i n_\phi n_a/N}, (-{\cal V_S}) \cap {\cal V})},
\label{200805.1727}
\end{equation}
where $\Omega_{\cal V}$ and ${\cal V_S}$ are 4- and 3-dimensional subspaces satisfying $\partial\Omega_{\cal V}={\cal V}$ and $\partial{\cal V_S}={\cal S}$, respectively.
Here, we have eliminated the 0-form and electric 1-form symmetry generators by redefining the integral variables of the path integral in \er{200805.1726} and \er{200805.1727}, respectively.
The minus sign in $-{\cal V}_{{\cal S}}$ in \er{200805.1727} 
is due to 
the minus sign in  
$\delta_2 ({\cal S}) = - d\delta_1 ({\cal V}_{{\cal S}})$.
This sign matches the choice of the background gauge fields
$B_2 = \fr{2\pi n_a }{N} \delta_2 ({\cal S}) $ 
and 
$B_1 = - 2\pi n_a  \delta_1 ({\cal V}_{{\cal S}_1}) $ 
in \er{200716.1848}.

A physical meaning of these relations in \ers{200805.1726} 
and \eqref{200805.1727}
is that the electric flux can be induced by the axionic domain wall
in the presence of the magnetic monopole inside the axionic domain wall.
Therefore, the correlation functions can be interpreted as 
the Witten effect of the axion~\cite{Hidaka:2020iaz}:
if the domain wall encloses a magnetic monopole, 
the domain wall induces the electric 
flux~\cite{Sikivie:1984yz,Wilczek:1987mv,Kogan:1992bq,Kogan:1993yw}.

Second, we consider the correlation function of the 1-form symmetry
generators, which leads to a 2-form symmetry generator:
\begin{equation}
\begin{split}
  \vevs{
U_{a E} (e^{2\pi i n_{a} /N}, {\cal S}_1) 
U_{aE} (e^{2\pi i n'_a/N}, {\cal S}_2) 
}
= \vevs{U_{\phi M} (e^{2\pi i n_a n'_a/N}, 
(-{\cal V}_{{\cal S}_1}) \cap {\cal S}_2)
U_{aE} (e^{2\pi i n'_a/N}, {\cal S}_2) }. 
\end{split}
\label{200807.1426}
\end{equation}
Here, we have eliminated 
$U_{a E} (e^{2\pi i n_{a} /N}, {\cal S}_1) $ by the same procedure.
The minus sign in $-{\cal V}_{{\cal S}_1}$ is due to 
the minus sign in  
$\delta_2 ({\cal S}_1) = - d\delta_1 ({\cal V}_{{\cal S}_1})$.
This sign matches the choice of the background gauge fields
$B_2 = \fr{2\pi }{N}(n_a \delta_2 ({\cal S}_1) 
+ n_a' \delta_2 ({\cal S}_2)) $ 
and 
$B_1 = - 2\pi (n_a \delta_2 ({\cal V}_{{\cal S}_1}) 
+ n_a' \delta_1 ({\cal V}_{{\cal S}_2}) )$ 
in \er{200716.1848}.

Physically, the relation in \er{200807.1426} means 
that the magnetic field can be induced by the electric field 
in the presence of the axionic string.
The correlation function represents the anomalous Hall effect 
for the axion~\cite{Hidaka:2020iaz}.
In the presence of the axionic string and the electric field, 
the electric current is induced~\cite{Qi:2008ew,Teo:2010zb,Wang:2012bgb}.
By the Maxwell-Amp\`ere law, the electric current induces the magnetic field.

Other correlation functions induce no further symmetry generators.
For example, one can evaluate the 
correlation function of the 0-form and 2-form symmetry generators,
\begin{equation}
 \vevs{U_{\phi E} (e^{2\pi i n_{\phi} /N}, {\cal V})
U_{\phi M} (e^{i\alpha_\phi}, {\cal C})}
=  \vevs{U_{\phi M} (e^{i\alpha_\phi}, {\cal C})}.
\label{200807.0644}
\end{equation}

We would like to discuss a mathematical structure 
behind these correlation functions.
One candidate is a 2-group, which is roughly given by two groups 
and maps between them.
In terms of a 2-group, we may describe the correlation function
of 0- and 1-form symmetry generators.
However, the correlation function of the 1-form symmetry generators 
that generate a 2-form symmetry generator
cannot be described by a 2-group, since 
there is no such a structure in a 2-group.
Fortunately, we can find an appropriate structure by 
extending the 2-group to a 3-group, which we explain below.

\subsection{Global 3-group symmetry for axion electrodynamics}

Here, we review the global 3-group symmetry 
for the axion electrodynamics~\cite{Hidaka:2020iaz}.
The detail of the axioms of the 3-group is explained in appendix \ref{3g}.
Here, we summarize 
the ingredients of the 3-group $(L\os{\der_2}{\to} H \os{\der_1}{\to} G, \trr, \{-,-\})$ are as follows:
\begin{enumerate}
\item Three groups $G$, $H$, and $L$:
They are not necessarily Lie groups.

\item Maps $\der_1$ and $\der_2$ between the three groups:
$\der_1: H \to G$, $\der_2: L \to H$.
These maps are group homomorphism, i.e.,
they are compatible with group compositions.
The composition of the maps satisfies $\der_1 \circ \der_2 (l) = 1_G$
for all $l \in L$, where $1_G\in G$ is the identity element in $G$.

\item Action $\trr$ of $G$ on $G$, $H$, $L$ by automorphism:
The actions are denoted as
$g \trr g' \in G$, 
$g \trr h \in H$,
and 
 $g \trr l \in L$
 for $g, g' \in G$, $h \in H$, and $l \in L$.
In particular the action of $G$ on $G$ is 
defined by conjugation:
$g \trr g' := g g' g^{-1}$.
The actions are compatible with the group compositions.

\item Peiffer lifting $\{-,-\}: H\times H \to L$. 
In terms of the elements, 
the Peiffer lifting is written as $\{h,h'\} \in L$ for $h,h' \in H$.
The action $\trr$ is compatible with 
the Peiffer lifting: 
\begin{equation}
 g \trr \{h,h'\} = \{g \trr h, g\trr h'\}.
\label{200807.1813}
\end{equation}
\end{enumerate}

By the discussion in section \ref{3gcorr}, 
let us specify the 3-group for the axion-photon system.
First, we identify the three groups
$G$, $H$, and $L$ 
as the 0-, 1-, 2-form symmetry groups, respectively:
\begin{equation}
 G = \bb{Z}_N,
\quad
H = \bb{Z}_N \times U(1),
\quad
L = U(1).
\end{equation}

Next, we define the maps $\der_1 $ and $\der_2$ 
for the axion electrodynamics.
In the correlation function, there are 
no maps which relate the 1-form symmetry generators 
to 0-form symmetry generators, 
or 2-form symmetry generators to 1-form symmetry 
generators.
Therefore, we define these maps as follows,
\begin{equation}
 \der_1 (e^{2\pi i m/N}, e^{i\alpha}) = 1,
\quad
\der_2  e^{i\beta} =(1,1)
\end{equation}
for all $(e^{2\pi i m/N}, e^{i\alpha}) \in H$
and $e^{i\beta} \in L$.
Note that the requirement $\der_1 \circ \der_2 =1$ is 
trivially satisfied.

Third, we consider the action $\trr$ of $G$ on 
$G$, $H$, and $L$.
Since $G = \bb{Z}_N$ is Abelian, 
the conjugation is trivial: $g g' g^{-1} = g'$.
Therefore, the action of $G$ on itself is defined by 
a trivial one: $g \trr g' = g'$.
However, the action of $G$ on $H$ should be nontrivial, 
since the correlation function in \er{200805.1726} 
implies that $G$ can act on $H$.
We define the action of $G$ on $H$ following 
the correlation function in \er{200805.1726}.
For $e^{2\pi in/N} \in G = \bb{Z}_N$, 
$(e^{2\pi i m/N}, e^{i\alpha}) \in H = \bb{Z}_N \times U(1)$, 
the action $\trr$ is given by
\begin{equation}
 e^{2\pi in/N} \trr (e^{2\pi i m/N}, e^{i\alpha})
 = 
 (e^{2\pi i m/N}, e^{-2\pi i n m/N}e^{i\alpha}).
\end{equation}
Since the action of the 0-form symmetry generator 
on the 2-form symmetry generator is trivial as in \er{200807.0644}, 
we define the action of $G$ on $L$ as
\begin{equation}
 e^{2\pi in/N} \trr e^{i\beta}
 = e^{i\beta}.
\label{200807.1823}
\end{equation}

Finally, we identify the Peiffer lifting.
Since the Peiffer lifting generates an element of $L$ 
from the two elements of $H$,
we can relate the Peiffer lifting to the correlation function in 
\er{200807.1426}.
The diagrammatic expression in \er{200909.0052} 
and the correlation function in \er{200807.1426} 
suggest that we define the Peiffer lifting 
such that it satisfies
\begin{equation}
 \{(e^{2 \pi i m/N}, e^{i \alpha}),(e^{2 \pi i m'/N}, e^{i\alpha'}) \}
\{(e^{2 \pi i m'/N}, e^{i \alpha'}),(e^{2 \pi i m/N}, e^{i\alpha}) \}
= e^{2\pi i mm'/N} \in L.
\label{200913.2020}
\end{equation}
Since $L = U(1)$ is Abelian and 
the right-hand side is symmetric under $m \corr m'$, 
we may introduce the Peiffer lifting as\footnote{We choose that a different definition of the Peiffer lifting from our previous paper to be consistent with the background gauging, although the previous definition, $\{(e^{2 \pi i m/N}, e^{i \alpha}),(e^{2 \pi i m'/N}, e^{i\alpha'}) \}
= e^{2\pi i mm'/N}$, is also consistent with the axiom of the 3-group~\cite{Hidaka:2020iaz}.} 
\begin{equation}
   \{(e^{2 \pi i m/N}, 
e^{i \alpha}),(e^{2 \pi i m'/N}, e^{i\alpha'}) \}
= e^{2\pi i mm'/2N}.
\end{equation}
Although this definition has ambiguities under the shift 
$m \to m+N$ and $m' \to m'+ N$, 
\er{200913.2020} is unambiguous.
As we discuss later, 
we can show that 
the gauge transformation law and field strength of the 3-form gauge field 
in the 3-group gauge theory 
match the one obtained by the background gauging procedure
in \er{200910.1735} by this definition.%
\footnote{In order to define the Peiffer lifting itself 
in an unambiguous way, 
we may need to treat the spin structure explicitly.
We leave this issue as future work.}

We have defined the three groups, actions, and Peiffer lifting.
We should confirm that these satisfy the 
axioms of the 3-group, summarized in \ref{3gaxiom}.
In particular, it is nontrivial 
is to confirm the compatibility between the
action and Peiffer lifting in \er{200807.1813}.
The right-hand side of \er{200807.1813} can be evaluated as 
\begin{equation}
\begin{split}
&  \{ e^{2\pi i n/N} \trr (e^{2\pi i m/N}, e^{i\alpha}),
 e^{2\pi i n/N} \trr(e^{2\pi i m'/N}, e^{i\alpha'}) \}
\\
&
= 
  \{ (e^{2\pi i m/N},  e^{-2\pi i nm/N}e^{i\alpha}),
 (e^{2\pi i m'/N},  e^{-2\pi i nm'/N}e^{i\alpha'}) \}
\\
&
= 
e^{2\pi i mm'/2 N}.
\end{split}
\end{equation}
Meanwhile, the left-hand side of \er{200807.1813} is 
\begin{equation}
 e^{2\pi i n/N} \trr e^{2\pi i mm'/2N} = e^{ 2\pi i mm'/2 N},
\end{equation}
where we have used \er{200807.1823}.
Therefore, we have confirmed that 
the compatibility in \er{200807.1813} is satisfied.

We can understand that the correlation functions between the 
symmetry generators
in \ers{200805.1726}--\eqref{200807.0644} 
as a 3-group generalization of the current algebra,
which imply that a conserved current can be 
a source of another current.
Note that the correlation functions between symmetry generators,
but not between the conserved currents, 
are physically meaningful in our case. 
This is because the conserved currents $j_{\phi E}$ and $j_{a E}$ are not gauge invariant.
For the detailed discussion, 
see appendix~\ref{corrsym} and 
the previous paper of the present authors~\cite{Hidaka:2020iaz}.

\subsection{Gauging 3-group symmetry}

Here, we consider the gauging the 3-group symmetry in terms of the 
3-group gauge theory, and show that the gauging based on 
the 3-group gauge theory is consistent with the gauging
which avoids the operator shifts in section \ref{allgauge}.

In order to establish the 3-group gauge theory, 
we need the Lie algebra of the 3-group in the axion electrodynamics.
The Lie algebra of $U(1)$ is $i \bb{R}$, 
but the Lie algebra of $\bb{Z}_N$ 
does not exist since $\bb{Z}_N$ is a discrete group.
However, we can introduce a $\bb{Z}_N$ $p$-form gauge field $\omega_p$
by embedding $\bb{Z}_N$ to $U(1)$, in which 
 a $p$-form gauge field satisfies the condition
$N A_p = d A_{p-1}$ with the normalization 
$\int_{\Sigma_{p-1}} d A_{p-1} \in 2\pi \bb{Z}$ 
for a $(p-1)$-dimensional subspace $\Sigma_{p-1}$.

By using $\bb{Z}_N$ gauge fields as well as $U(1)$ gauge fields,
we now establish the background gauging of the 3-group.
The 1-, 2-, 2-, and 3-form 
gauge fields $A_1$, $B^{E}_2$, $B^{M}_2$, and $C_3$ 
denote the gauge fields of $\bb{Z}_N$, $\bb{Z}_N \times U(1)$, 
and $U(1)$ of the 3-group 
$(U(1) \to \bb{Z}_N \times U(1)\to \bb{Z}_N, \trr, \{-,-\})$,
respectively.
Here, the gauge fields $A_1$ and $B^E_2$ are constrained by
 $N A_1 = d A_0$ and $NB^E_2 = dB^E_1$
 for the 0- and 1-form gauge fields $A_0$ and $B^E_1$ with 
 proper normalization, respectively.
The action $\trr$ and the Peiffer lifting $\{-,-\}$
determine the gauge transformation laws and field strengths.
For the gauge transformation laws, they are given in 
\ers{200715.2156}--\eqref{200715.2203} as
\begin{align}
A_1 &\to A_1 +  d \Lambda_0,\\
\begin{split}
\begin{pmatrix}
  B^E_2\\
  B^M_2
  \end{pmatrix}
&
\to
\begin{pmatrix}
  B^E_2\\
  B^M_2
  \end{pmatrix}
  +
  \begin{pmatrix}
  d\Lambda^E_1\\
  d \Lambda^M_1
  \end{pmatrix}
 - \Lambda_0 \trr 
 \begin{pmatrix}
 B^E_2\\
 B^M_2
 \end{pmatrix} 
+ (A_1 +  d \Lambda_0) \trr 
\begin{pmatrix}
d\Lambda^E_1\\
d \Lambda^M_1
\end{pmatrix}
\\
&
=\begin{pmatrix}
B^E_2 + d \Lambda^E_1 \\
B^M_2 + d \Lambda^M_1 + \fr{N}{2\pi}\Lambda_0 B_2^E 
-\fr{N}{2\pi } (A_1 +  d \Lambda_0) \wed \Lambda^E_1
\end{pmatrix},
\end{split}\\
\begin{split}
C_3 &
\to C_3 + d\Lambda_2 
- \{B^E_2 + d\Lambda^E_1, \Lambda^E_1 \}
- \{\Lambda^E_1, B^E_2 \}
\\
& =
C_3 + d\Lambda_2 
- \fr{N}{2\pi} \Lambda^E_1 \wed  B^E_2 
-\fr{N}{4\pi} \Lambda^E_1 \wed d\Lambda^E_1.
\end{split}
\label{200715.1602}
\end{align}
We note that it is necessary to treat the two-form gauge fields as a pair of gauge fields $(B^E_2, B^M_2)$ because $\Lambda_0 \trr (B^E_2, B^M_2)$ does not act on $B^E_2$ and $B^M_2$, independently.
The field strengths are given in \ers{200715.2206}--\eqref{200715.2207} as
\begin{align}
F_2 &= dA_1 = 0,\\
\begin{pmatrix}
H^E_3\\
H^M_3
\end{pmatrix}
&=
\begin{pmatrix}
dB^E_2\\
dB^M_2
\end{pmatrix}
+A_1 \trr
\begin{pmatrix}
B^E_2\\
B^M_2
\end{pmatrix}
=\begin{pmatrix}
0\\
dB^M_2 - \fr{N}{2\pi} A_1 \wed B^E_2
\end{pmatrix},
\\
G_4 &= d C_3 + \{B^E_2,B^E_2\}
= d C_3 + \fr{N}{4\pi} B^E_2 \wed B^E_2.
\label{200715.1507}
\end{align}
By the above structure of the gauge transformation laws and field strengths,
the gauge fields coincide with the gauging that avoids operator-valued shifts.

The gauge fields $(A_1, B_2^E, B_2^M, C_3) $ 
correspond to $(A^{\phi E}_1, B_2^{aE}, B_2^{aM}, C^{\phi M}_3) $
given in section \ref{allgauge},
since the gauge transformation laws and field strengths 
coincide with each other.
Therefore, we have confirmed that the higher-form symmetries 
of the axion electrodynamics possess the 3-group structure 
in the viewpoint of the background gauging of the 3-group symmetry.
We also have the same gauged action as the one in \er{200716.1848}:
the coupling of $A_1$ and $B^E_2$ with the axion and photon 
are give by the combinations $d\phi - A_1$ and $da -B^E_2 $, 
respectively.
Furthermore, the coupling of
 $B_2^M$ and $C_3$ with 
$da - B_2$ and $da- B^E_2$ are described as five dimensional actions, 
$\fr{1}{2\pi} \int_{X_5} (da - B_2) \wed H_3$ and 
$- \fr{1}{2\pi} \int_{X_5} (d\phi - A_1) \wed G_4$, respectively.
Since this gauging procedure gives the same action, 
the 't Hooft anomalies of the higher-form symmetries is 
also the same.

One comment is in order.
As we have seen in section \ref{bgphys}, 
the gauge invariance of the background fields 
corresponds to the invariance under the topological deformations, 
i.e., the conservation laws of the conserved currents.
The deformation of the gauge transformation laws corresponds to the deformation of the conservation laws of the conserved currents.
In our case, 
the deformation of the gauge transformation laws 
can be understood by the correlation functions of the symmetry generators
in \ers{200805.1726}--\eqref{200807.0644},
which are generalizations of the ordinary current algebra.
For example, the correlation function in \er{200805.1726} implies
that the conservation law of $j_{\phi E}$
in the presence of another current $j_{aE}$ is deformed.
Note that the conserved currents $j_{\phi E}$ and $j_{a E}$ themselves are not physical observable, and we have discussed the correlation functions of the gauge invariant symmetry generators.

\section{Summary and discussion\label{sum}}
In this paper, we have studied the higher group structure 
of the higher-form symmetries and 't Hooft anomalies
in the $(3+1)$-dimensional 
axion electrodynamics in detail by using the background gauging.
We have found that the axion electrodynamics offers a simple model exhibiting the 3-group structure and 't Hooft anomalies of the 3-group.

We have discussed two independent gauging procedures.
One is to formulate the gauged action where the background gauge fields 
are coupled with the symmetry generators.
This procedure does not rely on the existence of a 3-group structure.
We have determined the gauge transformation laws and 
gauge invariant field strengths by the requirement that
the gauged action preserves the gauge invariance
for the dynamical fields.
The other is the gauging based on the 3-group gauge theories.
This gauging can be established by using a global 3-group structure
 in Ref.~\cite{Hidaka:2020iaz} and 
a mathematical procedure in the 3-group gauge theory.
By comparing the gauge transformation laws and field strengths of the 
background gauge fields,
we have then shown that 
the above two independent gauging procedures give the same result.
Furthermore, we have determined 
the 't Hooft anomalies of the global 3-group symmetry.
In particular, we have found a 2-group anomaly, which 
forbids a simultaneous gauging of 
the $\bb{Z}_N$ 0-form and electric $\bb{Z}_N$ 1-form symmetries.

There are several avenues for future work.
One is to analyze physics in the background magnetic field, 
spatially varying axion field, and so on. 
It has been shown that in non-trivial backgrounds, 
mass spectra of the axion and photon are 
deformed~\cite{Yamamoto:2015maz,Ozaki:2016vwu,Brauner:2017mui,Sogabe:2019gif}.
We may understand such deformations of the phase structure by using
higher-form symmetries, 3-group, and their 't Hooft anomalies 
(see e.g., Ref.~\cite{Yamamoto:2020vlk} for recent discussion).
Another important direction is to discuss what happens 
for higher-form symmetries 
when the axion becomes massive, 
while in this work we have assumed that the axion is massless.
When the axion becomes massive by non-perturbative effects, 
there can be axionic domain walls, which have a topological domain wall 
charge.
In the absence of the photon, 
it has been shown that there can be a discrete 3-form symmetry whose 
charged object is a worldvolume of the axionic domain 
wall~\cite{Hidaka:2019mfm}.
Therefore, the higher-group structure may be deformed 
in the presence of the 3-form symmetry.

\subsection*{Acknowledgements}
R.~Y.~thanks Yuji Hirono, Taro Kimura,
and Naoki Yamamoto for discussions.
This work is supported in part by Japan Society of Promotion of Science 
(JSPS) Grant-in-Aid for Scientific Research 
(KAKENHI Grants No. 17H06462, 18H01211 (Y.~H.) and 18H01217 (M.~N.)).
\appendix

\section{'t Hooft loop and worldsheet of axionic string}
\label{dual}

Here, we summarize the expressions of the 
't Hooft loop and the worldsheet of the 
axionic string in terms of local fields
by using dual transformations.
The 't Hooft loop and the worldsheet of the 
axionic string can be expressed as line and surface integrals of 
a 1-form and a 2-form gauge fields, which are dual of $a$ and $\phi$,
respectively.

\subsection{'t Hooft loop}

First, we express the 't Hooft loop in terms of local fields.
In the original formulation based on $a$ and $\phi$, 
the configuration of the 't Hooft loop $T(q_{a M}, {\cal C})$
can be expressed as a singular part of $a$~\cite{Kleinert:2008zzb}.
We decompose $a$ into the singular part $a_{\rm S}$ 
and the regular part $a_{\rm R}$ as
\begin{equation}
 a = a_{\rm R} + a_{\rm S}.
\end{equation}
Here, we require the configuration of the $a$ due to the monopole 
is expressed by $a_{\rm S}$:
\begin{equation}
 \int_{\cal S} da =
 \int_{\cal S} d(a_{\rm R} + a_{\rm S})
 = 
 \int_{\cal S} d a_{\rm S}
 = 2\pi q_{ a M} \link ({\cal S,C}).
\end{equation}
The singular part $a_{\rm S}$ can be understood 
as the 1-form that breaks the Bianchi identity,
\begin{equation}
 dda =
 dd a_{\rm S} = 2 \pi q_{a M} \delta_3 ({\cal C}),
\label{201128.1747}
\end{equation}
which represents the existence of the magnetic monopole current.
In the path integral formalism, we can express the 't Hooft loop as
\begin{equation}
 \vevs{T (q_{a M}, {\cal C})}
 = \int {\cal D}[\phi, a_{\rm R}]
e^{i S[\phi, a_{\rm R} + a_{\rm S}]}.
\label{201130.1454}
\end{equation}

Now, we consider the explicit form of the 't Hooft loop in
terms of a local field.
This can be done by the dual transformation 
of the photon $a$~\cite{Witten:1995gf}.
We can rewrite \er{201130.1454} by using the Fourier transformation, 
\begin{equation}
  \vevs{T (q_{a M}, {\cal C})}
=
  \int {\cal D}[\phi, a_{\rm R}, f', g]
e^{i  S_{a, {\rm 1st} } [\phi, a_{\rm R}+ a_{\rm S}, f',g]},
\end{equation}
where $S_{a, {\rm 1st} } [\phi, a_{\rm R}+ a_{\rm S}, f',g]$
 is a first order derivative action for $a$:
\begin{equation}
 S_{a, {\rm 1st} }
= 
-\fr{1}{2e^2}\int_{M_4} f' \wed \star  f'
-\fr{v^2}{2}\int_{M_4} d\phi \wed \star d\phi 
+ \fr{N}{8\pi^2 }\int_{M_4} \phi f' \wed f'
- \fr{1}{2\pi} \int_{M_4} g \wed (f'-  da_{\rm R} - da_{\rm S}).
\end{equation}
Here, we have introduced new dynamical valuable $f'$ and $g$, 
which are 2-form fields independent of $a$ and $\phi$.

We can go back to the original action in \er{200628.1231}
by integrating out $g$ and $f'$, where 
the integral of $g$ 
gives us the delta function
$\delta [f' - d(a_{\rm R} + a_{\rm S})] $.
Instead, we can go to the dual action as follows.
By integrating out $a_{\rm R}$, we have the delta function 
$\delta [dg] $, which implies that $g$ 
can be locally given by a 1-form gauge field $w$,
\begin{equation}
 g= dw,
\end{equation}
with a gauge transformation by a 0-form gauge parameter $\lambda_w$,
\begin{equation}
 w \to w + d\lambda_w.
\end{equation}
Therefore, $\vevs{T(q_{a M}, {\cal C})}$ can be written as
\begin{equation}
   \vevs{T (q_{a M}, {\cal C})}
= 
  \int {\cal D}[\phi,v, f']
e^{i ( -\fr{1}{2e^2}\int_{M_4} f' \wed \star  f'
-\fr{v^2}{2}\int_{M_4} d\phi \wed \star d\phi 
+ \fr{N}{8\pi^2 }\int_{M_4} \phi f' \wed f'
- \fr{1}{2\pi} \int_{M_4} dw \wed (f'- da_{\rm S}))}.
\end{equation}
We find that the term given by the singular part, 
$e^{-\fr{i}{2\pi} \int dw \wed da_{\rm S}}$, 
can be expressed by a line integral of 
$w$ along ${\cal C}$:
\begin{equation}
e^{-\fr{i}{2\pi} \int_{M_4} dw \wed da_{\rm S}}
 = 
 e^{i q_{a M}\int_{M_4} w \wed \delta_3({\cal C})}
= 
e^{ i q_{a M}\int_{\cal C} w} ,
\end{equation}
where we have used \er{201128.1747} and then \er{eq:deltaFunctionalForm}.
This is the expression of the 
desired 't Hooft loop in terms of the local field.

\subsection{Worldsheet of axionic string}
Similarly, we can express the worldsheet of the axionic string 
in terms of a local field.
In the original formulation, 
the configuration of  $V(q_{\phi M}, {\cal S})$
can be expressed as a singular part of $\phi$.
We again decompose $\phi$ into the singular part $\phi_{\rm S}$ 
and the regular part $\phi_{\rm R}$ as
\begin{equation}
 \phi = \phi_{\rm R} + \phi_{\rm S},
\end{equation}
where we have assumed that 
$\phi_{\rm S}$ have non-trivial winding number,
\begin{equation}
 \int_{\cal C} d\phi  =
 \int_{\cal C} d(\phi_{\rm R} + \phi_{\rm S})
 = 
 \int_{\cal C} d \phi_{\rm S}
 = 2\pi q_{ \phi M} \link ({\cal C,S}).
\end{equation}
The singular part $\phi_{\rm S}$ can be understood 
as the function that breaks the Bianchi identity of the axion,
\begin{equation}
 dd\phi =
 dd \phi_{\rm S} = 2 \pi q_{\phi M} \delta_2 ({\cal S}).
\label{201128.1816}
\end{equation}
In the path integral formalism, we can express $V(q_{\phi M},{\cal S})$ 
as
\begin{equation}
 \vevs{V(q_{\phi M},{\cal S})}
 = \int {\cal D}[\phi_{\rm R}, a]
e^{i S[\phi_{\rm R}+ \phi_{\rm S}, a]}.
\label{201130.1708}
\end{equation}

Now, we consider an
alternative expression of $V(q_{\phi M}, {\cal S})$ 
by the dual transformation of the axion $\phi$ to a
2-form gauge field~\cite{Lee:1993ty}.
We can again rewrite \er{201130.1708}
 by using the Fourier transformation, 
\begin{equation}
 \vevs{V(q_{\phi M},{\cal S})}
 = 
  \int {\cal D}[\phi, a_{\rm R}, \zeta, h]
e^{i  S_{\phi, {\rm 1st} } [\phi_{\rm R}+ \phi_{\rm S}, a, \zeta, h ]}.
\end{equation}
Here, $S_{\phi, {\rm 1st} } [\phi_{\rm R}+ \phi_{\rm S}, a, \zeta, h ]$
 is a first order derivative action for $\phi$:
\begin{equation}
 S_{\phi , {\rm 1st} }
= 
-\fr{1}{2e^2}\int_{M_4} da \wed \star  da
-\fr{v^2}{2}\int_{M_4} \zeta \wed \star \zeta 
- \fr{N}{8\pi^2 }\int_{M_4} \zeta \wed a \wed f
+ \fr{1}{2\pi} \int_{M_4} h \wed (\zeta -  d\phi_{\rm R} - d\phi_{\rm S}).
\end{equation}
We have introduced new dynamical valuable $\zeta$ and $h$, 
which are 1-form and 3-form fields independent of $a$ and $\phi$.
Note that the 3-form $h$ should be shifted under the gauge transformation 
of the 1-form field as
\begin{equation}
a \to a + d\lambda, \quad
 h \to h - \fr{N}{4\pi} d\lambda \wed da
\label{201128.1842}
\end{equation}
in order to make the action to be gauge invariant.
As in the case of $S_{a, {\rm 1st}}$,
we can go back to the original action in \er{200628.1231}
by integrating out $\zeta$ and $h$, where 
the integral of $h$ 
gives us the delta function
$\delta [\zeta - d(\phi_{\rm R} + \phi_{\rm S})] $.
Instead, we can again go to the dual action.
By integrating out $\phi_{\rm R}$, we have the delta function 
$\delta [dh] $ implying that $h$ 
can be locally given by a 2-form gauge field $b$,
\begin{equation}
 h = db,
\end{equation}
with a gauge transformation by a 1-form gauge parameter $\lambda_b$,
\begin{equation}
 b \to b + d\lambda_b ,
\end{equation}
in addition to the one corresponding to \er{201128.1842},
\begin{equation}
 b \to b - \fr{N}{4\pi} \lambda da.
\label{201128.1849}
\end{equation}
Substituting the condition $h= db$,
 $ \vevs{V(q_{\phi M},{\cal S})}$ can be written as
\begin{equation}
 \vevs{V(q_{\phi M},{\cal S})}
= 
  \int {\cal D}[b, a, \zeta]
e^{i ( -\fr{1}{2e^2}\int_{M_4} da \wed \star  da
-\fr{v^2}{2}\int_{M_4} \zeta \wed \star \zeta 
- \fr{N}{8\pi^2 }\int_{M_4} \zeta \wed a \wed f
+ \fr{1}{2\pi} \int_{M_4} db \wed (\zeta - d\phi_{\rm S}))}.
\end{equation}
We find that the term given by the singular part, 
$e^{-\fr{i}{2\pi} \int_{M_4} db \wed d\phi_{\rm S}}$, 
can be expressed by a surface integral of 
$b$ along ${\cal S}$:
\begin{equation}
e^{-\fr{i}{2\pi} \int_{M_4} db \wed d\phi_{\rm S}}
 = 
 e^{i q_{a M}\int_{M_4} b \wed \delta_2 ({\cal S})}
= 
e^{ i q_{a M}\int_{\cal S} b},
\end{equation}
where we have again used \er{201128.1816} 
and \er{eq:deltaFunctionalForm}.
This is the expression of the worldsheet of the axionic string.

We remark that this expression of the axionic string is not invariant 
under \er{201128.1849}.
Therefore, the gauge invariance for the photon seems to be violated 
on the axionic string.
The inconsistency can be resolved by taking  into account 
the chiral mode on the axionic string.
The chiral mode charged under the $U(1)$ gauge symmetry cancels 
the violation.
This mechanism is nothing but 
the anomaly inflow mechanism~\cite{Callan:1984sa,Naculich:1987ci}.
Note that we do not need to consider the contribution from the chiral modes
in the discussion in this paper, 
since we focus on the bulk physics around the axionic string.

\section{Correlation functions\label{corr}}
Here, we summarize detailed calculations of the correlation functions.

\subsection{{$\bb{Z}_N$} 0-form transformation\label{corr0}}
We first evaluate the 0-form transformation law in 
\er{200713.1906}. 
The transformation can be written
by the following correlation function:
\begin{equation}
 \vevs{
U_{\phi E} (e^{2\pi i n_\phi/N},{\cal V})
 e^{i q_{\phi E} \phi ({\cal P})}
}
= 
{\cal N} \int {\cal D}[\phi,a]
e^{iS 
+ 
\fr{2\pi i n_\phi}{N}\int_{\cal V} j_{\phi E}
+
i q_{\phi E} \phi ({\cal P}) }.
\end{equation}
Here, ${\cal N}$ is the normalization factor 
such that $\vevs{1} =1$.
In order to evaluate 
the correlation function, 
we rewrite the point operator 
and symmetry generator in term of spacetime integral
by using delta-function forms.
The point operator $e^{i q_{\phi E} \phi ({\cal P})}$ 
can be rewritten as
\begin{equation}
e^{i q_{\phi E} \phi ({\cal P})}
 = e^{i q_{\phi E} \int_{M_4} \phi \wed \delta_4 ({\cal P})}.
\end{equation}
Here, the delta-function form is defined in Eq.~\eqref{eq:deltaFunctionalForm}.
Suppose  ${\cal V}$ can be expressed as the boundary of a 4-dimensional subspace $\Omega_{\cal V}$. 
In general, it may not be taken as a boundary, but it is necessary for discussing the transformation law.
The symmetry generator can be rewritten as
\begin{equation}
 \int_{\cal V} j_{\phi E}
=  
 \int_{\der \Omega_{\cal V}} j_{\phi E}
=  
 \int_{\Omega_{\cal V}} d j_{\phi E}
= \int_{M_4} d j_{\phi E} \delta_0(\Omega_{\cal V}).
\label{200803.2352}
\end{equation}
We can eliminate the symmetry generator by using
\begin{equation}
\begin{split}
&
S[\phi,a]
+
\fr{2\pi n_\phi}{N}\int_{M_4} 
d j_{\phi E}
\delta_0 (\Omega_{\cal V}) 
\\
&=
S[\phi - \fr{2\pi n_\phi}{N} \delta_0 (\Omega_{\cal V}),a]
+\fr{v^2}{2} \(\fr{2\pi n_{\phi}}{N}\)^2
\int_{M_4} 
\delta_1 ({\cal V})
\wed 
\star \delta_1 ({\cal V}) ,
\end{split}
\label{200803.2353}
\end{equation}
and the redefinition 
$\phi - \fr{2\pi n_\phi}{N} \delta_0 (\Omega_{\cal V}) \to \phi$ 
as
\begin{equation}
 \vevs{
U_{\phi E} (e^{2\pi i n_\phi/N},{\cal V})
 e^{i q_{\phi E} \phi ({\cal P})}
}
= 
e^{\fr{2\pi i q_{\phi E}\, n_\phi }{N} 
\int_{M_4} \delta_0 (\Omega_{\cal V}) \delta_4 ({\cal P})}
 \vevs{ e^{i q_{\phi E} \phi ({\cal P})}}.
\label{200702.0104}
\end{equation}
Here, we have regularized the 
trivial divergence 
$\int_{M_4} \delta_1 ({\cal V})
\wed 
\star \delta_1 ({\cal V}) $ 
by adding a local counter term.
The integral 
$\int_{M_4} \delta_0 (\Omega_{\cal V}) \delta_4 ({\cal P})$
in \er{200702.0104} is the intersection number 
of $\Omega_{\cal V}$ and ${\cal P}$, 
which is equal to the linking number of ${\cal V}$ and ${\cal P}$,
\begin{equation}
 \int_{M_4} \delta_0 (\Omega_{\cal V}) \delta_4 ({\cal P})
 = \link ({\cal V,P}) \in \bb{Z}.
\end{equation} 
Therefore we have 
\begin{equation}
 \vevs{
U_{\phi E} (e^{2\pi i n_\phi/N},{\cal V})
 e^{i q_{\phi E} \phi ({\cal P})}
}
= 
e^{\fr{2\pi i q_{\phi E} n_\phi }{N} \link ({\cal V,P})}
 \vevs{ e^{i q_{\phi E} \phi ({\cal P})}}.
\label{200702.0107}
\end{equation}

\subsection{{$\bb{Z}_N$} 1-form transformation\label{corr1}}
Second, we consider the correlation function, which 
represents the $\bb{Z}_N$ electric 1-form transformation:
\begin{equation}
 \vevs{
U_{a E} (e^{2\pi i n_a /N},{\cal S})
 e^{i q_a \int_{\cal C} a }
}
= 
\int {\cal D}[\phi,a]
e^{iS 
+ 
\fr{2\pi i n_a}{N}\int_{\cal S} j_{a E}
+
i q_a \int_{\cal C} a}.
\end{equation}
The Wilson loop  $e^{i q_a \int_{\cal C} a}$ 
can be rewritten as
\begin{equation}
e^{i q_a \int_{\cal C} a}
 = e^{i q_a \int_{M_4} a \wed \delta_3({\cal C})}.
\end{equation}
Similarly, 
the symmetry generator can be rewritten as
\begin{equation}
 \int_{\cal S} j_{a E}
=  
 \int_{\der {\cal V_S}} j_{a E}
=  
 \int_{{\cal V_S}} d j_{a E}
= \int_{M_4} d j_{a E} \wed  \delta_1 ({\cal V_S}),
\label{200804.0013}
\end{equation}
where we have assumed that ${\cal S}$ can be written as the boundary of a three dimensional subspace ${\cal V}_{\cal S}$.
We can eliminate  the symmetry generator by using
\begin{equation}
\begin{split}
&
S[\phi,a]
+
\fr{2\pi n_a }{N} \int_{M_4} 
d j_{a E} \wed 
\delta_1 ({\cal V_S}) 
\\
&=
S[\phi ,a + \fr{2\pi n_a}{N} \delta_1 ({\cal V_S})]
\\
&
\quad
-
\fr{N}{8\pi^2} \(\fr{2\pi n_{a}}{N}\)^2
\int_{M_4}\phi 
\delta_2 ({\cal S})
\wed 
\delta_2 ({\cal S}) 
+
\fr{1}{2 e^2} \(\fr{2\pi n_{a}}{N}\)^2
\int_{M_4}
\delta_2 ({\cal S})
\wed 
\star \delta_2 ({\cal S}) ,
\end{split}
\label{200804.0014}
\end{equation}
and the redefinition 
$a + \fr{2\pi n_a}{N} \delta_1 ({\cal V_S}) \to a$
as
\begin{equation}
 \vevs{
U_{a E} (e^{2\pi i n_a/N},{\cal V})
 e^{i q_a \int_{\cal C} a}
}
= 
e^{\fr{2\pi i q_{a} n_a }{N} 
\int_{M_4}\delta_3 ({\cal C}) \wed \delta_1 ({\cal V_S})}
 \vevs{e^{i q_a \int_{\cal C} a}}.
\label{200713.1846}
\end{equation}
Here, we have again regularized the 
trivial divergence 
$\fr{1}{2 e^2} \(\fr{2\pi n_{a}}{N}\)^2
\int_{M_4}
\delta_2 ({\cal S})
\wed 
\star \delta_2 ({\cal S}) $
by the local counter term.
Furthermore, the term $\int_{M_4}\phi 
\delta_2 ({\cal S})
\wed 
\delta_2 ({\cal S}) $
in \er{200804.0014} is equal to zero if we consider 
a closed surface without self-intersections.
The integral 
$\int_{M_4}\delta_3 ({\cal C})\wed \delta_1 ({\cal V_S})$
in \er{200713.1846} is the intersection number 
of ${\cal V_S}$ and ${\cal C}$, 
which is equal to the linking number of ${\cal S}$ and ${\cal C}$,
\begin{equation}
\int_{M_4}\delta_3 ({\cal C}) \wed \delta_1 ({\cal V_S})
= 
\int_{\cal V_S} \delta_3 ({\cal C})
 = \link ({\cal S,C}) \in \bb{Z}.
\end{equation} 
Therefore, we have 
\begin{equation}
 \vevs{
U_{a E} (e^{2\pi i n_a/N},{\cal S})
 e^{i q_a \int_{\cal C}a }
}
= 
e^{\fr{2\pi i q_a n_a }{N} \link ({\cal S,C})}
 \vevs{ e^{i q_a \int_{\cal C} a}}.
\label{200713.1850}
\end{equation}

\subsection{Correlation functions of symmetry generators\label{corrsym}}

We here show the derivations of the correlation functions
of symmetry generators.
\subsubsection{Correlation function of 
0-form and 1-form symmetry generators\label{corr01}}
First, we consider the correlation function of 
0- and 1-form symmetry generators
\begin{equation}
\vevs{
U_{\phi E}(e^{2\pi i n_\phi/ N}, {\cal V})
U_{a E}(e^{2\pi i n_a/N}, {\cal S})
}
 = 
{\cal N}
\int {\cal D}[\phi,a]
e^{
iS [\phi,a]
+
 \fr{2\pi i n_\phi}{N} \int_{\cal V} j_{\phi E}
+
 \fr{2\pi i n_a}{N} \int_{\cal S} j_{a E}
},
\end{equation}
 which 
corresponds to the Witten effect~\cite{Hidaka:2020iaz}.
We first eliminate the 0-form symmetry generator
$U_{\phi E}(e^{2\pi i n_\phi/ N}, {\cal V})$.
By the same procedures as \ers{200803.2352} and \eqref{200803.2353}, 
we obtain 
\begin{equation}
\begin{split}
&
\vevs{
U_{\phi E}(e^{2\pi i n_\phi/ N}, {\cal V})
U_{a E}(e^{2\pi i n_a/N}, {\cal S})
}
\\
&
 = 
\vevs{U_{a E}(e^{2\pi i n_a/N}, {\cal S})
e^{-\fr{ i n_\phi n_a }{ N} \int_{\cal S} \delta_0 (\Omega_{\cal V}) da}}
\\
&
= 
\vevs{U_{a E}(e^{2\pi i n_a/N}, {\cal S})
U_{aM} (e^{- 2\pi i n_\phi n_a / N} ,\Omega_{\cal V} \cap {\cal S})
}.
\end{split}
\end{equation}
Alternatively, one can eliminate 
the 1-form symmetry generator as follows.
Here, we assume that ${\cal V}$ and ${\cal S}$ are 
not intersected to each other.
By using \ers{200804.0013} and \eqref{200804.0014},
we obtain
\begin{equation}
  \begin{split}
&
\vevs{
U_{\phi E}(e^{2\pi i n_\phi/ N}, {\cal V})
U_{a E}(e^{2\pi i n_a/N}, {\cal S})
}
\\
&
 = 
\vevs{
U_{\phi E}(e^{2\pi i n_\phi /N}, {\cal V})
e^{- \fr{n_\phi n_a}{N} \int
da
\wed 
\delta_1({\cal V})
\wed  \delta_1 ({\cal V_S})
- \fr{i \pi n_a^2 n_\phi }{N^2}
\int
\delta_1 ({\cal V_S})
\wed
d \delta_1 ({\cal V_S}) \wed \delta_1 ({\cal V})
)
}
}
\\
&
= 
\vevs{
U_{\phi E}(e^{2\pi i n_\phi /N}, {\cal V})
U_{a M}(e^{-2\pi i n_\phi n_a /N}, {\cal V} \cap {\cal V_S})
}
e^{-
\fr{i \pi n_a^2 n_\phi }{N^2}
\int
\delta_1 ({\cal V_S})
\wed
d \delta_1 ({\cal V_S}) \wed \delta_1 ({\cal V})
}.
 \end{split}
\end{equation}
The last term $e^{-
\fr{i \pi n_a^2 n_\phi }{N^2}
\int
\delta_1 ({\cal V_S})
\wed
d \delta_1 ({\cal V_S}) \wed \delta_1 ({\cal V})}$
is equal to zero, 
since we have assumed that ${\cal V}$ and ${\cal S}$ 
do not intersect. 
Therefore, we obtain 
 \begin{equation}
\vevs{
U_{\phi E}(e^{2\pi i n_\phi/ N}, {\cal V})
U_{a E}(e^{2\pi i n_a/N}, {\cal S})
}
=
\vevs{
U_{\phi E}(e^{2\pi i n_\phi /N}, {\cal V})
U_{a M}(e^{-2\pi i n_\phi n_a /N}, {\cal V} \cap {\cal V_S})
}.
 \end{equation}
The correlation function implies that 
the EOM of the photon, i.e., the conservation law of $j_{aE}$ 
is deformed in the presence of the current 
$j_{\phi E}$.
Here, the EOM of the photon corresponds to the finite redefinition 
which has been applied in \er{200914.0309}.
Note that 
we have discussed 
the deformation in terms of the gauge invariant symmetry generators,
since the currents $j_{\phi E}$ and $j_{a E}$ are not gauge invariant.

\subsubsection{Correlation function of 
1-form symmetry generators\label{corr11}}
 Similarly, we can evaluate the correlation function between 
  1-form symmetry generators,
  \begin{equation}
  \vevs{
  U_{a E}(e^{2\pi i n_a/ N}, {\cal S}_1)
  U_{a E}(e^{2\pi i n'_a/N}, {\cal S}_2)
  }
   = 
  {\cal N}
  \int {\cal D}[\phi,a]
  e^{iS [\phi,a]
  +
   \fr{2\pi i n_a}{N} \int_{{\cal S}_1} j_{a E}
  +
   \fr{2\pi i n'_a}{N} \int_{{\cal S}_2} j_{a E}
  },
  \end{equation}
  which corresponds to the anomalous Hall effect~\cite{Hidaka:2020iaz}.
  We eliminate $U_{a E}(e^{2\pi i n_a/ N}, {\cal S}_1)$
  by the same procedures as \ers{200804.0013} and \eqref{200804.0014}, 
  and obtain 
  \begin{equation}
  \begin{split}
  &
  \vevs{
  U_{a E}(e^{2\pi i n_a/ N}, {\cal S}_1)
  U_{a E}(e^{2\pi i n'_a/N}, {\cal S}_2)
  }
  \\
  &
   = 
  \vevs{
  U_{a E}(e^{2\pi i n'_a/N}, {\cal S}_2)
  e^{-\fr{ i n_a' n_a }{ N} 
  \int_{{\cal S}_2} d \phi \wed \delta_1({\cal V}_{{\cal S}_1})
  }}
  \\
  &
  = 
  \vevs{U_{a E}(e^{2\pi i n'_a/N}, {\cal S}_2)
  U_{\phi M} (e^{2\pi i n_a n'_a / N} ,-{\cal V}_{{\cal S}_1} \cap {\cal S}_2)
  }.
  \end{split}
\label{200914.0304}
  \end{equation}
We can consider another property of the correlation function.
Remarking the following relation,
\begin{equation}
   U_{a E}(e^{2\pi i n_a/ N}, {\cal S}_1)
  U_{a E}(e^{2\pi i n'_a/N}, {\cal S}_2)
 = e^{\fr{2\pi i }{N} \int_{M_4} j_{aE} \wed 
(n_a \delta_2 ({\cal S}_1) + n_a'\delta_2 ({\cal S}_2))},
\end{equation}
we can simply evaluate the
1-form symmetry generators by 
the redefinition 
$a + \fr{2\pi n_a}{N} \delta_1 ({\cal V}_{{\cal S}_1}) 
+ \fr{2\pi n_a'}{N} \delta_1 ({\cal V}_{{\cal S}_2})\to a$ 
as 
\begin{equation}
\begin{split}
& \vevs{   U_{a E}(e^{2\pi i n_a/ N}, {\cal S}_1)
  U_{a E}(e^{2\pi i n'_a/N}, {\cal S}_2)}
\\
&
 = 
\vevs{
U_{\phi M}(
e^{\fr{2 \pi i n_an_a' }{2 N}}, 
-{\cal V}_{{\cal S}_1} \cap
{\cal S}_2)
U_{\phi M}(
e^{\fr{2 \pi i n_an_a' }{2 N}}, 
-{\cal V}_{{\cal S}_2} \cap
{\cal S}_1)
} 
\\
&
 = 
\vevs{
U_{\phi M}(
e^{\fr{2 \pi i n_an_a' }{N}}, 
-{\cal V}_{{\cal S}_1} \cap
{\cal S}_2)}.
\end{split}
\label{200914.0309}
\end{equation}
This result coincides with \er{200914.0304} 
after eliminating $U_{a E}(e^{2\pi i n'_a/N}, {\cal S}_2)$
by a topological deformation.
The second line of \er{200914.0309} shows that 
the correlation function of two 1-form symmetry generators 
linking with each other
leads to two 2-form symmetry generators, 
which is consistent with the diagrammatic expression 
in \er{200909.0052}.
Note that the linking of surfaces is 
called a surface link~\cite{Carter:2001,NAKAMURA:2014}.

The correlation functions in \ers{200914.0304} and \eqref{200914.0309} 
again imply that the conservation law of $j_{aE}$ 
is deformed in the presence of another $j_{aE}$.
Note again that
we have discussed the deformation in terms of the gauge invariant symmetry generators.
\subsubsection{Correlation function of 
0-form and 2-form symmetry generators\label{corr02}}

  We also show the correlation function of the 
  0-form and 2-form symmetry generators:
  \begin{equation}
   \vevs{U_{\phi E}(e^{2\pi i n_\phi/N}, {\cal V}) 
  U_{\phi M}(e^{i\alpha_\phi},{\cal C})}
  =
  {\cal N} 
  \int {\cal D}[\phi,a]
  e^{iS + \fr{2\pi i n_\phi}{N} \int_{\cal V} j_{\phi E} 
  + \fr{i \alpha_{\phi}}{2\pi} \int_{\cal C} d\phi}.
  \end{equation}
  The redefinition
   $\phi - \fr{2\pi n_\phi}{N} \delta_0 (\Omega_{\cal V}) \to \phi$ 
  leads to
  \begin{equation}
   \vevs{U_{\phi E}(e^{2\pi i n_\phi/N}, {\cal V}) 
  U_{\phi M}(e^{i\alpha_\phi},{\cal C})}
  =
  \vevs{U_{\phi M}(e^{i\alpha_\phi},{\cal C})}
  e^{- \fr{i \alpha_{\phi}}{2\pi} \fr{2\pi n_\phi}{N} \int_{\cal C} \delta_1 ({\cal V})}.
\label{201129.1725}
  \end{equation}
  The integral $\int_{\cal C} \delta_1 ({\cal V})$ 
  in the numerical factor 
  $e^{- \fr{i \alpha_{\phi}}{2\pi} \fr{2\pi n_\phi}{N} \int_{\cal C} \delta_1 ({\cal V})}$ 
  is the transversally intersecting number of ${\cal C}$ and 
  ${\cal V}$, which is equal to zero since 
  both of ${\cal C}$ and ${\cal V}$ are closed.
  Therefore, we obtain
  \begin{equation}
   \vevs{U_{\phi E}(e^{2\pi i n_\phi/N}, {\cal V}) 
  U_{\phi M}(e^{i\alpha_\phi},{\cal C})}
  =
  \vevs{U_{\phi M}(e^{i\alpha_\phi},{\cal C})}.
\label{201129.1732}
  \end{equation}
Note that the resulting correlation function 
implies that the EOM of the axion, 
i.e., the conservation of $j_{\phi E}$ is not deformed in the presence of 
$U_{\phi M} (e^{i\alpha_\phi}, {\cal C})$.
\footnote{Technically, the correlation between $d j_{\phi E}$
and $j_{\phi M}$ is non-zero, but the correlation between 
$U_{\phi E}$ and $U_{\phi M}$ is zero due to the integration of $\delta_1({\cal V})$ over ${\cal C}$.
Since $j_{\phi E}$ is not gauge invariant, the correlation of 
the symmetry generators is physically meaningful.}

\section{3-group gauge theory\label{3g}}

In this section, we review 
the semistrict 3-group or 2-crossed module.
Hereafter, we simply refer to the semistrict 3-group as the 3-group.
We first present the axiom and a simple example of the 3-group.  
Next, we show a Lie algebra of the 3-group, and the 3-group gauge theory based on it.
Since the axioms of the 3-group are complicated compared to 
that of ordinary groups, we give a diagrammatic 
explanation of the axiom of the 3-group in Appendix~\ref{diag}.

\subsection{Axiom of 3-group\label{3gaxiom}}

A 3-group $(L\os{\der_2}{\to} H \os{\der_1}{\to} G, \trr, \{-,-\})$
satisfies the following axioms~\cite{CONDUCHE1984155} 
(see also \cite{Martins:2009evc,Saemann:2013pca,Wang:2013dwa,Palmer:2014jma,Radenkovic:2019qme}):
\begin{enumerate}
\item $G$, $H$, and $L$ are groups.
\item The maps
\begin{equation}
 \der_1 : H \to G,
\quad
\der_2: L \to H
\end{equation}
 are group homomorphisms 
$\der_{1}(h_1h_2) = (\der_1 h_1) (\der_1 h_2)$
and $\der_{2}(l_1 l_2) = (\der_2 l_1) (\der_2 l_2)$
for $h_{1,2} \in H$ and $l_{1,2} \in L$, 
respectively.
They satisfy 
\begin{equation}
 \der_1 \circ \der_2 (l)= 1_G
\end{equation}
for all $l \in L$, where $1_G\in G$ is the identity element in $G$.
 \item The symbol 
$\trr$ is an action of $g \in G$ on $g' \in G$, $h \in H$, 
and $l \in L$ by automorphisms, 
$g \trr g' \in G$,
$g\trr h \in H$, and $g \trr l \in L$.
In particular, the action $g \trr g'$ is defined by conjugation,
\begin{equation}
 g \trr g' :=  g g' g^{-1}.
\label{200622.1143}
\end{equation}

\item The maps $\der_{1,2} $ are $G$-equivalent, that is,
for all $g\in G$, $h \in H$, and $l\in L$,
\begin{equation}
 g \trr (\der_1 h) = \der_1 (g \trr h), 
\quad
 g \trr (\der_2 l) = \der_2 (g \trr l).
\label{200622.1529}
\end{equation}
\item The Peiffer lifting $\{-,-\}$ is a map $H \times H \to L$.
In terms of the elements,
\begin{equation}
 \{h_1,h_2\} \in L,
\end{equation} 
for $h_{1,2} \in H$.
The Peiffer lifting satisfies
\begin{align}
 \der_2 \{h_1,h_2\} 
 &= h_1 h_2 h_1^{-1} (\der_1 h_1) \trr h_2^{-1},
\label{200622.1536}\\
g \trr \{h_1,h_2\} 
&=  \{g \trr h_1,g \trr h_2\},
\label{200623.0212}\\
\{\der_2 l_1, \der_2 l_2 \}
&= l_1 l_2 l_1^{-1} l_2^{-1},
\label{200623.0213}\\
 \{h_1h_2, h_3\}
&=
\{h_1,h_2 h_3h_2^{-1}\}
(\der_1 h_1) \trr \{h_2, h_3\},
\label{200623.0214}\\
 \{h_1, h_2 h_3\}
 &= 
 \{h_1, h_2\}
 \{h_1, h_3\}
 \{ 
\der_2\{h_1,h_3\}^{-1}
, (\der_1 h_1) \trr h_2\},
\label{200623.0215}\\
 \{\der_2 l, h \}\{h, \der_2 l\}
 &= l (\der_1 h) \trr l^{-1},
\label{200623.0216}
\end{align}
for $h_{1,2,3} \in H$ and $l_{1,2} \in L$.
\end{enumerate}

We note that the 3-group contains a 2-group $(L \os{\der_2}{\to} H, \trr')$ as the subgroup, 
where the actions $\trr': H\to H$ and $\trr': H\to L$ are defined as
 $h\trr' h':=h h' h^{-1} $ and  $h\trr' l:=l\{\partial_2 l^{-1}, h\}$ for $h,h'\in H$ and $l\in L$, respectively.
The (strict) 2-group $(L \os{\der_2}{\to} H, \trr')$ is a set of two groups $H$ and $L$, 
a group homomorphism $\der_2: L \to H$,
and an action $\trr' $ of $H$ on $H$ and $L$.
The map $\der_2$ is compatible with the action,
$h \trr \der_2 (l) = \der_2 (h \trr l)$
for $h \in H$ and $l \in L$.
The action of $\der_2(l) \in H$ on $l'\in L$ satisfies $\der_2 (l) \trr' l' = l l' l^{-1}$, which is called the Peiffer identity.
One can easily check that $(L \os{\der_2}{\to} H, \trr')$ satisfies these axioms.
On the other hand, $(H \os{\der_1}{\to} G, \trr)$ is generally not  a 2-group.
From Eq.~\eqref{200622.1536}, $(\der_1 h) \trr h'=hh'h^{-1}\der_2 \{h,h'^{-1}\} $ follows. In this sense, the Peiffer lifting measures the failure of the Peiffer identity.

\subsection{Example of 3-group}
\label{3-groupex}

Before explaining the 3-group gauge theory,
we give a simple and non-trivial example of the 3-group.
It is given by 
an $n$-dimensional Euclidean group (or isometry group) $ISO(n)$,
\begin{equation}
ISO(n) = \left\{
\mtx{A & \bs{a}\\ \bs{0}_n^T & 1} \in M(n+1, \bb{R})| 
A \in O(n), \bs{a} \in \bb{R}^n
\right\},
\end{equation}
and we show that we can decompose the Euclidean group 
as a 3-group.
Hereafter, we abbreviate $\bs{0}^T_n $ to $0$.
Note that the product of two elements $h_1, h_2 \in ISO(n)$,
\begin{equation}
h_1 = \mtx{A_1 & \bs{a}_1 \\ 0 & 1}, 
\quad
h_2 = \mtx{A_2 & \bs{a}_2 \\ 0 & 1} ,
\end{equation}
is 
\begin{equation}
 h_1 h_2 
= 
\mtx{A_1 A_2 & A_1 \bs{a}_2 + \bs{a}_1 \\ 0 & 1} \in ISO(n). 
\end{equation}

The Euclidean group can be decomposed into the orthogonal group $O(n)$
and translation group $\bb{R}^n$.
In other words, 
there are a projection map $\der_1$ and an embedding map $\der_2$,
\begin{equation}
 \bb{R}^n \overset{\der_2}{\to} ISO(n) \overset{\der_1}{\to} O(n).
\end{equation}
The actions of $\der_2$ and $\der_1$ 
on $\bs{a}_1 \in \bb{R}^n $ and $h_2 \in ISO(n) $ are
\begin{equation}
 \der_2(\bs{a}_1) := \mtx{1_{n} & \bs{a}_1 \\ 0 & 1},
\end{equation}
and 
\begin{equation}
 \der_1 (h_2) := A_2,
\end{equation}
respectively.
These maps are compatible with products.
In fact, the embedding map $\der_2$ satisfies
\begin{equation}
 \der_2 (\bs{a}_1 + \bs{a}_2 )
 = \der_2 (\bs{a}_1) \der_2 (\bs{a}_2)
 = \der_2 (\bs{a}_2)\der_2 (\bs{a}_1) ,
\end{equation}
and $\der_1$ satisfies
\begin{equation}
 \der_1 (h_1 h_2) = \der_1 (h_1) \der_1 (h_2).
\end{equation}
The maps $\der_{1,2}$ satisfy
\begin{equation}
 \der_1 \circ \der_2 (\bs{a}) =1_n.
\label{200810.0358}
\end{equation}
Note that we would decompose $ISO(n) $ as
$ O(n) \overset{}{\to} ISO(n) \overset{}{\to} \bb{R}^n$,
but this decomposition is not compatible with the product.

The elements of $\bb{R}^n$, $ISO(n)$, and $O(n)$ are 
transformed under an action generated by $O(n)$.
We denote and define the action as
\begin{align}
 A_1 \triangleright \bs{a}_2 &:= A_1 \bs{a}_2,\\
 A_1 \trr h_2 &:= 
\mtx{A_1 & \bs{0} \\ 0 & 1 } h_2 \mtx{A_1^{-1} & \bs{0} \\ 0 & 1 }
= 
\mtx{A_1 A_2 A^{-1}_1& A_1 \bs{a}_2 \\ 0 & 1 }, 
\end{align}
and 
\begin{equation}
 A_1 \trr A_2 := A_1 A_2 A_1^{-1}.
\end{equation}
One can check that 
the actions are compatible with $\der_{1,2}$, 
\begin{equation}
\der_2( A_1 \triangleright \bs{a}_2 ) 
= A_1 \trr \der_2(\bs{a}_2),
\end{equation}
and 
\begin{equation}
 \der_1(A_1 \trr h_2) 
=
A_1 \trr \der_1(h_2) . 
\end{equation}

Finally we determine the Peiffer lifting.
Since we can obtain an element of $O(n)$ from $ISO(n)$ by
$\der_1$, we can construct an action of $ISO(n)$ itself by 
using $\der_1$ as
\begin{equation}
 \der_1 (h_1) \trr h_2
=
\mtx{A_1 & \bs{0} \\ 0 & 1 } h_2 \mtx{A_1^{-1} & \bs{0} \\ 0 & 1 }
= 
\mtx{A_1 A_2 A^{-1}_1& A_1 \bs{a}_2 \\ 0 & 1 }.
\end{equation}
On the other hand, the element of $ISO(n)$ can act on 
itself as
\begin{equation}
\begin{split}
 h_1 h_2 h_1^{-1}
& = 
\mtx{A_1 & \bs{a}_1 \\ 0 & 1 } 
\mtx{A_2 & \bs{a}_2 \\ 0 & 1 }
 \mtx{A_1^{-1} & - A^{-1}_1\bs{a}_1 \\ 0 & 1 }
\\
& = 
 \mtx{A_1 A_2 A_1^{-1} & - A_1 A_2 A^{-1}_1\bs{a}_1 + A_1 \bs{a}_2 + \bs{a}_1\\ 0 & 1 } .
\end{split}
\end{equation}
Therefore, two actions $ h_1 h_2 h_1^{-1}$ and $ \der_1 (h_1) \trr h_2$ 
are different.
The difference is the lack of $\bs{a}_1$ due to the projection 
$\der_1$.
We can measure the difference by 
\begin{equation}
h_1 h_2 h_1^{-1} (\der_1 (h_1) \trr h_2)^{-1}
 = 
 \mtx{1_n & - A_1 A_2 A^{-1}_1\bs{a}_1 + \bs{a}_1\\ 0 & 1 } ,
\end{equation}
which can be an image of $\der_2$.
Therefore, we can define the Peiffer lifting,
\begin{equation}
\{-,-\}: ISO(n)\times ISO(n) \to \bb{R}^n
\end{equation}
as
\begin{equation}
 \{h_1, h_2\}
 := - A_1 A_2 A^{-1}_1\bs{a}_1 + \bs{a}_1,
\end{equation}
which satisfies
\begin{equation}
 \der_2( \{h_1, h_2\}) = h_1 h_2 h_1^{-1} (\der_1 (h_1) \trr h_2)^{-1}.
\end{equation}
One can explicitly check that these definitions satisfy other axioms of the 3-group~\eqref{200623.0212}--\eqref{200623.0216}.

\subsection{Lie algebra of 3-group}

We consider the (background) gauging of the 3-group.
In order to introduce the gauge fields and their gauge transformation 
laws, we need the Lie algebra of the 3-group.
We denote the Lie algebra of $G$, $H$, and $L$ 
as $\frak{g}$, $\frak{h}$, and $\frak{l}$, respectively.
The Lie 3-group (differential 2-crossed module) is defined by 
the following axioms:
\begin{enumerate}
\item $\frak{g}$, $\frak{h}$, and $\frak{l}$ are Lie algebras.
\item The maps
\begin{equation}
 \der_1 : \frak{h} \to \frak{g},
\quad
\der_2: \frak{l} \to \frak{h}
\end{equation}
are $\frak{g}$-equivalent homomorphisms 
\begin{equation}
\der_{1}[\ul{h}_1, \ul{h}_2] = [\der_1 \ul{h}_1, \der_1 \ul{h}_2],
\quad
\der_{2}[\ul{l}_1 ,\ul{l}_2] = [\der_2 \ul{l}_1, \der_2 \ul{l}_2],
\end{equation}
for $\ul{h}_{1,2} \in \frak{h}$ and $\ul{l}_{1,2} \in \frak{l}$, 
respectively.
They satisfy 
\begin{equation}
 \der_1 \circ \der_2 \ul{l} = 0.
\end{equation}

 \item 
$\trr$ is an action of $\ul{g} \in \frak{g}$ 
on $\ul{g}' \in \frak{g}$, $\ul{h} \in \frak{h}$, 
and $\ul{l} \in \frak{l}$ by automorphisms, 
$\ul{g} \trr \ul{g}' \in \frak{g}$, 
$\ul{g} \trr \ul{h} \in \frak{h}$, 
and 
$\ul{g} \trr \ul{l} \in \frak{l}$.
The action $\ul{g} \trr \ul{g}' $ is defined by the commutator,
\begin{equation}
\ul{g} \trr \ul{g}' :=  [\ul{g},\ul{g}'].
\label{200713.0523}
\end{equation}

\item $\der_{1,2} $ are $\frak{g}$-equivalent, that is,
\begin{equation}
 \ul{g} \trr (\der_1 \ul{h}) = \der_1 (\ul{g} \trr \ul{h}), 
\quad
 \ul{g} \trr (\der_2 \ul{l}) =   \der_2 (\ul{g} \trr \ul{l}).
\label{200713.0543}
\end{equation}
\item The Peiffer lifting $\{-,-\}$ is a map 
$\frak{h} \times \frak{h} \to \frak{l}$.
In terms of the elements,
\begin{equation}
 \{\ul{h}_1,\ul{h}_2\} \in \frak{l}
\end{equation} 
for $\ul{h}_{1,2} \in \frak{l}$.
The Peiffer lifting satisfies
\begin{align}
 \der_2 \{\ul{h}_1,\ul{h}_2\} 
 &= [\ul{h}_1, \ul{h}_2] - 
 (\der_1 \ul{h}_1)  \trr \ul{h}_2,
\label{200713.0545}\\
\ul{g} \trr \{\ul{h}_1,\ul{h}_2\}  
&= 
\{\ul{g} \trr  \ul{h}_1,\ul{h}_2\}  
+
 \{\ul{h}_1,\ul{g} \trr \ul{h}_2\},
\label{200713.0547}\\
\{\der_2 \ul{l}_1, \der_2 \ul{l}_2 \}
& = [\ul{l}_1 ,\ul{l}_2],
\label{200713.0549}\\
\begin{split}
 \{[\ul{h}_1,\ul{h}_2], h_3\}
& =
\{\ul{h}_1,[\ul{h}_2, \ul{h}_3]\}
+
(\der_1 \ul{h}_1) \trr \{\ul{h}_2, \ul{h}_3\}
\\
&\quad
-
\{\ul{h}_2,[\ul{h}_1, \ul{h}_3]\}
-
(\der_1 \ul{h}_2) \trr \{\ul{h}_1, \ul{h}_3\},
\end{split}
\label{200713.0556}\\
 \{ \ul{h}_1, [\ul{h}_2, \ul{h}_3]\},
& = 
 \{\der_2\{\ul{h}_1,\ul{h}_2\}, \ul{h}_3\}
-
 \{\der_2\{\ul{h}_1,\ul{h}_3\}, \ul{h}_2\},
\label{200713.0609}\\
 \{\der_2 \ul{l}_1, \ul{h}_2 \}
 +
 \{\ul{h}_2, \der_2 \ul{l}_1\}
&= -(\der_1 \ul{h}) \trr \ul{l}_1,
\label{200713.0616}
\end{align}
for $\ul{h}_{1,2} \in H$ and $\ul{l}_{1,2} \in L$.
\end{enumerate}

\subsection{3-group gauge theory}
Now, we formulate a 3-group gauge 
theory~\cite{Martins:2009evc,Saemann:2013pca,Wang:2013dwa}.
We introduce 1-, 2-, and 3-form gauge fields,
$A_1$, $B_2$, and $C_3$, which are $\frak{g}$-, $\frak{h}$-, 
and $\frak{l}$-valued differential forms, respectively.
If we write the basis of the Lie algebras $\frak{g}$, 
$\frak{h}$, and $\frak{l}$ as
$\{ u_A\}$,
$\{ v_a\}$, 
and
$\{ w_\alpha \}$, respectively, 
the gauge fields can be written as 
$A_1 = A_1^A u_A$, 
$B_2 = B^a v_a$,
and 
$C_3 = C_3^\alpha w_\alpha$, 
respectively.

By the structure of the Lie algebra, 
the gauge transformation laws are given as follows.
Let $g$, $h_1 = h_1^a v_a$, and $l_2 = l_2^\alpha w_\alpha$ be 
the $G$-, $\frak{h}$-, and $\frak{l}$-valued 0-, 1-, and 2-form gauge parameters, respectively. 
Then the gauge transformations are  given by
\begin{align}
  A_1 \to A_1' &=  g \trr A_1 +  g d g^{-1} + \der_1 h_1,
  \label{200715.2156}\\
  B_2\to B_2' &=  g \trr B_2 + d h_1 - h_1 \wed h_1 + A'_1 \trr h_1
+ \der_2 l_2,\\
  C_3 \to C_3' &=  g \trr C_3+ d l_2 + A'_1 \trr l_2
+
\{\partial_2 l_2, h_1\}- \{ B'_2, h_1 \} -  \{ h_1, g^{-1} \trr B_2 \}.
  \label{200715.2203}
\end{align}
Here, we have used the following notations:
\begin{align}
 g \trr A_1
  &= A^A_1 g \trr u_A   = A^A_1 (g u_A g^{-1}), \\
    g \trr B_2 
    &= B^a_2 (g \trr v_a),\\
    g \trr C_3 
    &= C^\alpha_3 (g \trr w_\alpha),\\
    h_1 \wed h_1 
    &= h_1^a \wed h_1^b v_a v_b 
    = \fr{1}{2} h_1^a \wed h_1^b [v_a, v_b],\\
    A_1' \trr h_1
    &= (A_1')^A  \wed h_1^a (u_A \trr v_a),\\
    \der_1 h_1 &= h_1^a (\der_1 v_a),\\
    \der_2 l_2 &= l_2^\alpha (\der_2 w_\alpha),\\
    \{B_2' , h_1\} &= (B_2')^a \wed h_1^b \{ v_a, v_b\} .
\end{align}

The field strengths are defined by
\begin{align}
 F &:= dA_1 + A_1\wedge A_1,
 \label{200715.2206} \\
 H &:= d B_2 + A_1 \trr B_2,\\
 G &:= dC_3 + A_1 \trr C_3 + \{B_2, B_2\}.
\label{200715.2207} 
\end{align}
Note that the 3-group gauge theory for 
the axion electrodynamics can be obtained by 
substituting $u_A = -i$, $v_a = -i$, $w_\alpha = -i$,
$g  = e^{i\Lambda_0}$,
$h_1 = -i \Lambda_1$, and $l_2 = -i\Lambda_2$.

\section{Diagrammatic expression of 3-group\label{diag}}

Here, we explain the 3-group diagrammatically.
The definition of the 3-group based on the axioms 
seems abstract, but we show that the 
axioms of the 3-group can be understood in 
a more intuitive way.
In particular, we show that all of the axioms of the 
3-group can be translated to the only one simple statement 
``the group elements are topological''
in the viewpoint of the higher-form symmetries.

\subsection{Elements of groups as topological objects}
We identify the elements of groups $G$, $H$, and $L$
as
$(D-1)$-, $(D-2)$-, and $(D-3)$-dimensional 
topological objects 
 respectively.
By the identification, we may relate the groups $G$, $H$, and $L$
as symmetry groups of 0-, 1-, and 2-form symmetries.
Unlike ordinary higher-form symmetries, the topological 
objects may exit
as the boundaries of one-dimensional higher topological objects. 
Hereafter, we take $D = 3$ for simplicity, 
which is sufficient to describe all of their objects.
In this case, the group elements of $G$, $H$, and $L$ 
are expressed by surfaces, lines, and points, respectively. 
Note that we can easily have the $D = 4$ case by 
extending the objects such as worldvolumes, worldsurfaces, and worldlines
along the fourth direction, e.g., temporal direction.

The elements $g\in G$, $h \in H$, and $l \in L$
can be graphically expressed as follows: 
\begin{equation}
\ig[scale=0.33]{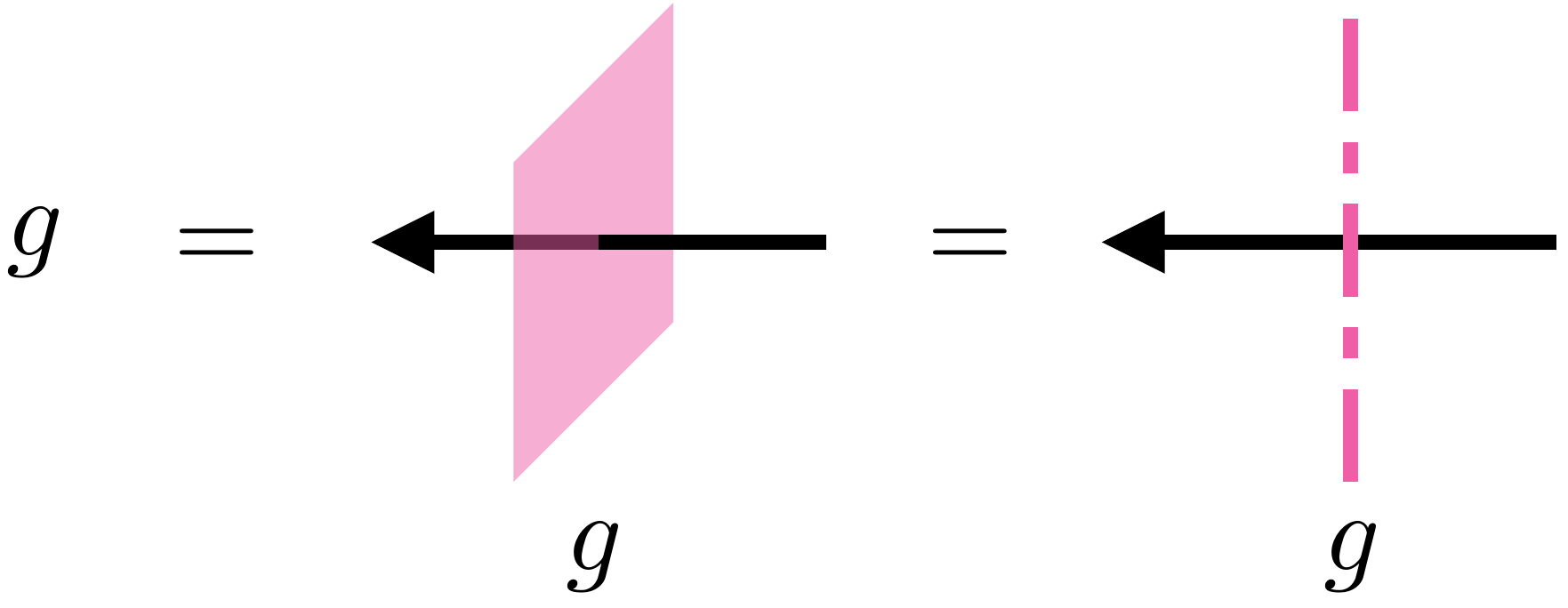}
\end{equation}
\begin{equation}
\ig[scale=0.33]{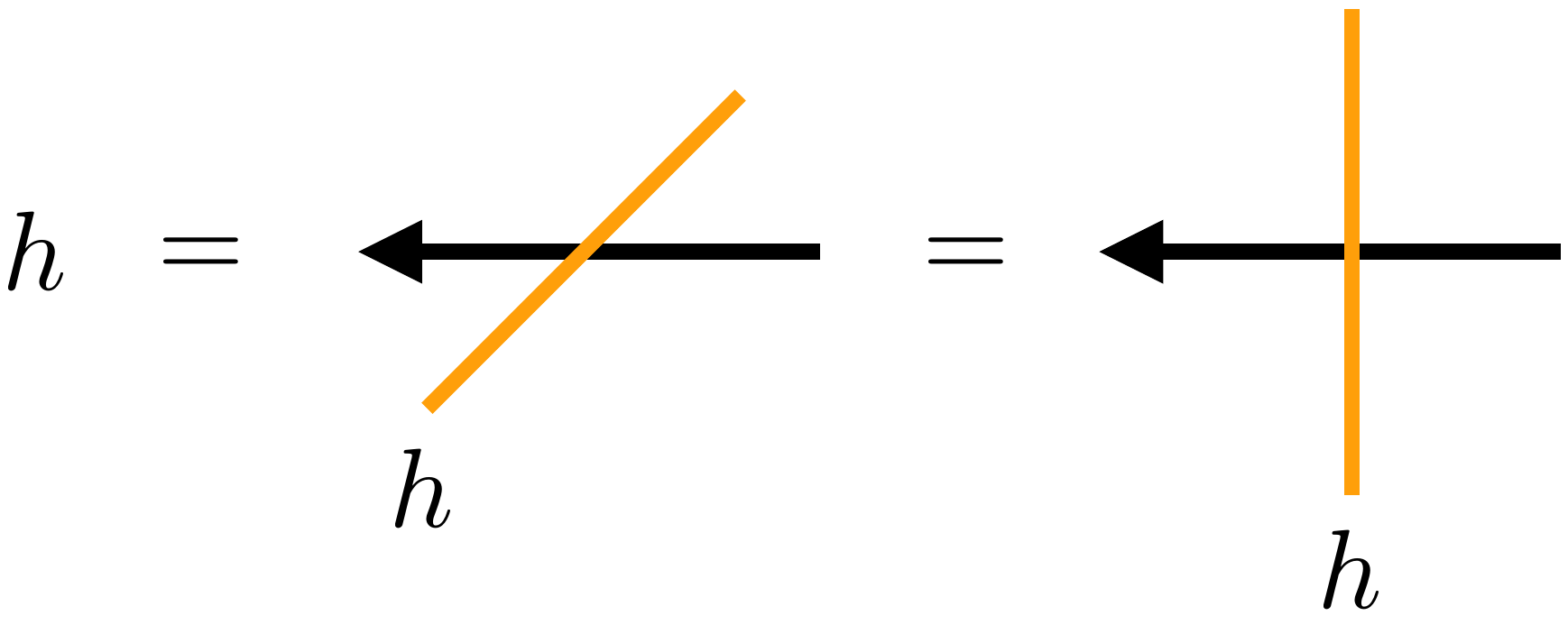}
\end{equation}
\begin{equation}
\ig[scale=0.33]{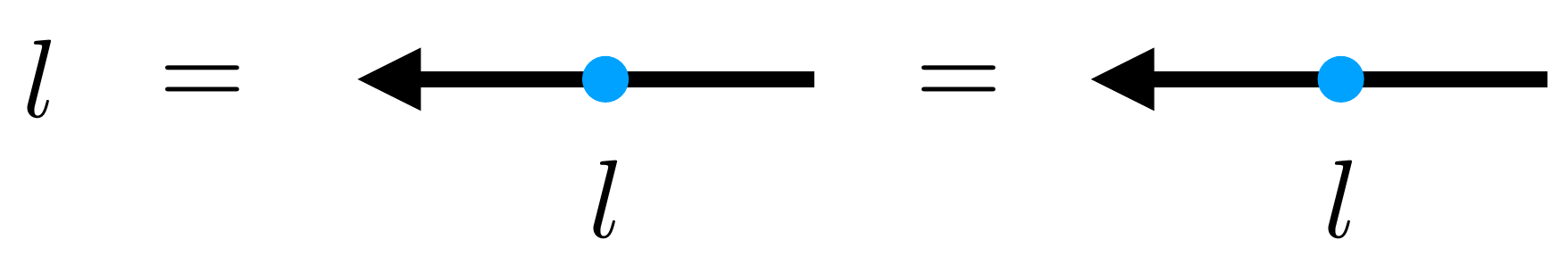}
\end{equation}
Here, the right-hand sides of the above equations 
are projections of the diagrams.
The black left arrows represent the order of the products.
We require that the elements of the groups can freely 
move as long as they intersect with the left arrow.

By using the left arrow, 
the group operations can be expressed as follows:
\begin{equation}
\ig[scale=0.33]{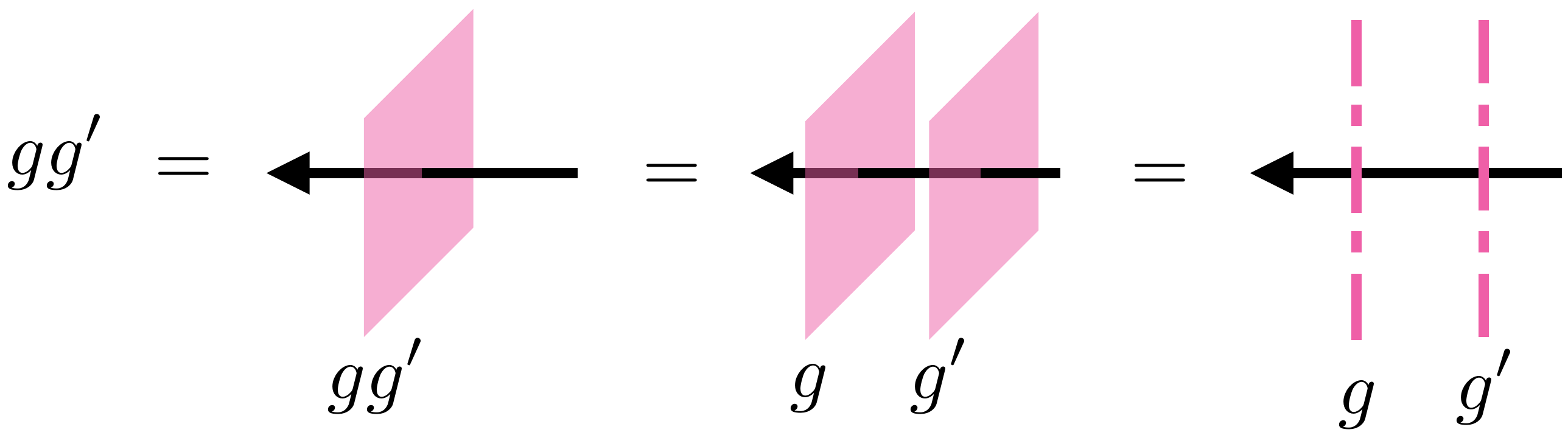} 
\end{equation}
\begin{equation}
\ig[scale=0.33]{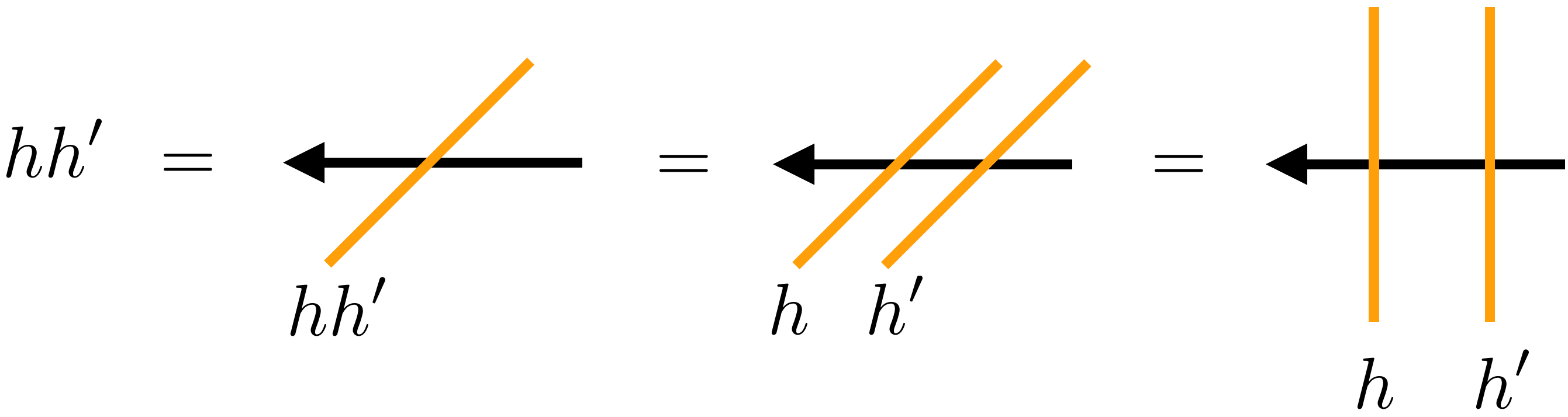}
\end{equation}
\begin{equation}
\ig[scale=0.33]{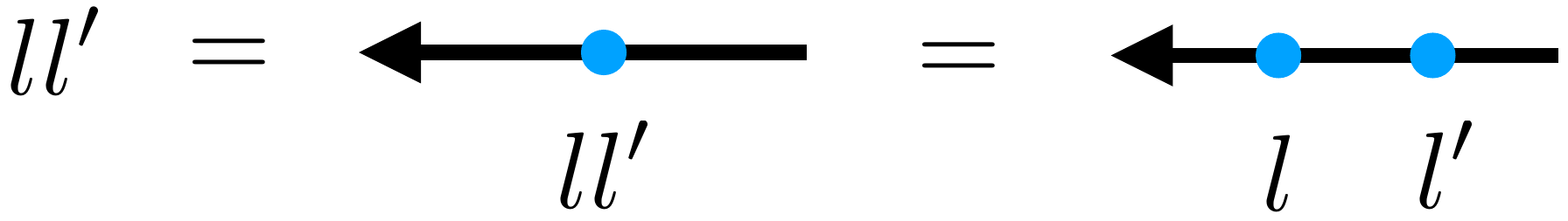}
\end{equation}
The identity elements, $1_G \in G$, $1_H \in H$, and $1_L \in L$
are represented as a dotted surface, line, and point, respectively.
Frequently, we abbreviate these identity elements to nothing.
They are explicitly described as follows:
\begin{equation}
\ig[scale=0.33]{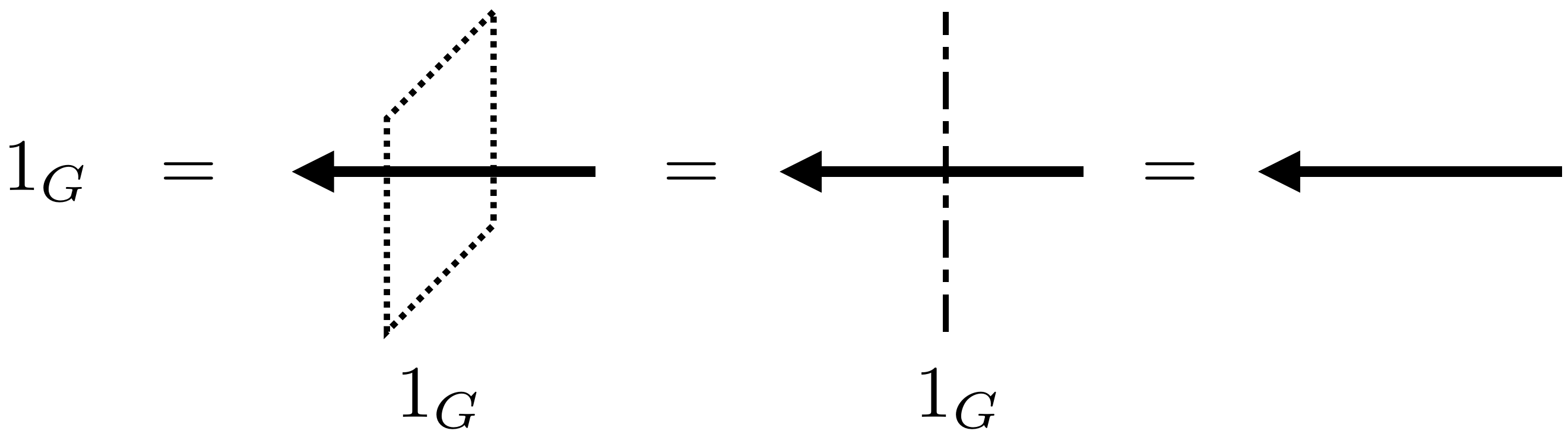} 
\end{equation}
\begin{equation}
\ig[scale=0.33]{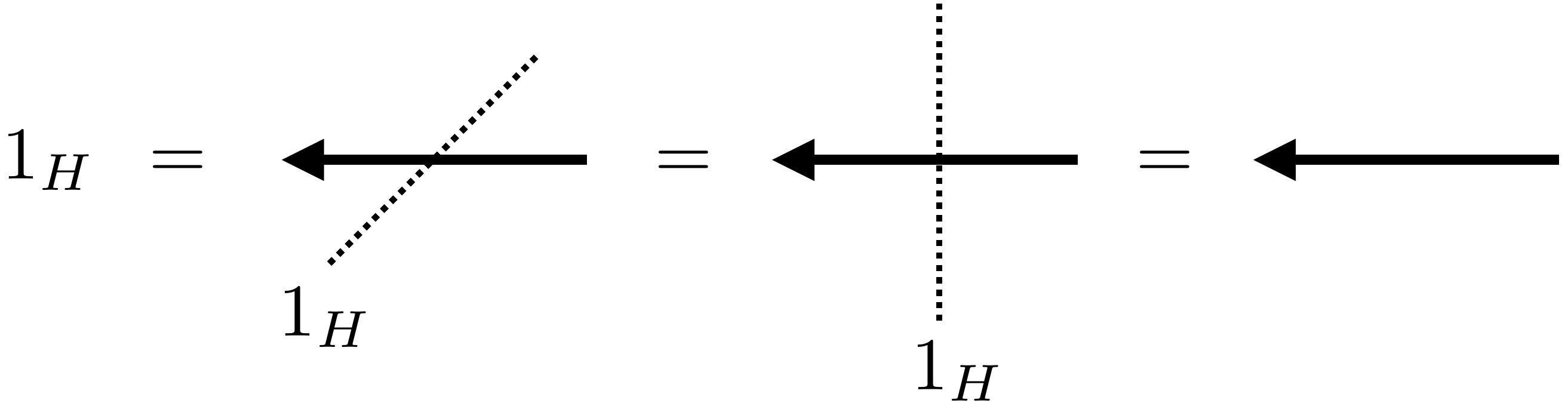}
\end{equation}
\begin{equation}
\ig[scale=0.33]{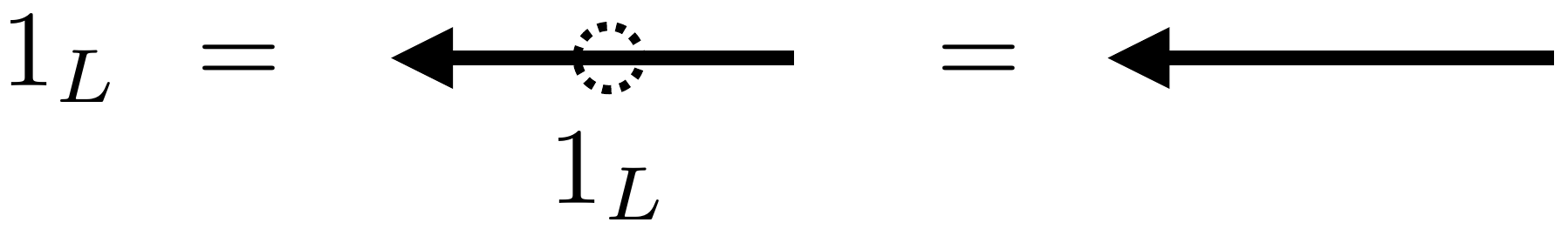}
\end{equation}

Finally, we express the inverses of the elements,
$g^{-1} \in G$, $h^{-1} \in H$, and $l^{-1} \in L$
 as objects which 
annihilate $g$, $h$ and $l$, respectively.
One of the properties of the inverses is 
that we can connect the object $g$ and $h$ 
with the inverses $g^{-1}$ and $h^{-1}$ as 
intermediate states of the annihilation, respectively:
\begin{equation}
\ig[scale=0.33]{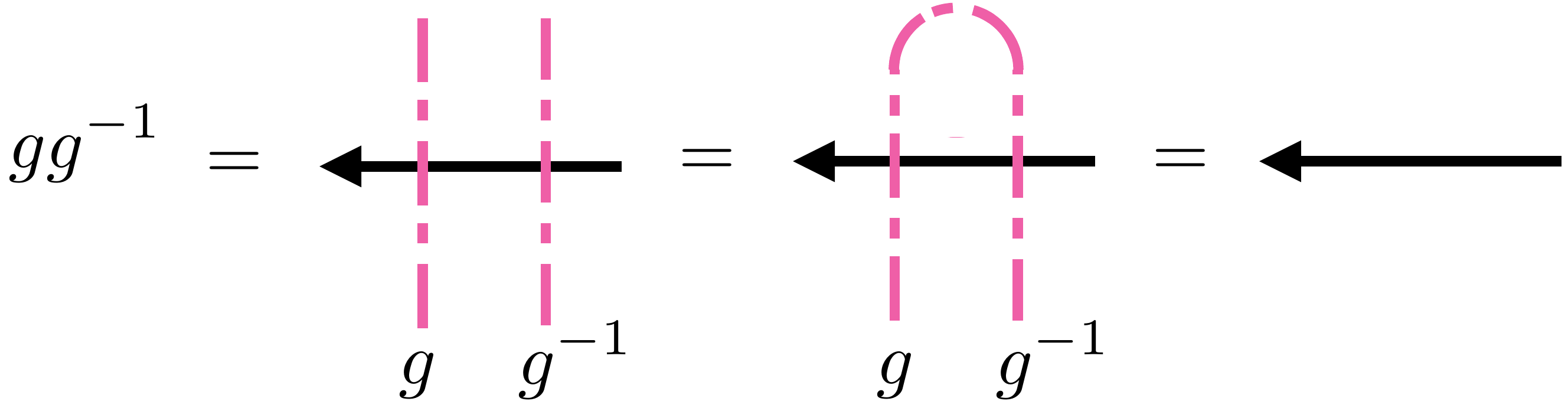} 
\label{200811.1343}
\end{equation}
\begin{equation}
\ig[scale=0.33]{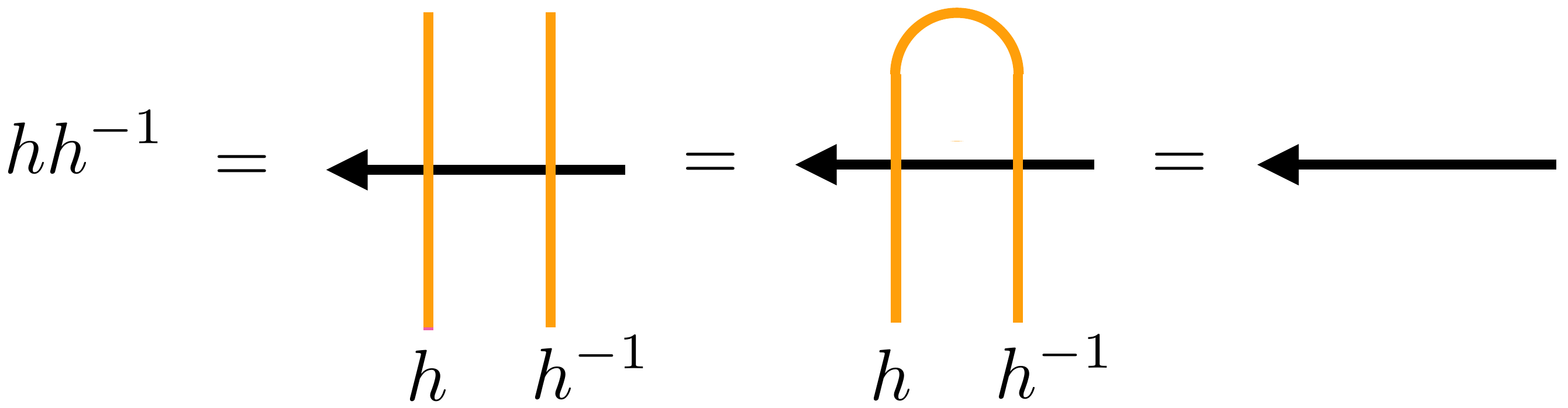}
\end{equation}
\begin{equation}
\ig[scale=0.33]{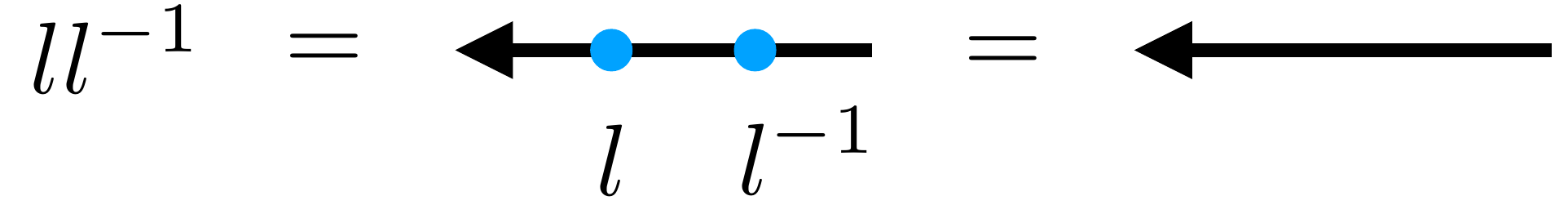}
\end{equation}

\subsection{{$\der_{1}$} and {$\der_2$}: taking interior of topological objects}
Next, we consider diagrammatic expression of the maps $\der_{1,2}$.
Since we have regarded the elements of the groups
as generally extended objects, 
the elements can be boundaries of the other objects.
The maps $\der_{1,2}$ give 
the elements of the interior from the boundary elements:
\begin{equation}
\ig[scale=0.33]{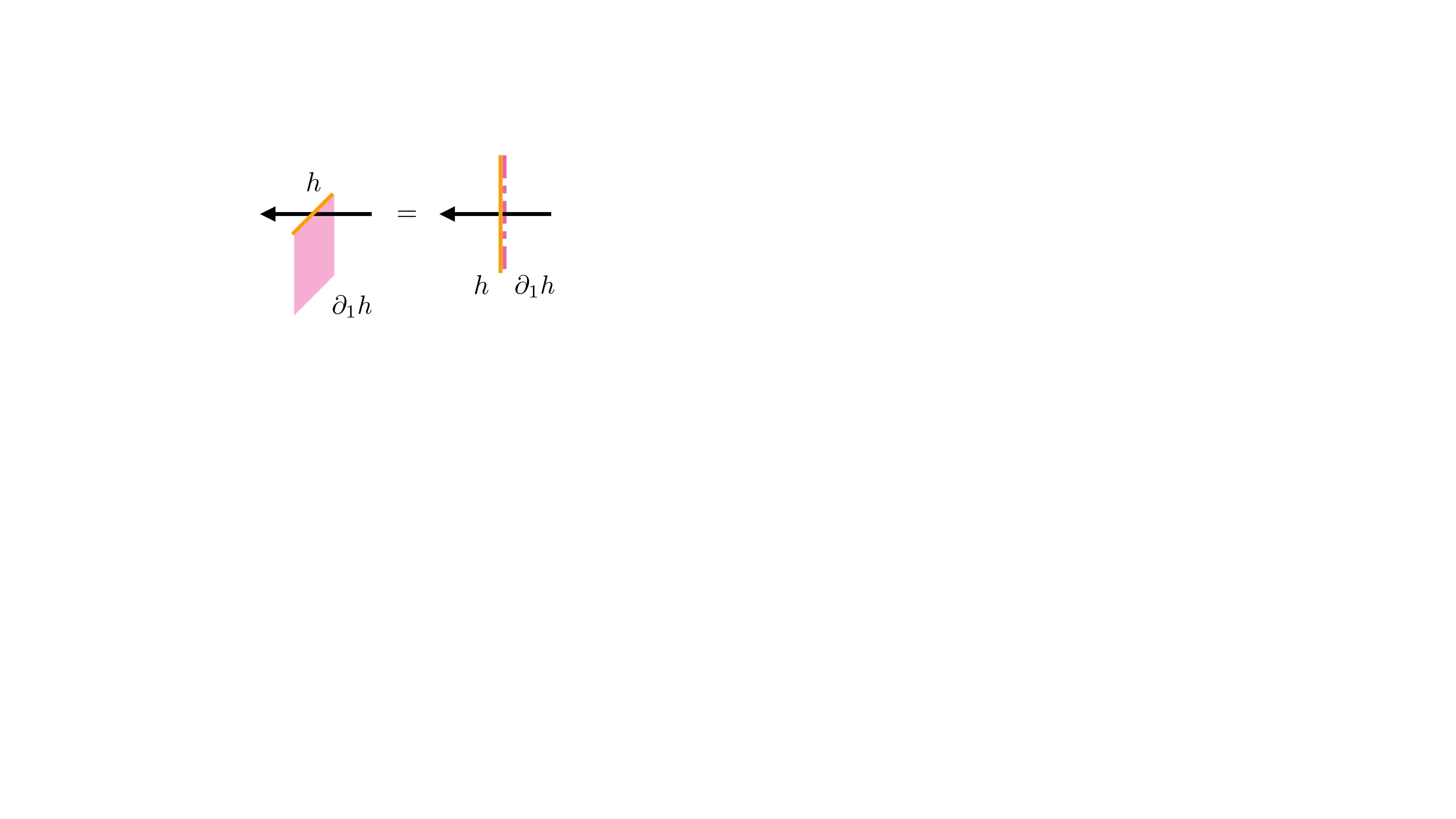}
\end{equation}
\begin{equation}
\ig[scale=0.33]{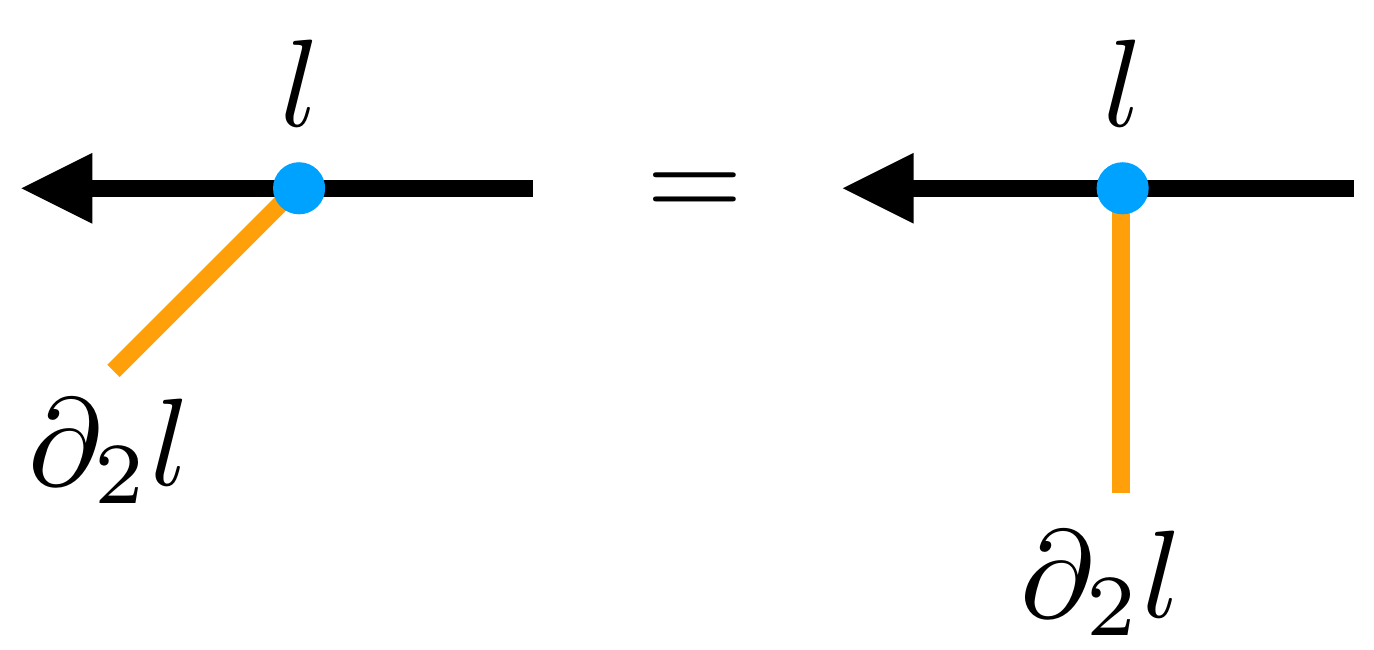}
\end{equation}
The right-hand sides of the above equations are
projected diagrams.
By the expression, the axiom 
$\der_1 \circ \der_2 l = 1_{G}$ 
in \er{200810.0358} is manifest, since
 an interior of an interior is 
nothing (conversely, the boundary of a boundary is nothing).
The property of group homomorphism is just saying that 
the product of the elements is compatible with 
the product of the interior of the elements:
\begin{equation}
\ig[scale=0.33]{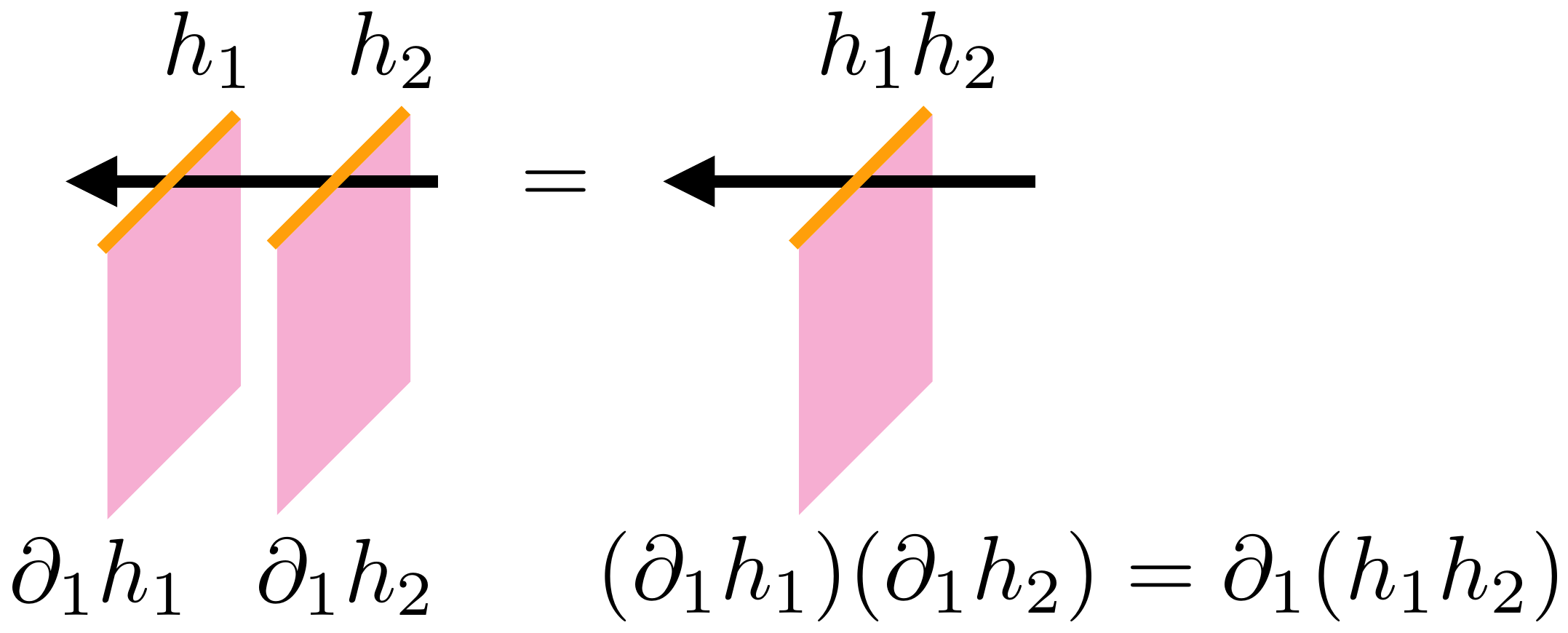}
\end{equation}
\begin{equation}
\ig[scale=0.33]{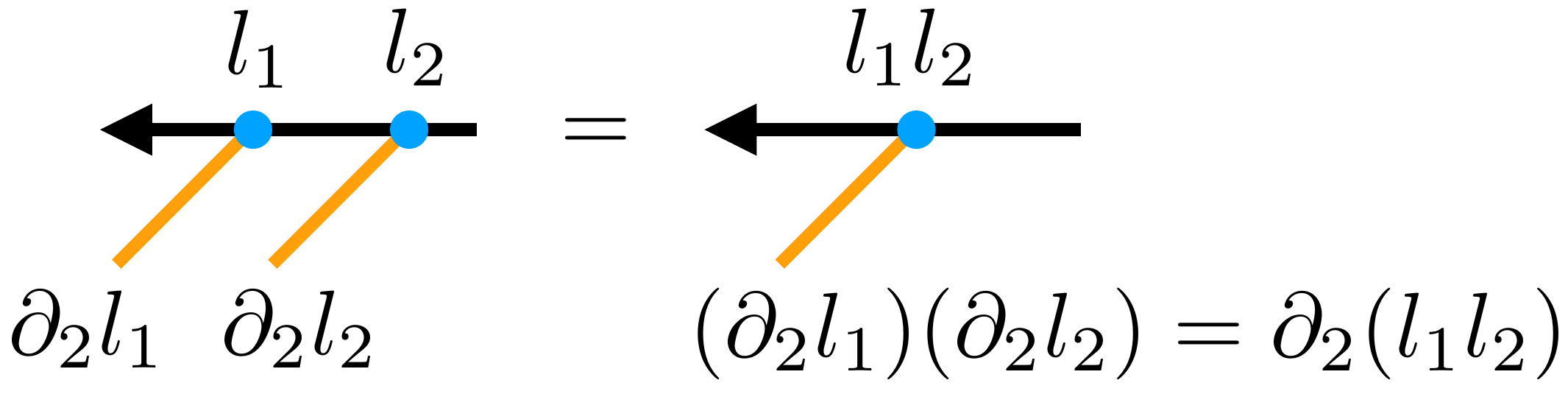}
\end{equation}

While the elements should intersect with the left arrow, 
we allow $l \in L$ to move vertically 
as long as $\der_2 l \in H$ intersects with the left arrow:
\begin{equation}
 \ig[scale=0.33]{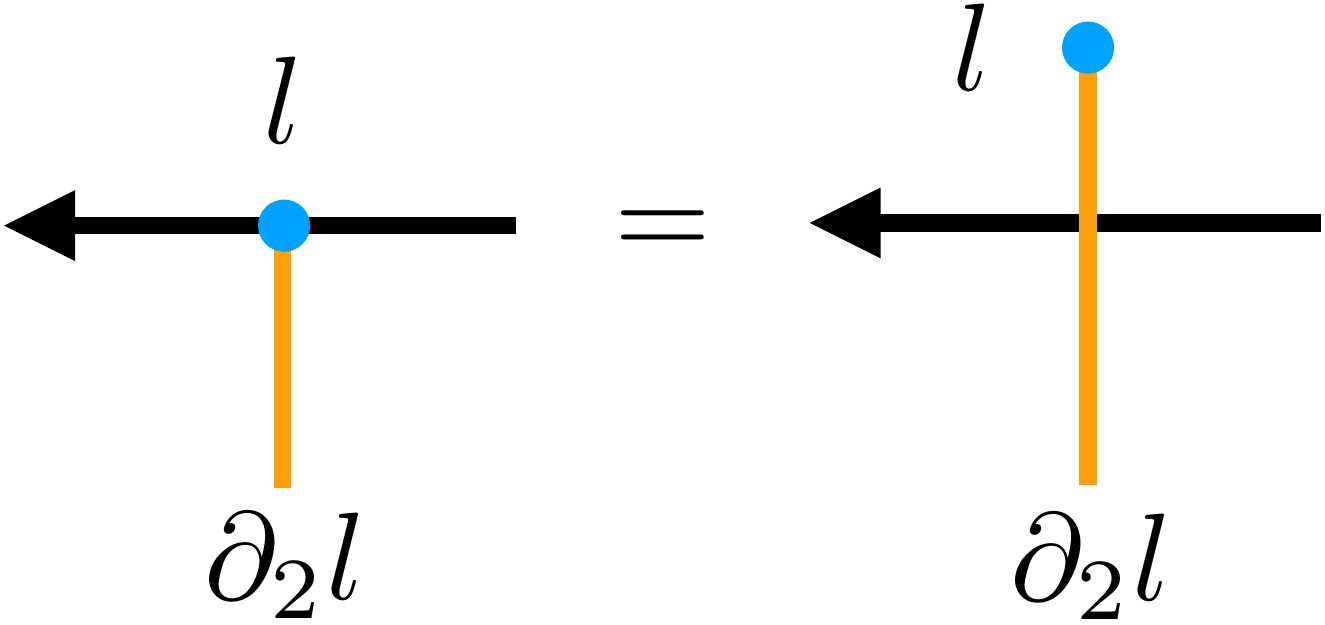}
\label{200830.1858}
\end{equation}
This property implies that $L$ and $H$ have a 2-group structure, 
which we use in section \ref{lh2g}.
As an application, we can deform $l$ and $l^{-1}$ as follows:
\begin{equation}
 \ig[scale=0.33]{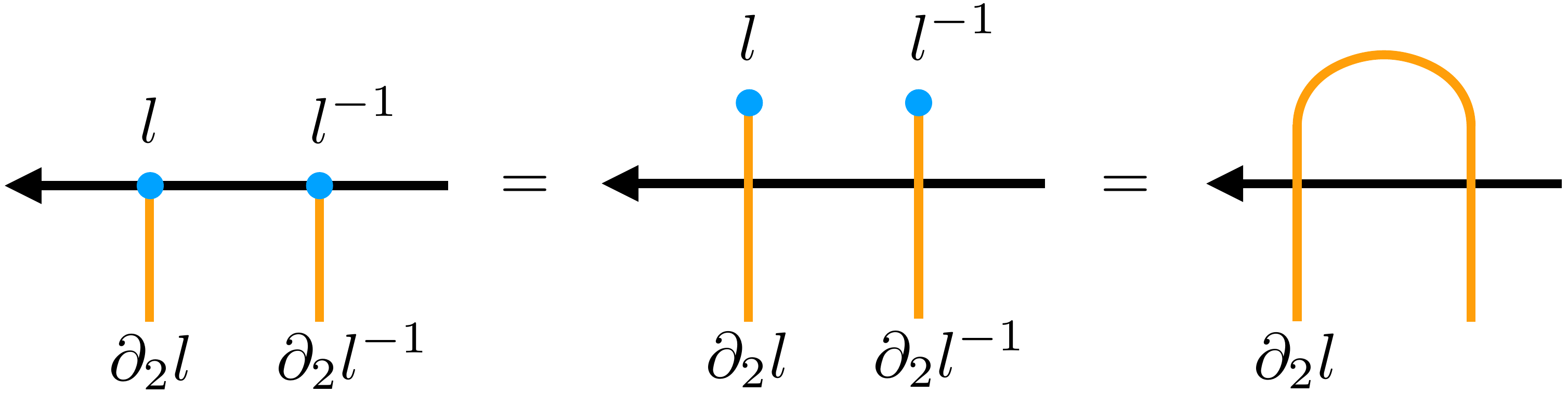}
\label{200830.1859}
\end{equation}

\subsection{Action of {$G$}: enclosing elements by surfaces}

Third, we express the actions of $G$ on $G$, $H$, and $L$, 
following the above diagrammatic expressions.
The action $\trr$ of $g \in G$ on $g'\in G$, $h \in H$
and $l \in L$ can be simply described as the enclosing 
by $g$ and $g^{-1}$:
\begin{equation}
\ig[scale=0.33]{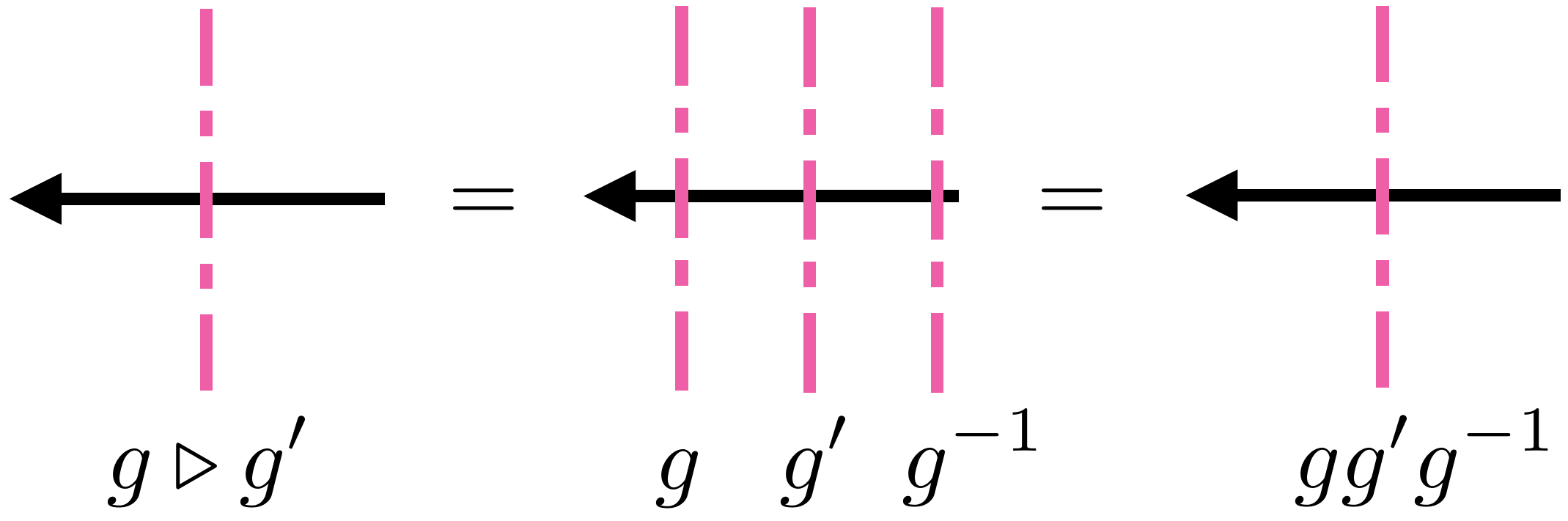} 
\end{equation}
\begin{equation}
\ig[scale=0.33]{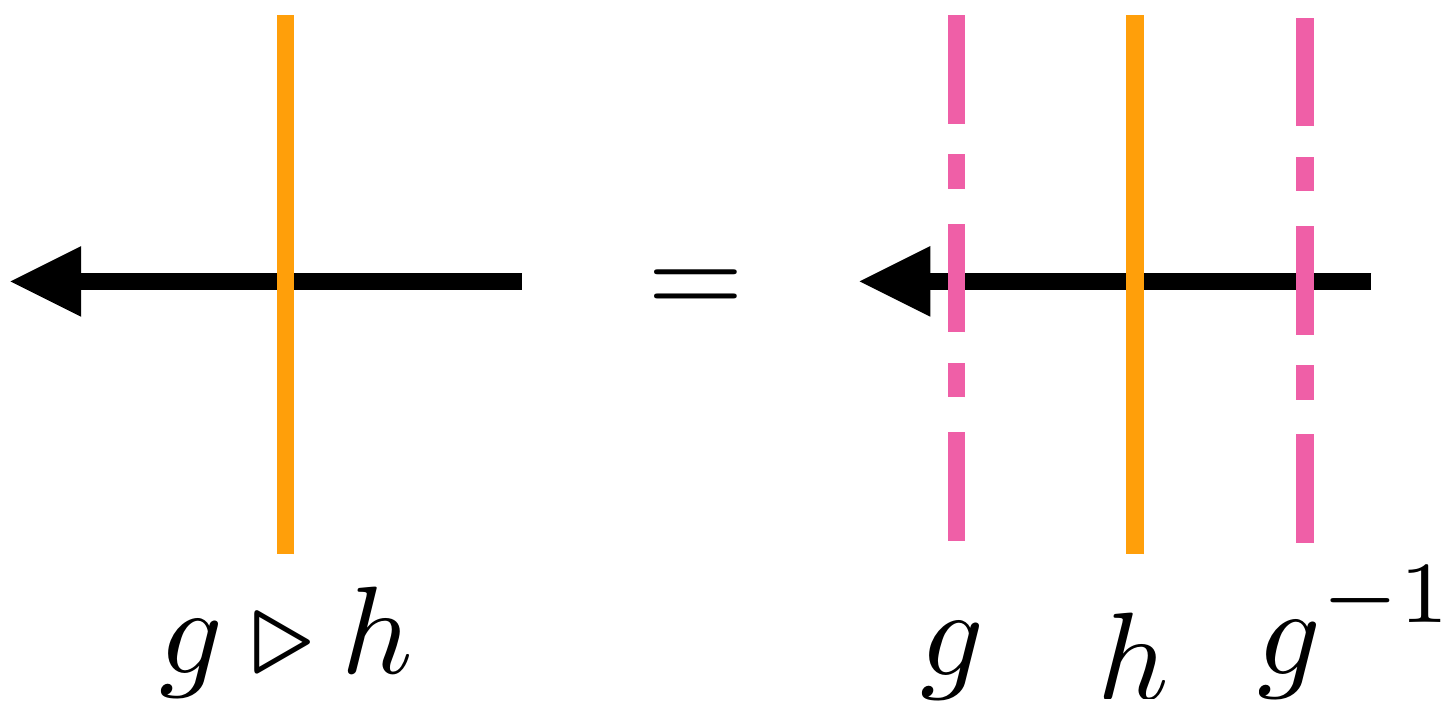}
\end{equation}
\begin{equation}
\ig[scale=0.33]{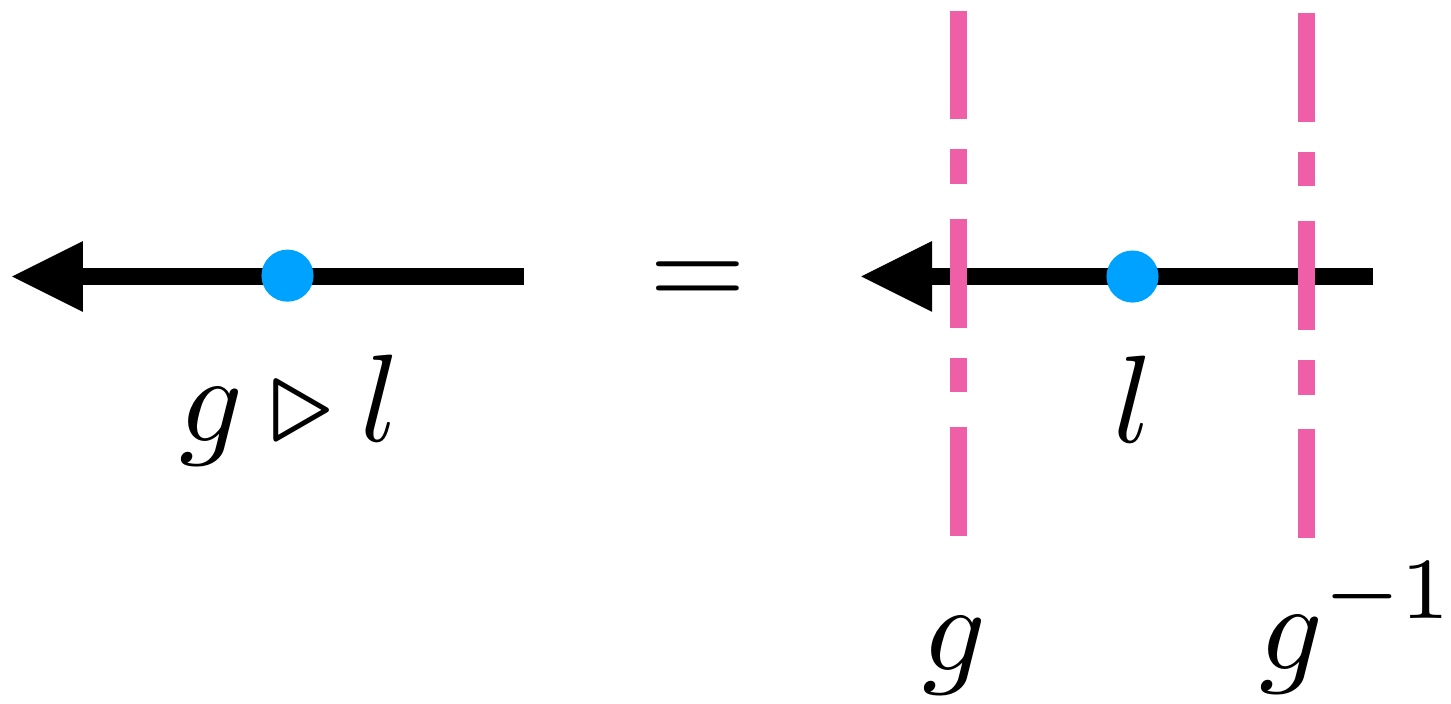}
\end{equation}

In particular, the axiom $g \trr g' = g g' g^{-1}$ 
given in \er{200622.1143} is manifest in our diagram.
Furthermore, the $G$-equivalence of $\der_{1,2}$ in \er{200622.1529} can be 
simply understood as the compatibility of $\der_{1,2}$ with 
the action:
\begin{equation}
\ig[scale=0.33]{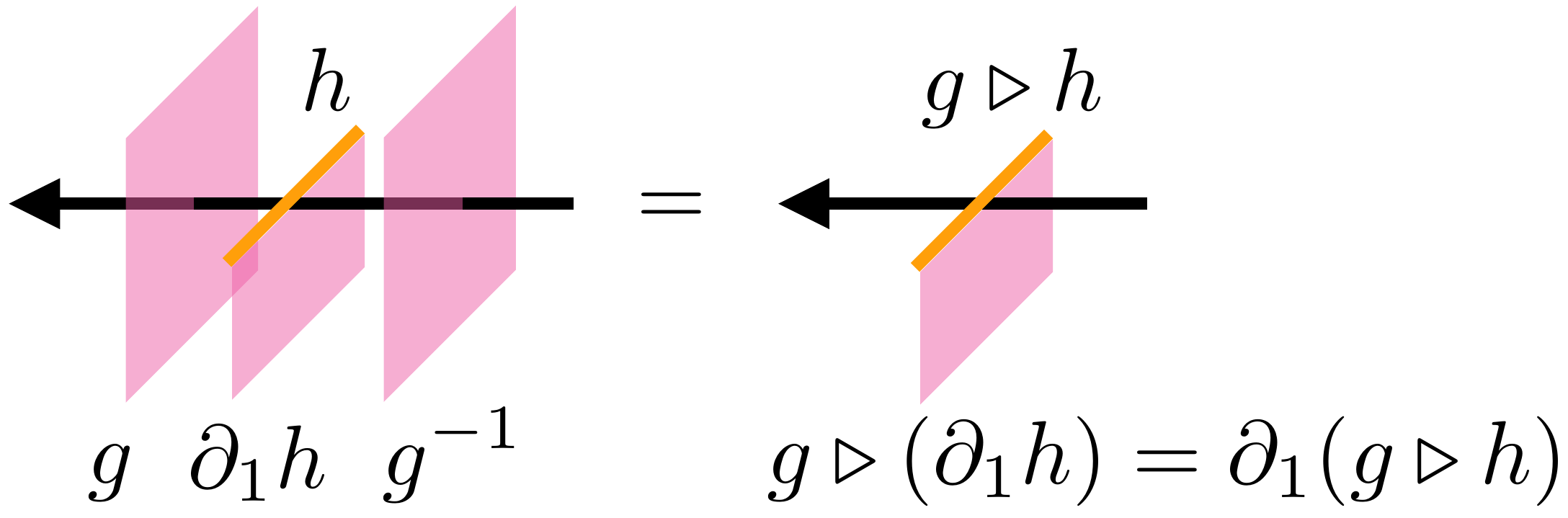}
\end{equation}
\begin{equation}
\ig[scale=0.33]{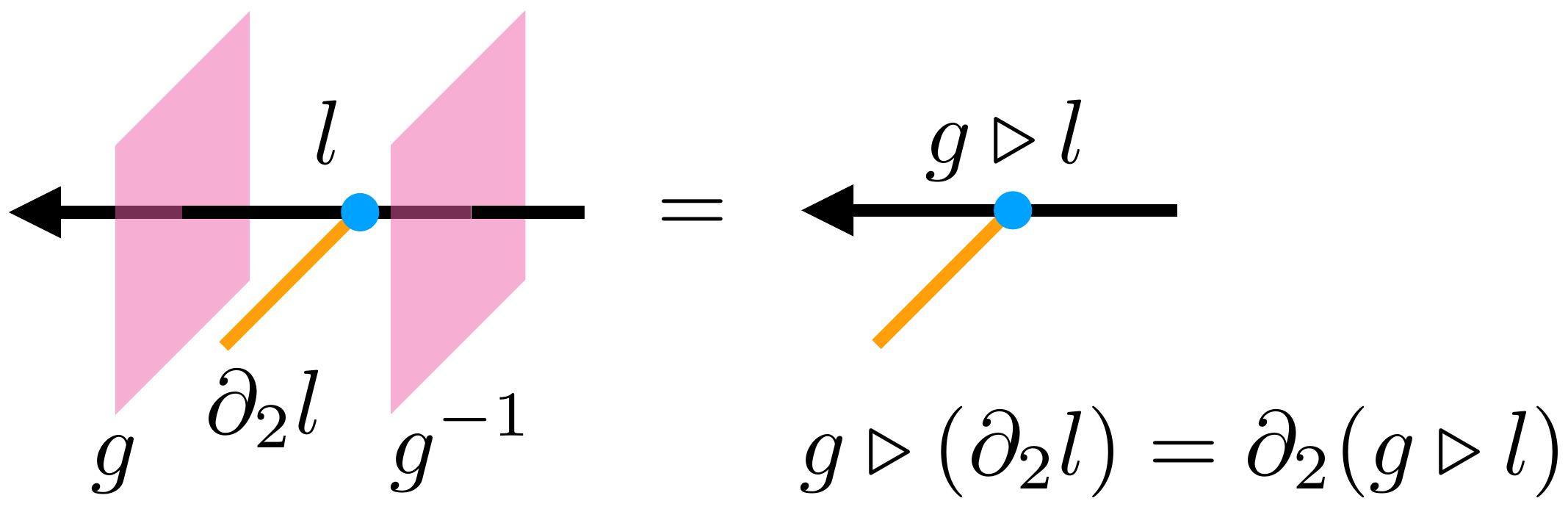}
\end{equation}

\subsection{Peiffer lifting: braiding of elements of {$H$}\label{peiffer}}

Finally, we express the Peiffer lifting diagrammatically.
We determine the expression of it 
as a braid of two elements in $H$
such that the axiom in \er{200622.1536} is satisfied:
\begin{equation}
\ig[scale=0.33]{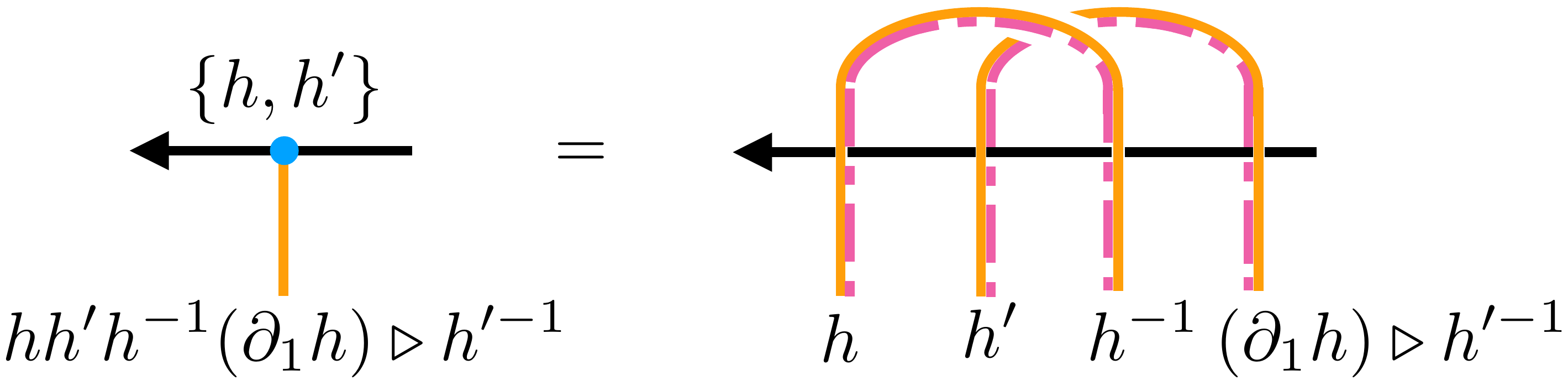}
\label{200908.2008}
\end{equation}
In the right-hand side, 
the line of $ h $ braids with the line of $h'$.
Since $h'$ intersects with $\der_1 h$, 
the surface of $\der_1 h$ acts on the line $h'$.
Therefore, the line of $h'$ ends on 
$(\der_1 h) \trr h'^{-1} = ((\der_1 h) \trr h')^{-1}$.
For the relation between the 3-group and braids, 
see, e.g., Refs.~\cite{brown1989algebraic,Arvasi:2007aaa}.

While we have expressed the Peiffer lifting diagrammatically, 
it is non-trivial whether the other axioms are satisfied 
in terms of the diagram or not.
We confirm that our diagram of the 
Peiffer lifting satisfies all of the axioms of the Peiffer lifting
as follows:
\begin{itemize}
 \item Equation \eqref{200623.0212}:
\begin{equation}
\ig[scale=0.33]{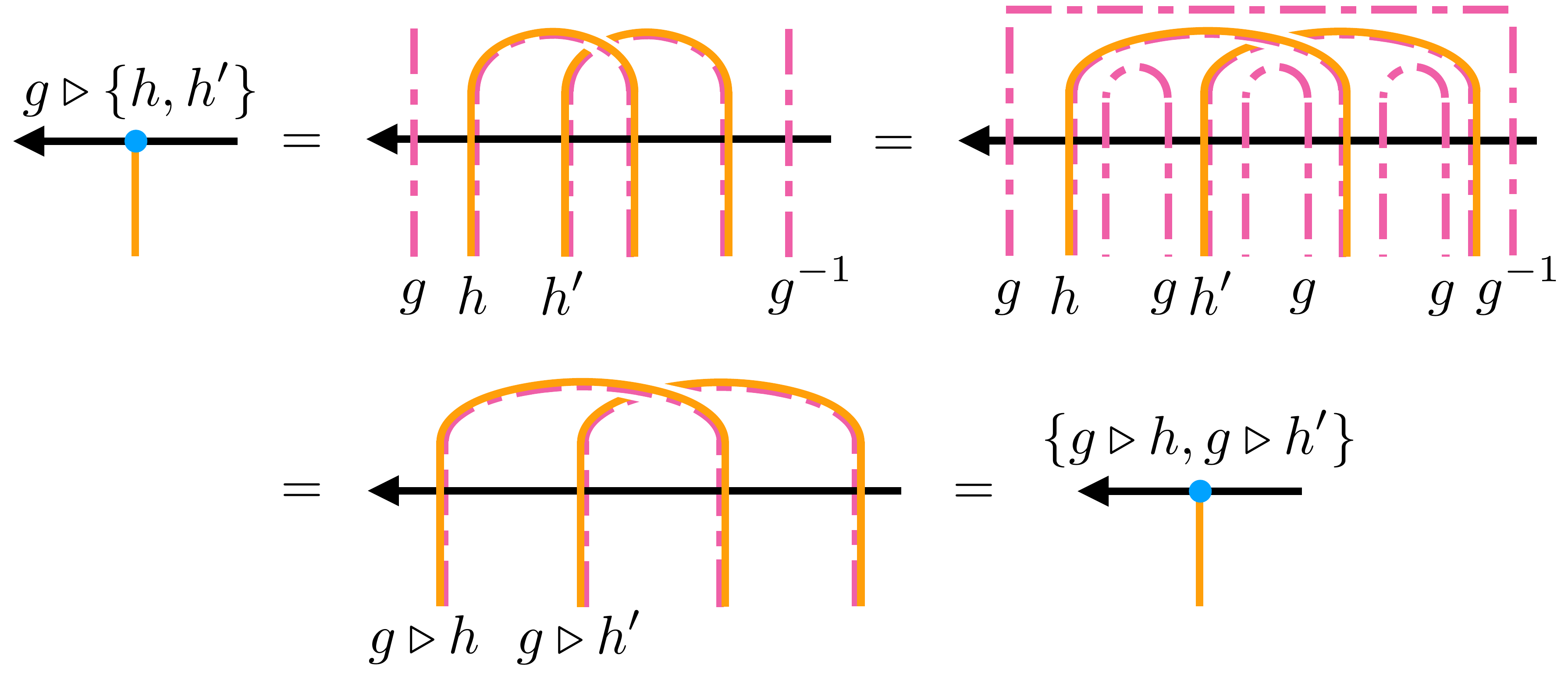}
\end{equation}
Here, we have used the property in \er{200811.1343}
for the third expression.
 \item Equation \eqref{200623.0213}:
\begin{equation}
\ig[scale=0.33]{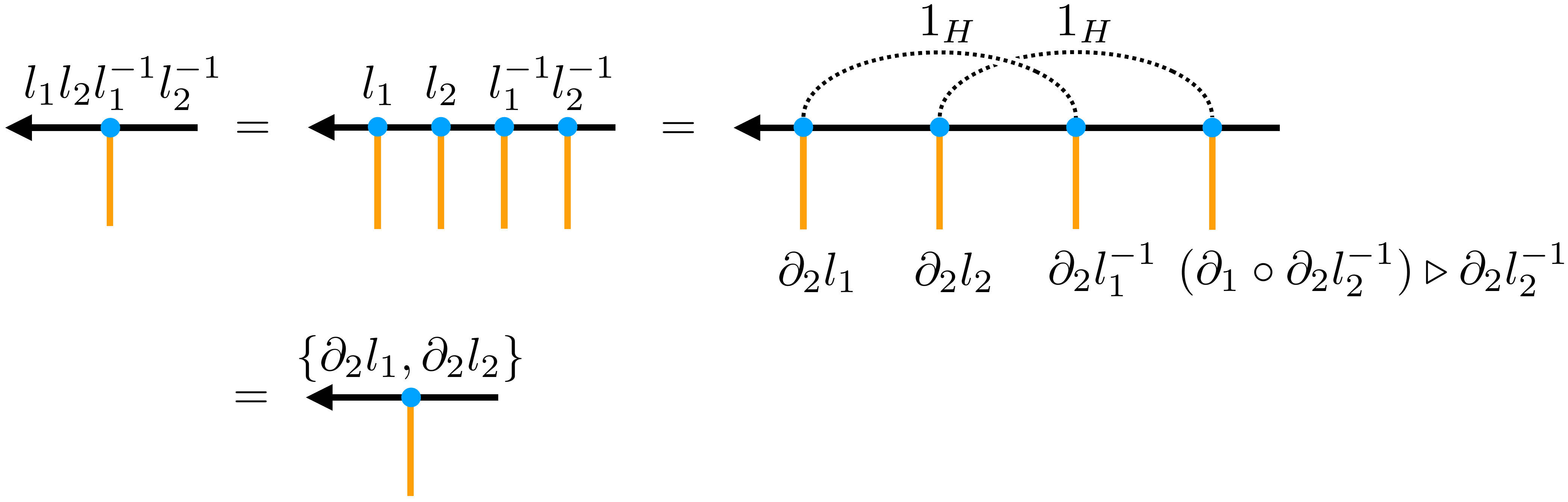}
\end{equation}
Here, we have used $\der_2 l_2^{-1} = 
(\der_1 \circ \der_2 l_1) \trr \der_2 l_2^{-1}$, 
since $\der_1 \circ \der_2 l_1  = 1_G$.
 \item Equation \eqref{200623.0214}: 
\begin{equation}
\ig[scale=0.33]{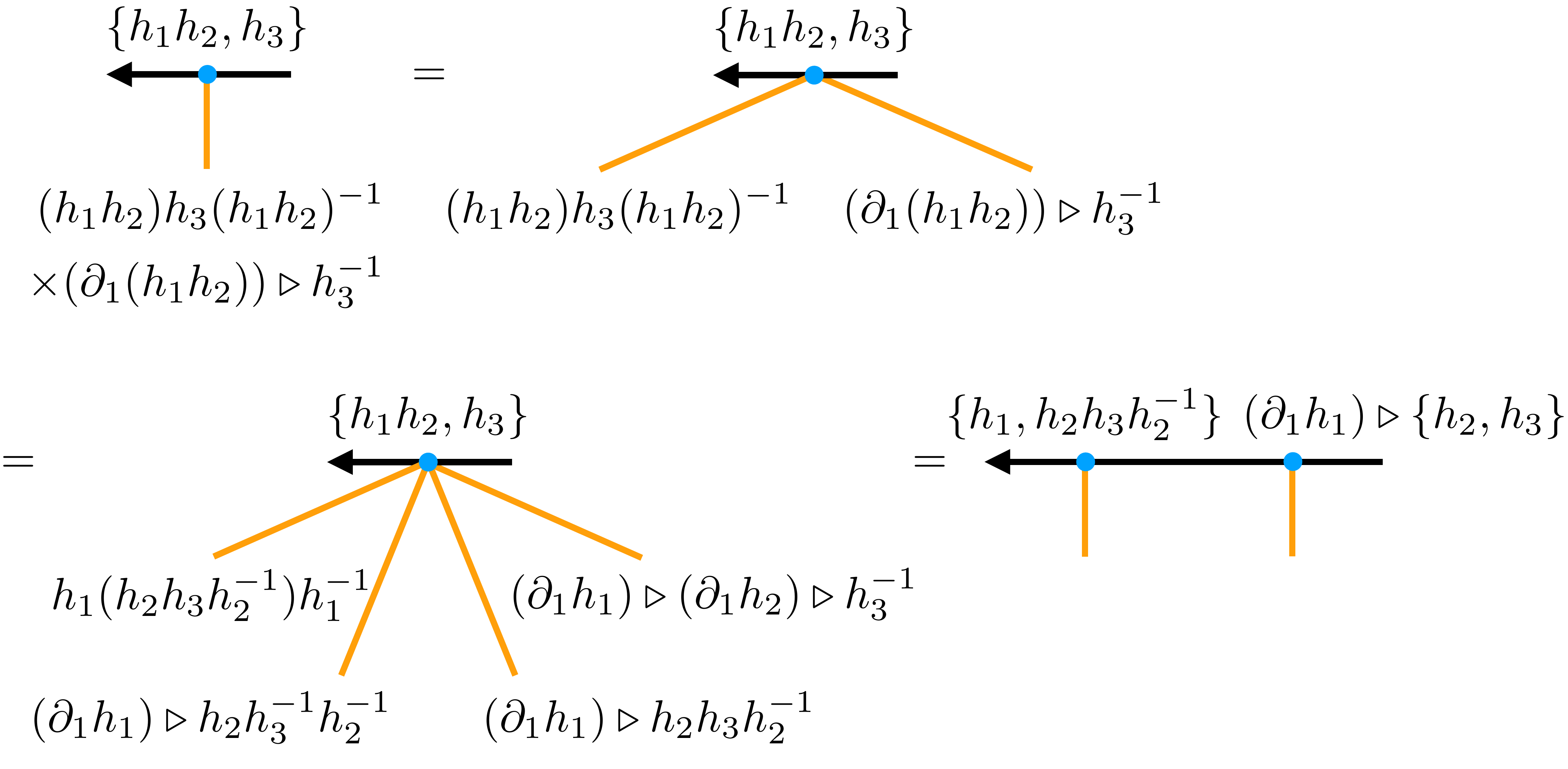}
\end{equation}
Here, we have used 
$(\der_1 (h_1h_2))\trr h_3^{-1} = ((\der_1 h_1) (\der_2 h_2)) \trr h_3^{-1}= 
(\der_1 h_1) \trr (\der_2 h_2) \trr h_3^{-1}$.
 \item Equation \eqref{200623.0215}: 
\begin{equation}
\ig[scale=0.33]{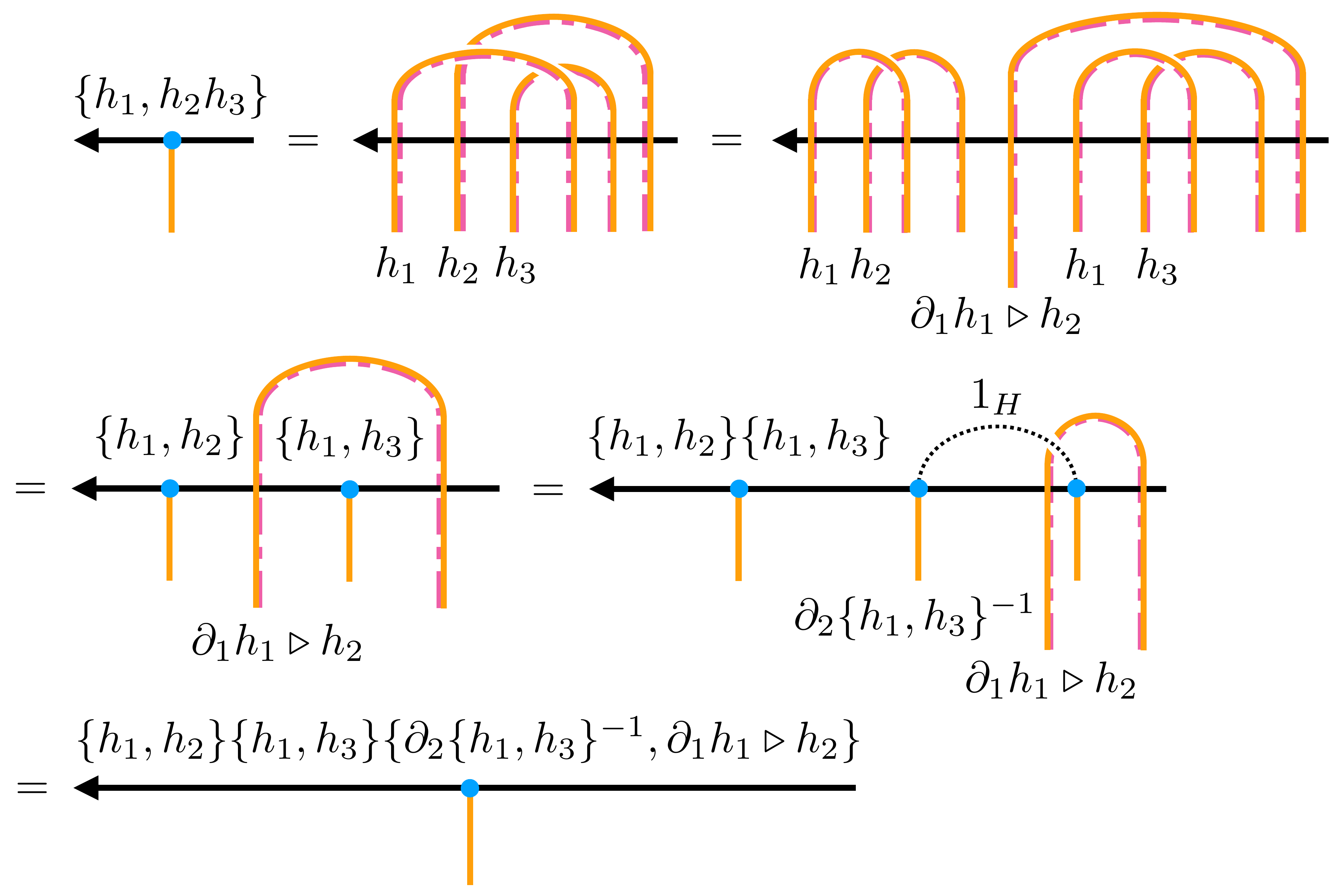}
\end{equation}
 \item Equation \eqref{200623.0216}: 
\begin{equation}
\ig[scale=0.33]{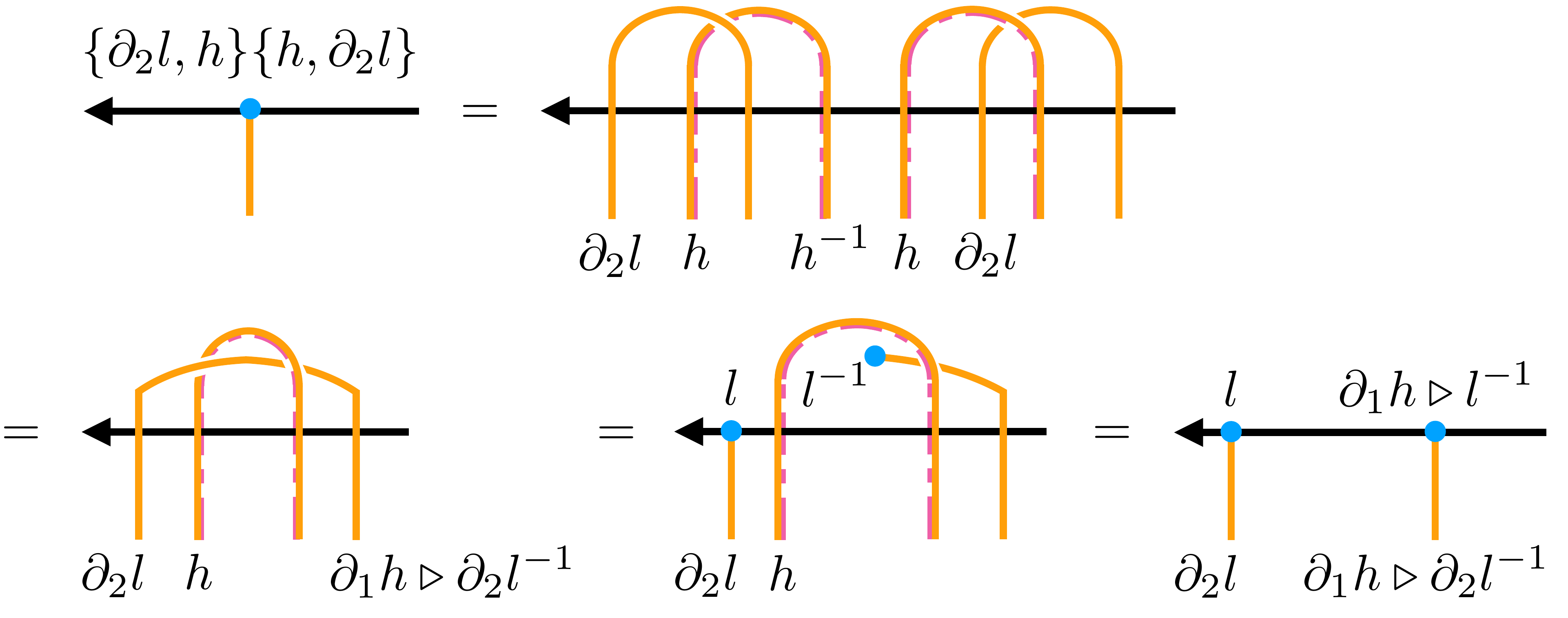}
\end{equation}
In order to obtain the last equation, we have used \er{200830.1858}.
\end{itemize}
\subsection{$(L,H)$ as 2-group\label{lh2g}}
We have shown that the 3-group can be diagrammatically expressed.
By using them, 
we can also describe the fact that 
the set 
$(L\os{\der_2}{\to} H, \trr')$ is a 2-group
(see appendix \ref{3gaxiom}).
Here, the action $\trr'$ of $H$ on $H$ and $L$ are defined by 
conjugation $h \trr' h' = hh'h^{-1}$ and
by the Peiffer lifting $ h\trr'l = l \{\der_2 l^{-1}, h\}$, 
respectively.
In order to reproduce the 2-group structure, we 
should diagrammatically show the action $\trr'$,
the compatibility of $\der_2$ with $\trr'$, and 
the Peiffer identity $l_1 l_2l_1^{-1} = (\der_2 l_1) \trr' l_2$.

One of the advantages of the diagrammatic expression is 
that we can straightforwardly reproduce them, in particular 
the action $\trr'$, which may be complicated.
Let us express the definition of the action $\trr'$.
As in the definition of the action $\trr$ for the 3-group, 
we can describe the action $h \trr' h'$ and $h\trr' l$ 
by enclosing $h \in H$ and $l \in L$ with $h\in H$,
respectively.
First, we consider the action of $h \in H$ on $h' \in H$
defined by conjugation $h \trr' h' = hh'h^{-1}$,
which can be expressed as follows:
\begin{equation}
 \ig[scale=0.33]{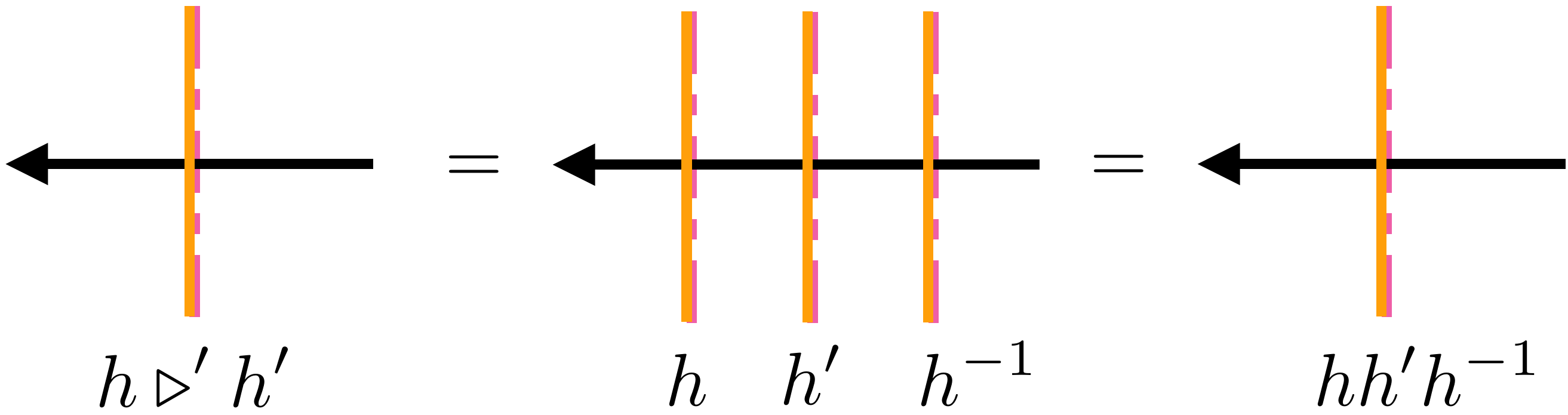}
\end{equation}
Second, we can simply reproduce 
the definition $h\trr' l = l \{\der_2 l^{-1}, h\}$ 
by using the following the deformations:
\begin{equation}
 \ig[scale=0.33]{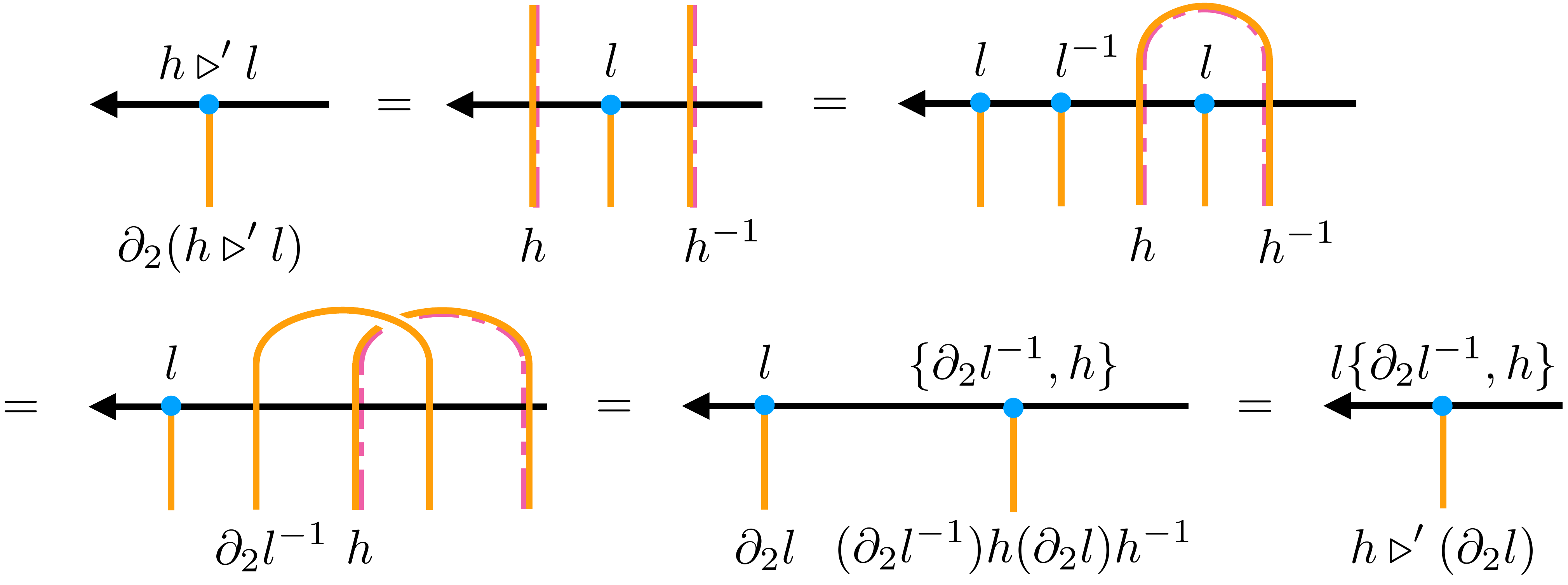}
\end{equation}
Here, we have used \er{200830.1859} in order to 
have the fourth expression.
The above deformations automatically show the compatibility of $\der_2$
with the action of $H$: 
$\der_2 (h \trr' l) = h \trr' (\der_2 l) = h (\der_2 l) h^{-1}$.

The 2-group should satisfy the Peiffer identity
$l_1 l_2 l_1^{-1}= \der_2 l_1 \trr' l_2$,
which can now be shown as follows:
\begin{equation}
 \ig[scale=0.33]{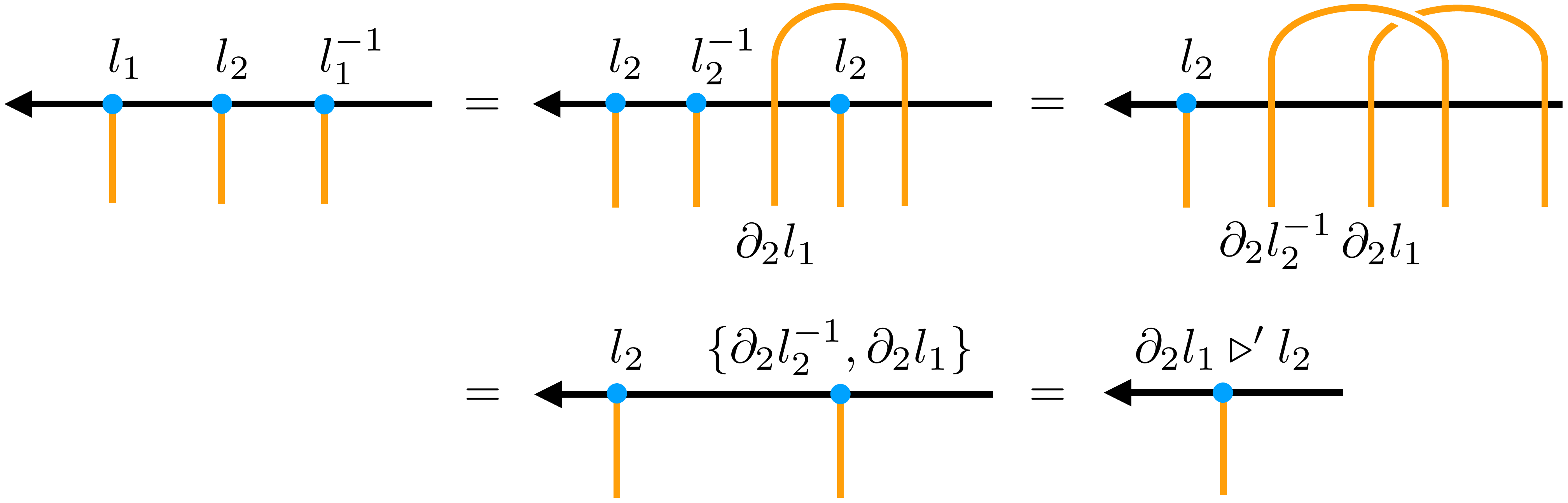}
\end{equation}
Here, we have used \er{200830.1859} in the first line.

\subsection{Global 3-group symmetry and symmetry generators}

Finally, we consider how to relate the above 
diagrammatic expressions to the symmetry generators of the 
higher-form global symmetries given by the 3-group.
In the following, 
we identify $G$, $H$, and $L$ as symmetry groups 
which parameterize the 0-, 1-, and 2-form symmetries, respectively.
In the following, we discuss symmetry generators for the 
higher-form symmetries which do not have interiors,
and are not boundaries of other objects.
The assumptions restrict the symmetry group 
that non-trivially parameterizes the symmetry generators.
Further, we assume that the restricted group also has a 3-group 
structure.
As we will show in appendix~\ref{symgroup},
the assumptions require that 
the symmetry groups $G$, $H$, and $L$ are reduced to  
\begin{equation}
G_{\rm gl.} := G/\im \der_1,
\quad
H_{\rm gl.} := H_{\rm Ab.}/\im \der_2 ,
\quad
L_{\rm gl.} := \ker \der_2,
\end{equation}
respectively.
Here, $H_{\rm Ab.}$ is the Abelian part of $\ker \der_1 \subset H$.
This assumption is sufficient to consider the symmetry 
generators of the axion electrodynamics.

\subsubsection{Symmetry transformations}

Let us recall the symmetry transformations in the higher-form symmetries.
For elements of the groups
$g \in G_{\rm gl.}$, 
$h \in H_{\rm gl.}$ 
and 
$l \in L_{\rm gl.}$, 
the corresponding symmetry generators are expressed by
topological objects
$U_0 (g, {\cal S})$, $U_1 (h, {\cal C})$, and 
$U_2 (l, ({\cal P,P'}))$, 
respectively.
Here, ${\cal S}$, ${\cal C}$, and $({\cal P,P'})$ 
are a closed surface, a closed line and two points.
The symmetry generators 
can act on the 0-, 1-, and 2-dimensional charged objects 
$\Phi ({\cal P}_\Phi)$, $W ({\cal C}_W)$, and $V ({\cal S}_V)$ as
unitary representations:
\begin{equation}
 \vevs{U_0 (g, {\cal S}) \Phi ({\cal P}_\Phi )}
 = R_0 (g) 
 \vevs{\Phi ({\cal P}_\Phi)}\qtq{if} 
\link ({\cal S}, {\cal P}_\Phi) = 1,
\label{200904.2057}
\end{equation}
\begin{equation}
 \vevs{U_1 (h, {\cal C}) W ({\cal C}_W )}
 =
 R_1 (h) 
 \vevs{W ({\cal C}_W )}
\qtq{if}
\link ({\cal C}, {\cal C}_W) = 1,
\label{200904.2058}
\end{equation}
\begin{equation}
 \vevs{U_2 (l, ({\cal P,P'})) V ({\cal S}_V )}
 = R_2 (l) 
 \vevs{V ({\cal S}_V )}
\qtq{if}
\link (({\cal P,P'}), {\cal S}_V) = 1,
\label{200904.2059}
\end{equation}
respectively.
Here, ${\cal P}_\Phi$, ${\cal C}_W$, and ${\cal S}_V$ 
are a point, a closed line, and a closed surface.
We denote $R_0 (g)$, $R_1(h)$, and $R_2(l)$ as 
unitary representation matrices (c-number) of $g$, $h$, 
and $l$, respectively.

\subsubsection{Diagrammatic expressions of symmetry transformations}
We now show the diagrammatic expressions of the
symmetry generators and their symmetry transformations.
The charged objects can be diagrammatically expressed as 
follows.
\begin{equation}
 \ig[scale=0.33]{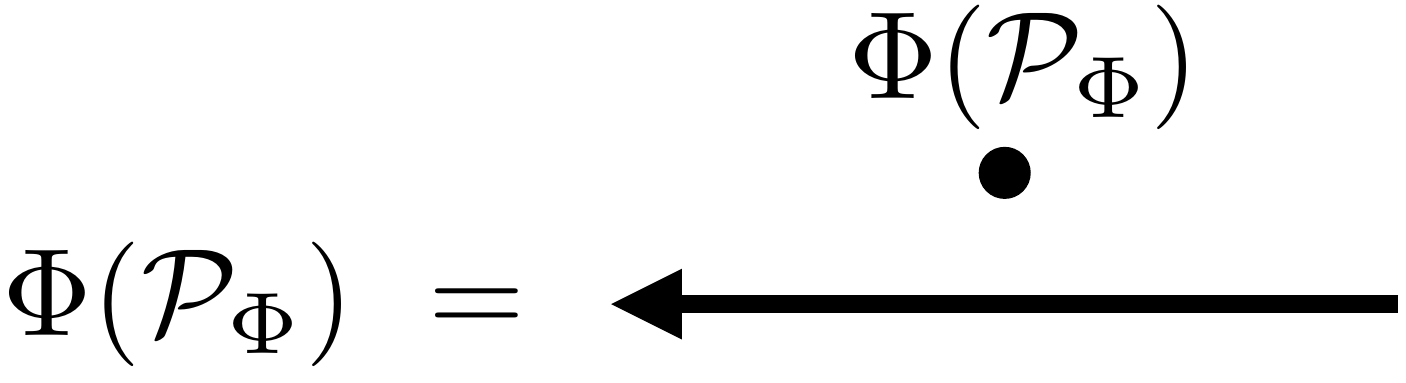} 
\end{equation}
\begin{equation}
 \ig[scale=0.33]{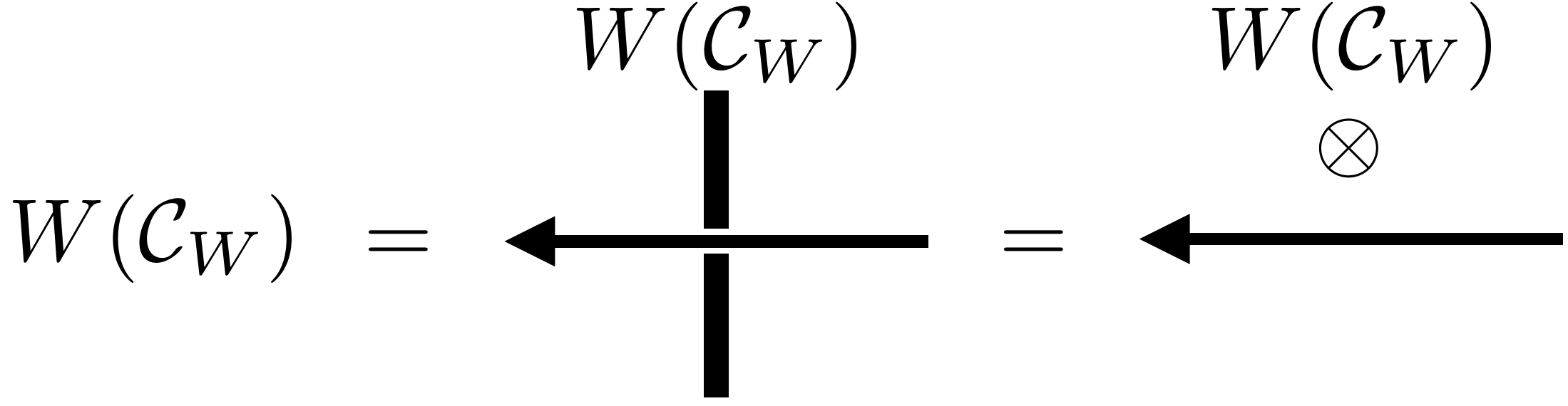} 
\label{200904.2305}
\end{equation}
\begin{equation}
 \ig[scale=0.33]{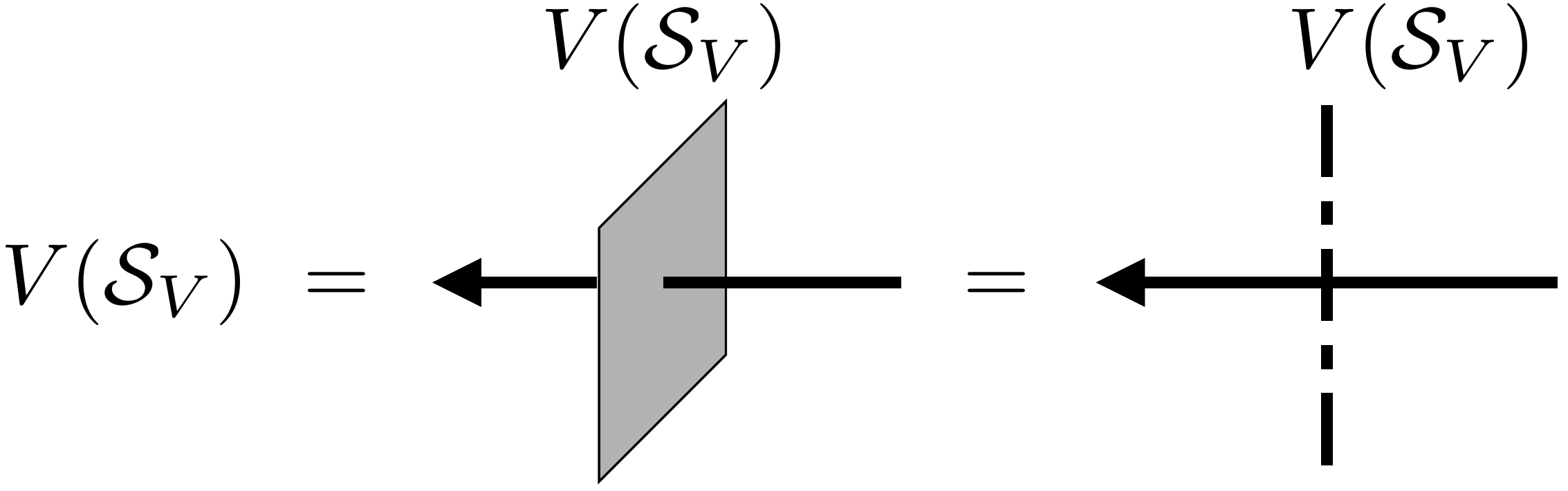} 
\label{200904.2306}
\end{equation}
The right-hand sides of \ers{200904.2305} and \eqref{200904.2306} 
are projected diagrams.

As in the ordinary quantum mechanics, 
the symmetry generators can be unitary representations of
 the symmetry groups which preserve the group structures.
This implies that we can simply replace the diagrammatic 
expressions of the $G_{\rm gl.}$, $H_{\rm gl.}$, and $L_{\rm gl.}$ with the 
 0-, 1-, and 2-form symmetry generators.
For the 0-form symmetry, the unitary representation 
in \er{200904.2057} can be 
described as follows:
\begin{equation}
 \ig[scale=0.33]{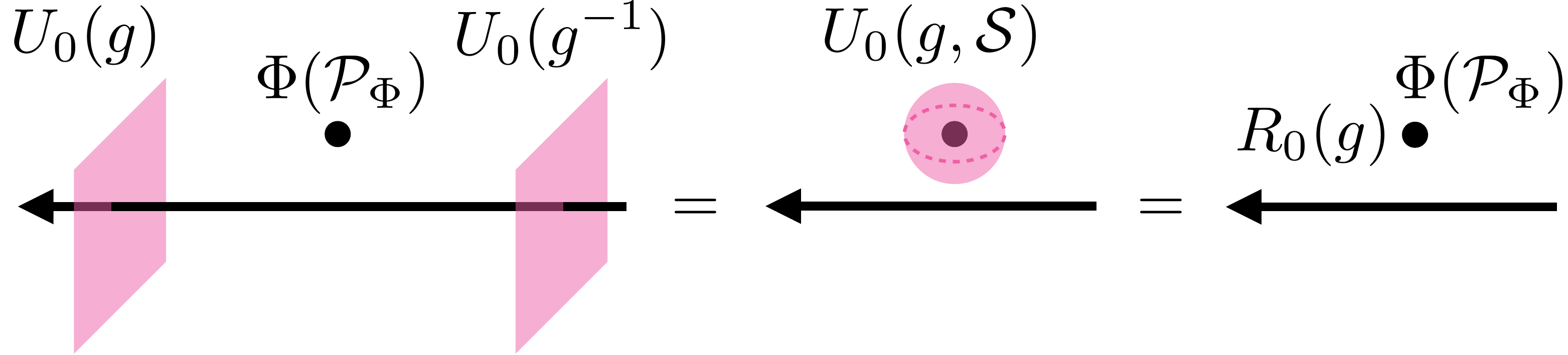} 
\label{200908.1832}
\end{equation}
Here, we have enclosed $\Phi ({\cal P}_\Phi )$
by surfaces $U_0 (g)$ and $U_0(g^{-1})$
parameterized by $g$ and $g^{-1}$, respectively.
Since $U_0 (g)$ are topological, 
we can deform $U_0 (g)$ and $U_0(g^{-1})$ to $U_0 (g,{\cal S})$
given by a surface ${\cal S}$.
The projections of the diagram can be shown as follows:
\begin{equation}
 \ig[scale=0.33]{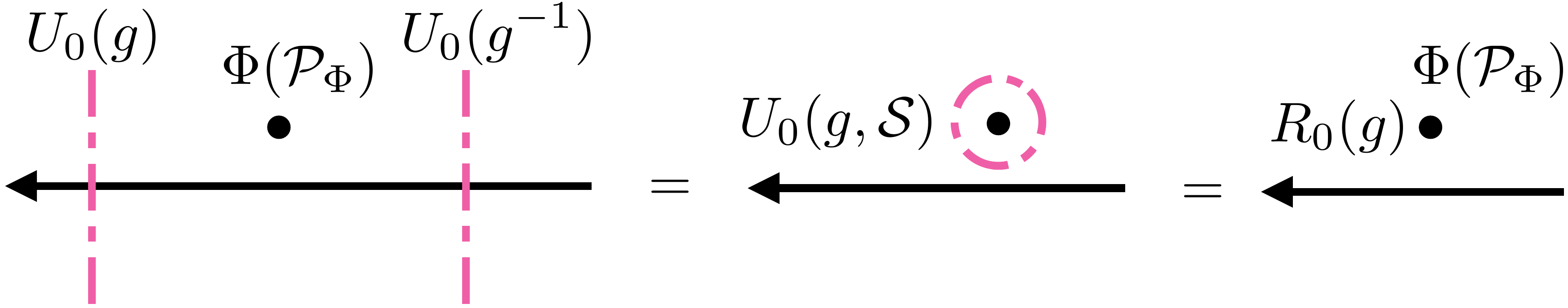} 
\end{equation}

For the 1-form symmetry, 
the diagram for the unitary representation in \er{200904.2058}
and the projection of the diagram are respectively given as follows:
\begin{equation}
 \ig[scale=0.33]{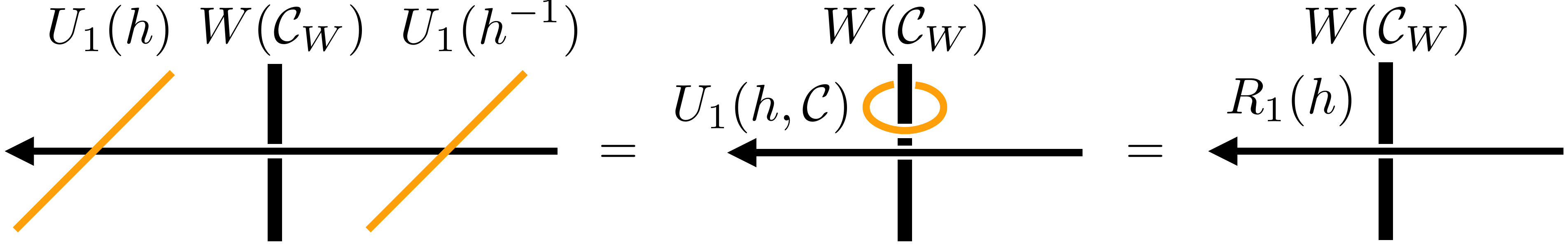}  
\end{equation}
\begin{equation}
 \ig[scale=0.33]{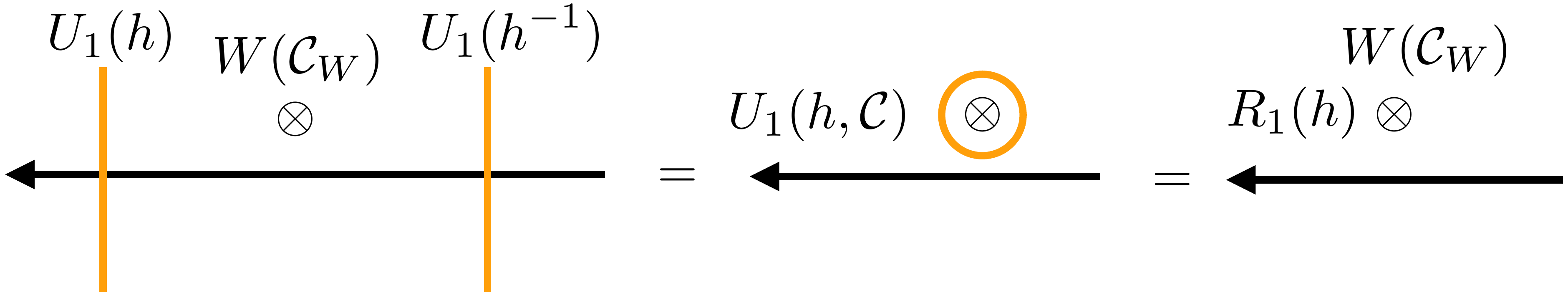}  
\end{equation} 
For the 2-form symmetry, the unitary representation in \er{200904.2059}
can be visualized by the following diagram and its projection:
\begin{equation}
 \ig[scale=0.33]{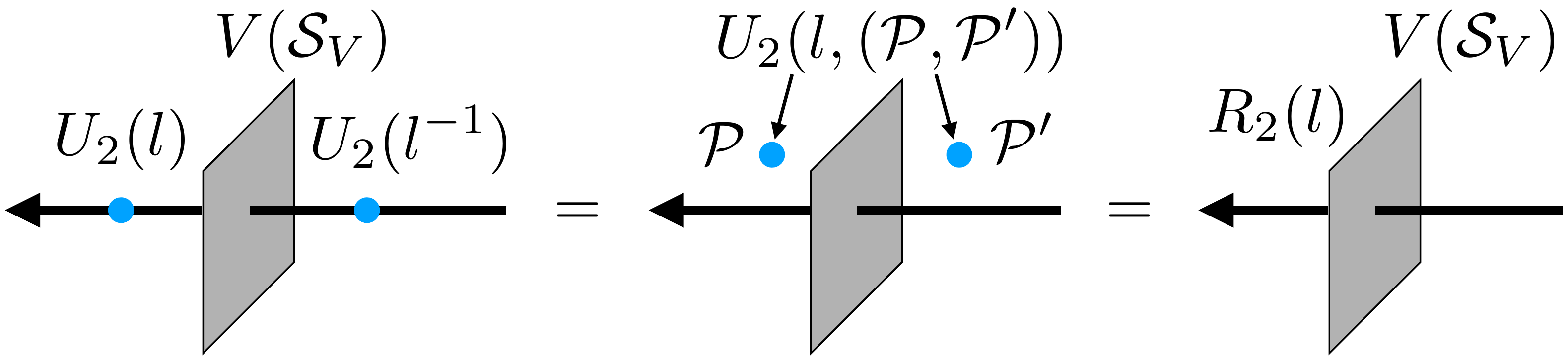}  
\label{200908.2331}
\end{equation}
\begin{equation}
 \ig[scale=0.33]{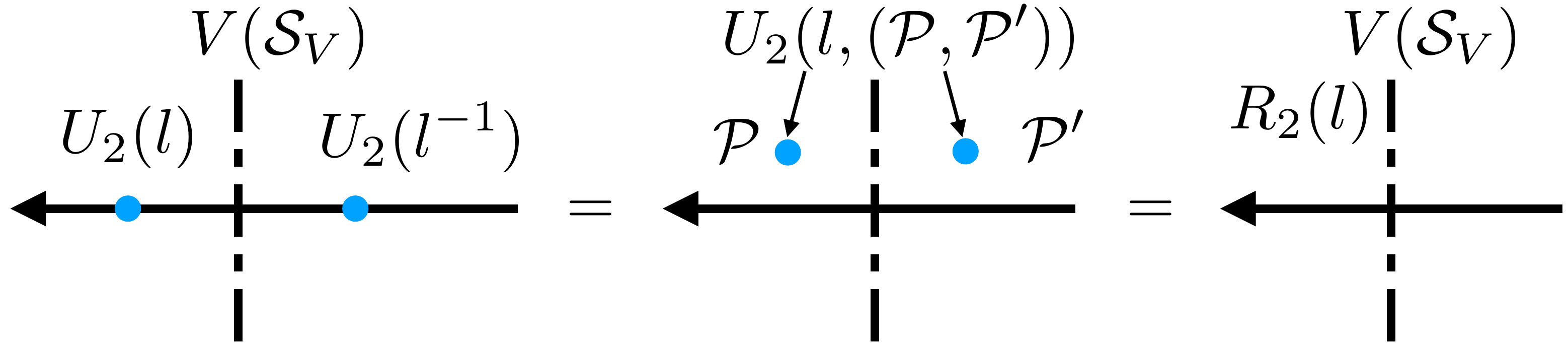}
\label{200908.2332}
\end{equation} 
Here, the ${\cal P}'$ has an orientation opposite to the point ${\cal P}$.

\subsubsection{Diagrammatic expression of actions}

Now we consider the diagrammatic expression of 
the symmetry generators parameterized by the action $\trr$.
For the 0-form symmetry, 
the symmetry generator $U_0 (g\trr g',{\cal S})$ is 
equal to $U_0 (g g' g^{-1},{\cal S})$, and 
the diagram is reduced to \er{200908.1832}.
The symmetry generators $U_1(g\trr h, {\cal C}) $
and $U_2(g \trr l, ({\cal P,P'}))$ can be respectively 
expressed as follows:
\begin{equation}
 \ig[scale=0.33]{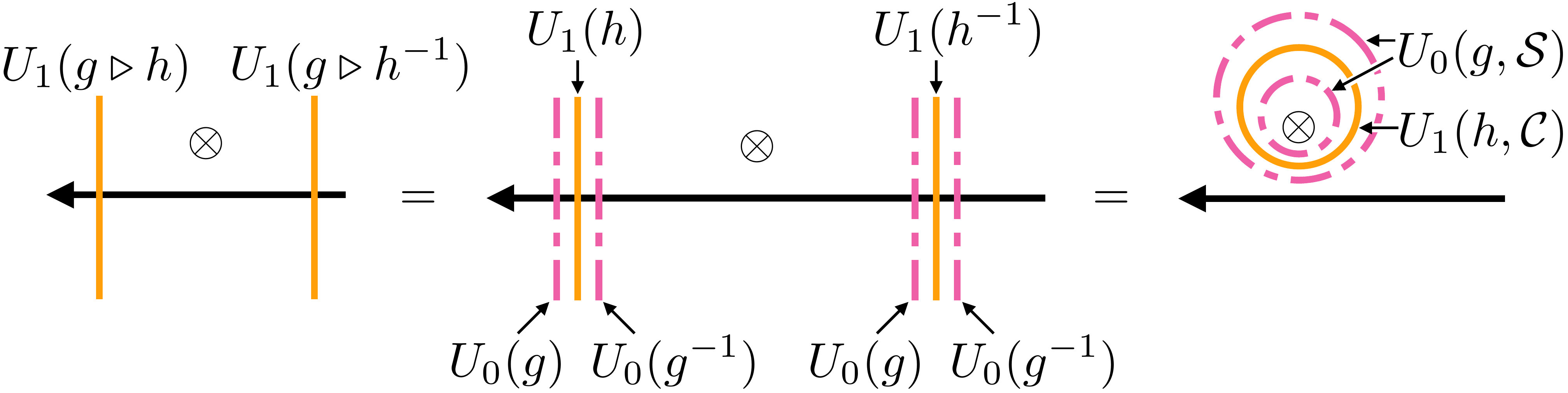} 
\label{200928.2319} 
\end{equation}
\begin{equation}
 \ig[scale=0.33]{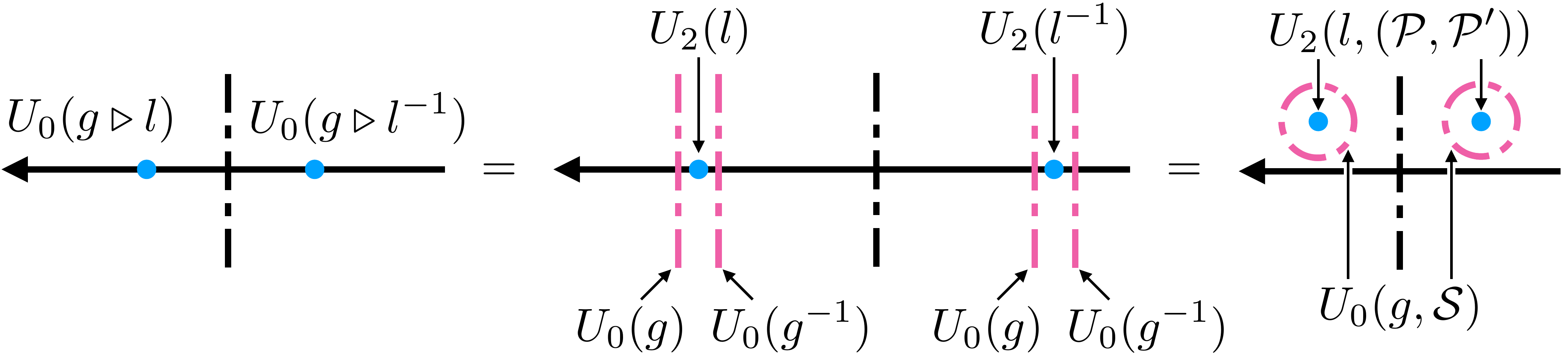}  
\end{equation} 
In the diagram in \er{200928.2319}, 
${\cal S}$ is a set of two cylinders, and the orientation of 
the cylinder inside $U_1(h,{\cal C})$ 
is opposite to the one outside $U_1(h,{\cal C})$.
Note that this diagram can express 
the Witten effect for the axionic domain walls~\cite{Hidaka:2020iaz},
which has been discussed in \er{200805.1726}.

\subsubsection{Diagrammatic expression of Peiffer lifting}
Here, we show the diagrammatic expression of the 
Peiffer lifting for the symmetry generators.
We can describe the symmetry generator
 $U_2 (\{h,h'\}, ({\cal P,P'}))$ 
for the elements $h,h' \in  H_{\rm gl.} = H_{\rm Ab}/ \im \der_1$
by using \ers{200908.2331} or \eqref{200908.2332}.

As we discussed in appendix \ref{peiffer},
the Peiffer lifting is related to the braiding of the 
elements $h,h' \in H_{\rm gl.}$.
In the case of the symmetry generators, 
the Peiffer lifting is related to the linking of 
two 1-form symmetry generators.
Before discussing the symmetry generators, 
we consider the linking of group elements.
For the elements $h, h' \in H_{\rm gl.}$,
we can construct linking of $h$ and $h'$ 
from the Peiffer lifting $\{h,h'\}\{h',h\}$:
\begin{equation}
\ig[scale=0.33]{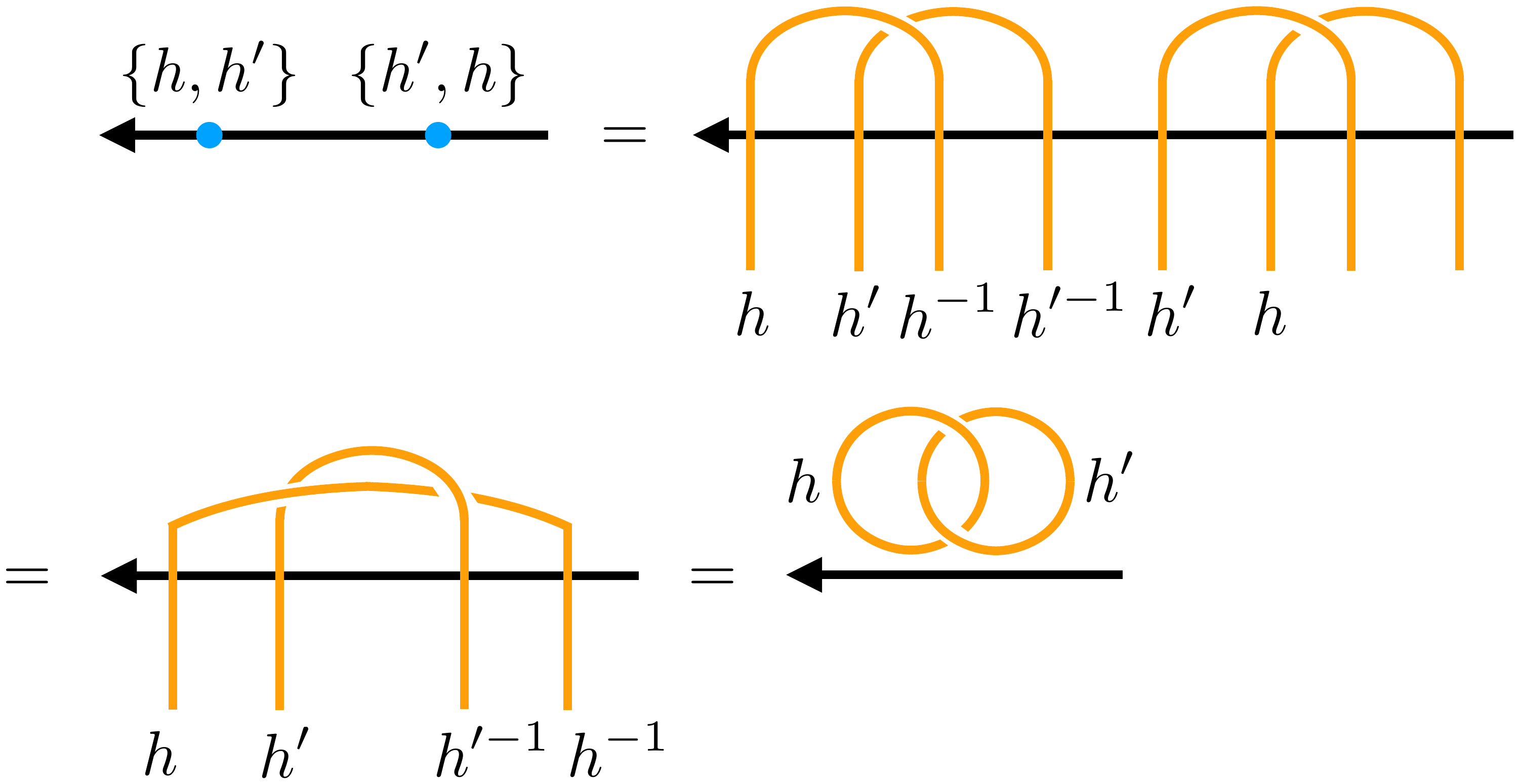}
\label{200909.0041}
\end{equation}
Now, we discuss the linking of the symmetry generators.
By the diagram in \er{200909.0041}, 
we find that the linking of the 1-form symmetry generators 
leads to 2-form symmetry generators, which can act on $V({\cal S}_V)$:
\begin{equation}
\ig[scale=0.33]{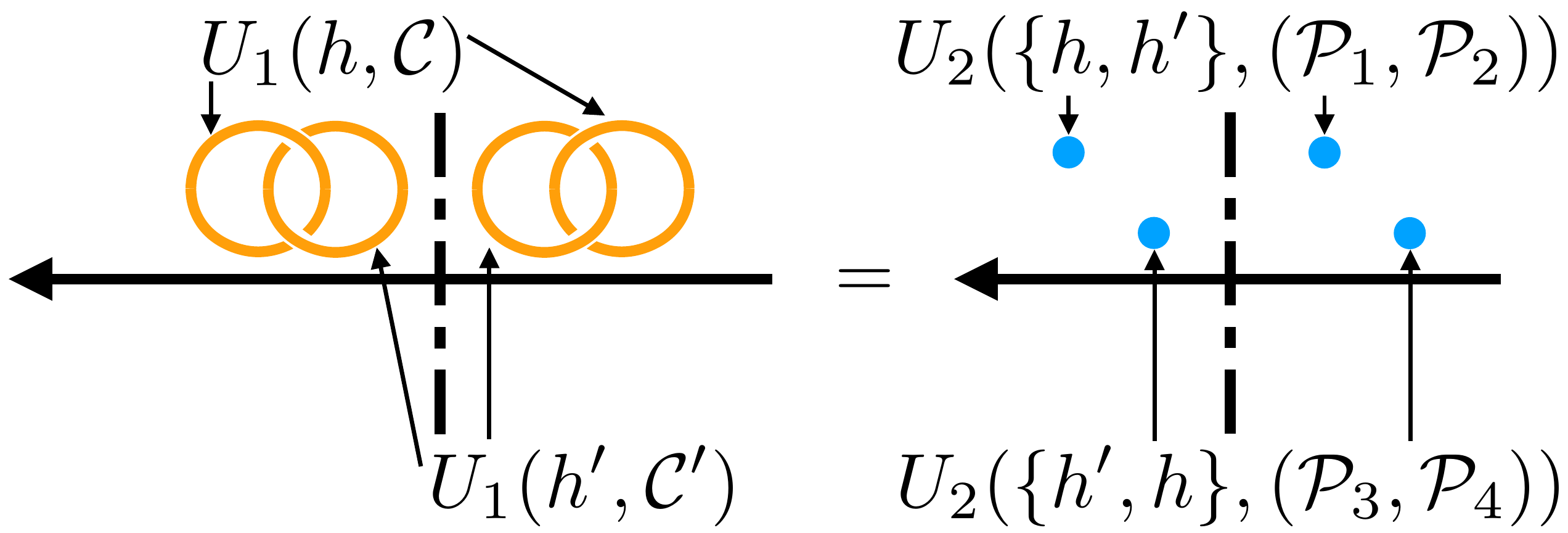}
\label{200909.0052}
\end{equation}
Note that the diagrammatic expression can describe
the anomalous effect around the axionic strings~\cite{Hidaka:2020iaz},
which has been discussed in \er{200807.1426}.

\subsubsection{Symmetry groups parameterizing symmetry generators\label{symgroup}}

Here, 
we show that 
the groups which parameterize the symmetry generators 
are not the groups $G$, $H$, and $L$ themselves 
but subgroups of them.
The subgroups are specified as
 $G/ \im \der_1$, $H_{\rm Ab.}/\im \der_2$ and 
$\ker \der_2$,
where $H_{\rm Ab.}$ is an Abelian part of $\ker \der_1 \subset H$.

We require three assumptions for the symmetry generators.
One is that the non-trivial symmetry generators 
are not boundaries of other objects.
Another is that the non-trivial symmetry generators do 
not have boundaries.
The last one is that the restricted groups also have 
the 3-group structure.

Let us specify the subgroups by using the assumptions.
By the first assumption, the symmetry generators are 
parameterized by the elements of $L$, $H$, 
and $G$ satisfying $\der_2 l = 1_H$ and $\der_1 h = 1_G$.
Therefore, the symmetry groups are reduced to 
$G$, $\ker \der_1 \subset H$, $\ker \der_2 \subset L$.
Note that
we can consistently 
define the actions of $G$ on $G$, $\ker \der_1$ and $\ker \der_2 $,
since the elements $h\in \ker \der_1$ and $l \in \ker \der_2$ 
satisfy $\der_1 (g \trr h ) = g \trr \der_1 h = 1_G$
and $\der_2 (g \trr l) = g \trr \der_2 l =1_H$, and 
$\der_2 (h \trr' l) = h \trr' \der_2 l =1_H$
for $g \in G$.

The axiom of the Peiffer lifting gives us some properties 
of the subsets.
In particular, the subset 
$\ker \der_2$ must be Abelian
by the axiom in \er{200623.0213}.
The last assumption requires 
that the Peiffer lifting $\{h, h'\}$ for
the elements $h,h' \in \ker \der_1$ should belong to 
$\ker \der_2$,
i.e., $\der_2 \{h,h'\}  = 1_H$. 
By the axiom in \er{200622.1536}, we have $h h' = h' h$.
Therefore, the Abelian part of $\ker \der_1$ contributes to 
the higher-form symmetries.
In the following, we denote the Abelian part of 
$\ker \der_1$ as $H_{\rm Ab}$.
Note that these restrictions are consistent with the 
fact that the $p$-form symmetries ($p >0$) must be Abelian (except for 
theories on manifolds with non-trivial topology)~\cite{Gaiotto:2014kfa}.

The second assumption implies that the images of $\der_1$ and 
$\der_2$ cannot parameterize non-trivial symmetry generators. 
The reason can be simply understood by using our diagrams:
the symmetry generators given by $\der_1 h$ and $\der_2 l$ 
can be annihilated by pair creations of $ h$ and $l$.
For example, the annihilation of $U_1 (\der_2 l, {\cal C}) $ can 
be seen as follows:
\begin{equation}
 \ig[scale=0.33]{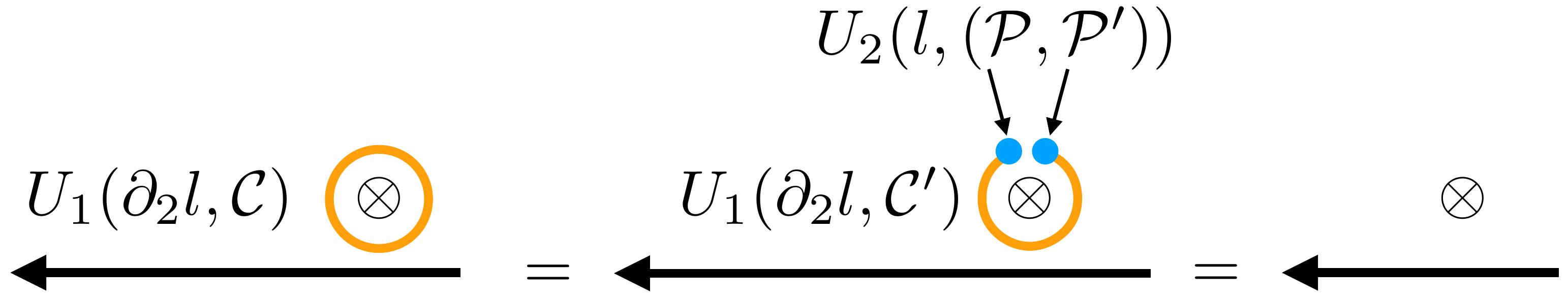}  
\end{equation}
Here, ${\cal C'}$ is a line whose boundaries are ${\cal P}$ 
and ${\cal P'}$.
The groups parameterizing the symmetry generators
are therefore 
$G/ \im \der_1$ and 
$H_{\rm Ab.}/\im \der_2$, $\ker \der_2$.
Note that $G/ \im \der_1$ and $H_{\rm Ab.}/ \im \der_2$ 
are groups, since $\im \der_1$ and $\im \der_2$ are 
normal subgroups of $G$ and $H$, 
$g (\der_1 h) g^{-1} = \der_1 (g\trr h) \in \im \der_1$ 
and $h (\der_2 l) h^{-1} = \der_2 (h\trr' l) \in \im \der_2$ 
for $g \in G$ and $h \in H$,
respectively.

In summary, the 0-, 1-, and 2-form symmetry generators are 
parameterized by $G/ \im \der_1$, $H_{\rm Ab.}/\im \der_2$ and 
$\ker \der_2$.

\providecommand{\href}[2]{#2}

\end{document}